\documentclass{aa}
\usepackage[latin1]{inputenc}
\usepackage{rotating}
\usepackage{fancyhdr}
\usepackage{amsmath}
\usepackage{amssymb}
\usepackage{txfonts}
\usepackage{multirow}
\usepackage{longtable}
\usepackage{color}

\usepackage{graphicx}
\usepackage{url}
\usepackage{xspace}

\usepackage[round]{natbib}
\bibpunct{(}{)}{;}{a}{}{,}
\usepackage{aas_macros} 

\newcommand{\survey}{\textit{400d}}
\newcommand{\megacam}{\textsc{Megacam}}

\newcommand{\tr}{\operatorname{tr}}

\begin{document}

\title{The \textit{400d} Galaxy Cluster Survey weak lensing programme:\\
II: Weak lensing study of seven clusters with MMT/Megacam
\thanks{Observations reported here were obtained at the MMT Observatory, a 
joint facility of the Smithsonian Institution and the University of Arizona.}
\thanks{Reduced and coadded MMT image files are available in electronic form
at the CDS via anonymous ftp to \texttt{cdsarc.u-strasbg.fr} (130.79.128.5).}
}
\titlerunning{The \textit{400d} weak lensing survey II}

\author{Holger Israel\inst{1} 
   \and Thomas Erben\inst{1} 
   \and Thomas H. Reiprich\inst{1}
   \and Alexey Vikhlinin\inst{2}
   \and Craig L. Sarazin\inst{3}
   \and Peter Schneider\inst{1}
    }
\institute{Argelander-Institut f\"ur Astronomie, Auf dem H\"ugel 71, 53121 Bonn,
Germany
 \and Harvard-Smithsonian Center for Astrophysics, 60 Garden Street, Cambridge, MA 02138, USA
 \and Department of Astronomy, University of Virginia, 530 McCormick Road, Charlottesville, VA 22904, USA}

\date{Accepted for publication in Astronomy \& Astrophysics}

\abstract{
Evolution in the mass function of galaxy clusters sensitively traces both the
expansion history of the Universe and cosmological structure formation.
Robust cluster mass determinations are a key ingredient for a reliable 
measurement of this evolution, especially at high redshift. Weak gravitational 
lensing is a promising tool for, on average, unbiased mass estimates.
}{
This weak lensing project aims at measuring reliable weak lensing masses
for a complete X-ray selected sample of $36$ high redshift 
($0.35\!<\!z\!<\!0.9$) clusters. The goal of this paper is to demonstrate the 
robustness of the methodology against commonly encountered problems, including 
pure instrumental effects, the presence of bright ($8$--$9$ mag) stars close to
the cluster centre, ground based measurements of high-$z$ ($z\!\sim\!0.8$) 
clusters, and the presence of massive unrelated structures along the line-sight.
}{
We select a subsample of seven clusters observed with
MMT/\textsc{Megacam}. Instrumental effects are checked in detail by 
cross-comparison with an archival CFHT/\textsc{MegaCam} observation. We derive mass
estimates for seven clusters by modelling the tangential shear with an
NFW profile, in two cases with multiple components to account for
projected structures in the line-of-sight. 
}{
We firmly detect lensing signals from all seven clusters at more than 
$3.5\sigma$ and determine their masses, ranging from
$10^{14}\,M_{\odot}$ to $10^{15}\,M_{\odot}$, despite the presence of
nearby bright stars. We retrieve the lensing signal of more than one
cluster in the CL~1701+6414 field, while apparently observing
CL~1701+6414 through a massive foreground filament. We also find a
multi-peaked shear signal in CL~1641+4001. Shear structures measured
in the MMT and CFHT images of CL~1701+6414 are highly correlated.
}{
We confirm the capability of MMT/\textsc{Megacam} to infer weak lensing masses 
from high-$z$ clusters, demonstrated by the high level of consistency between 
MMT and CFHT results for CL\,1701+6414. 
This shows that, when a sophisticated analysis is applied,
instrumental effects are well under control.
}


\maketitle

\section{Introduction}

Cosmological observables probing different physics
are found to agree within their uncertainties on a $\Lambda$CDM cosmological 
model dominated by Dark Energy and Dark Matter
\citep[e.g.,][]{2008ApJ...686..749K,2010A&A...516A..63S,2011ApJS..192...16L}.
Investigating the unknown physical nature of Dark Energy ranks among the 
foremost questions for cosmologists.
In particular, the presence or absence of evolution in Dark Energy density is
expressed by the equation-of-state parameter $w_{\mathrm{DE}}$.
State-of-the-art measurements 
\citep[e.g.,][]{2009ApJ...692.1060V,2010MNRAS.406.1759M,2011ApJS..192...18K} 
are consistent with $w_{\mathrm{DE}}\!=\!-1$, and hence with Dark Energy being an 
Einsteinian cosmological constant \citep[e.g.][]{2010A&ARv..18..595B}.
The tightest constraints on $w_{\mathrm{DE}}$ can be achieved only by combining
results from different probes showing complementary dependencies on cosmological
parameters.

The evolution of galaxy clusters is understood to be determined by cosmological
parameters through both cosmic expansion and hierarchical structure formation.
Thus, clusters provide information to constrain cosmological parameters that is
complementary to other tests (using e.g., the cosmic microwave background, type
Ia supernovae, or baryonic acoustic oscillations).
Information about the expansion history of the Universe and structure formation
is encoded in the
\emph{cluster mass function} $n(M,z)$; different cosmological models imply a 
different cluster mass function at high $z$, compared to the local mass function
\citep[e.g.][]{1996MNRAS.282..263E,2002ARA&A..40..539R,2005RvMP...77..207V,2005RvMA...18...76S,2006A&A...453L..39R}.
Refining the original analytic model of \citet{1974ApJ...187..425P}, various
fitting formulae have been developed based on numerical simulations
\citep[see][for an overview]{2010MNRAS.402..191P}.
The measured cluster mass function not only gives strong evidence for the
existence of Dark Energy \citep[e.g.][]{2009ApJ...692.1060V} but also adds
valuable information to the joint constraints of cosmological parameters
\citep[for a recent review, see][]{2011arXiv1103.4829A}.

Galaxy cluster cosmology relies on the accurate determination of cluster masses and a
thorough understanding of the different mass proxies in use. Cluster masses are
most commonly inferred from a variety of X-ray observables (X-ray luminosity
$L_{\mathrm{X}}$, gas mass $M_{\mathrm{gas}}$, the quantity 
$Y_{\mathrm{X}}\!=\!T_{\mathrm{X}}M_{\mathrm{gas}}$) or gravitational lensing (both
weak and strong), but as well using the Sunyaev-Zel'dovich effect or the
motions of member galaxies.
For distant galaxy clusters which have low masses owing to the early epoch of
structure formation they represent, the small number of available photons 
prohibits a detailed (spectral) X-ray analyses available for local clusters.
Nevertheless, weighing a large number of high-redshift clusters will yield the 
best constraints on cosmological parameters.
This strategy will be adopted by the upcoming generation of cluster surveys, 
e.g. using \textsl{eROSITA} \citep{2010SPIE.7732E..23P,2011arXiv1111.6587P}.
Hence, scaling relations connecting quantities like $L_{\mathrm{X}}$ or 
$M_{\mathrm{gas}}$ with the total cluster mass will continue to play important
roles and need to be understood and calibrated thoroughly.

Rooted in the thermodynamics of the intracluster medium (ICM), X-ray methods 
rely on assumptions of hydrostatic equilibrium, elemental composition, and, 
to a large extent, sphericity of the cluster's gravitational potential well
\citep{1988xrec.book.....S,2010A&ARv..18..127B}.
Weak gravitational lensing \citep[WL; e.g.][]{2006glsw.book.....S} offers an 
alternative avenue for determining cluster masses, which is completely 
independent of these assumptions, directly mapping the projected mass
distribution of matter, Dark and luminous.

Merging clusters deviate strongly from
thermal and hydrostatic equilibrium, with a significant amount of the internal
energy being present as kinetic energy of bulk motions or turbulent processes, 
e.g.\ merger shocks. Thus, the expectation from numerical simulations is that X-ray
masses for merging systems might be biased after a significant merger, with a
relaxation timescale of $\mathcal{O}(1\,\mathrm{Gyr})$. Simulations agree with
the expectation from the hierarchical structure formation picture that mergers
are more frequent at higher redshifts than in the local Universe
\citep[e.g.][]{2005APh....24..316C}.
There is no consensus between different simulations yet which of several
suggested physical effects dominates after the
phase in which a disturbed morphology can be seen
\citep[e.g.,][]{2006ApJ...650..128K,2007ApJ...668....1N,2010ApJ...715.1508S}.
Most importantly, bulk motions induce non-thermal pressure, 
providing support for the gas against gravity, thus possibly leading the
hydrostatic mass to underestimate the true mass by $5$--$20$\%
even in \emph{relaxed} clusters
\citep{2006MNRAS.369.2013R,2007ApJ...655...98N,2010A&A...514A..93M}.

Therefore, studying scaling relations of X-ray observables with weak lensing
masses has become an important ingredient in refining cluster masses from X-ray observations
\citep[e.g.,][]{2008A&A...482..451Z,2010ApJ...711.1033Z,2010A&A...514A..93M}.
Relative uncertainties of the individual WL cluster masses are higher than 
those from X-rays, largely due to intrinsic shape noise.
But the power of weak lensing comes through the statistical analysis of 
$M_{\mathrm{wl}}/M_{\mathrm{X}}$ for the whole sample, under the assumption that
WL mass estimates are, on average, unbiased.
This means, while WL mass estimates for \emph{individual clusters} are subject
to an error due to the projection of filaments or voids along the line-of-sight,
the stochastic nature of these errors makes them cancel out when averaging 
over a well-defined cluster sample.
Statistical comparisons to X-ray masses 
\citep[e.g.,][]{2010A&A...514A..93M} help us to investigate WL systematic 
uncertainties, i.e.\ triaxiality \citep{2009MNRAS.396..315C} 
and projection of unrelated LSS \citep{2003MNRAS.339.1155H}, to which X-ray 
observables are far less sensitive.

This article presents the second part of a series on weak lensing analyses
following up the 400 Square Degree Galaxy Cluster (\emph{400d}) Survey, 
initiated in \citet[hereafter Paper I]{2010A&A...520A..58I}.
The \emph{400d} Survey presents a flux-limited sample of galaxy clusters
detected serendipitously in a re-analysis of all suitable \textsl{ROSAT} PSPC
pointings \citep{2007ApJS..172..561B}. 
From the resulting catalogue, \citet[V09]{2009ApJ...692.1033V} drew the 
\emph{cosmological subsample} of $36$ X-ray luminous and distant clusters, 
for which high-quality \textsc{Chandra} X-ray observations were obtained and 
analysed.
The \textsc{Chandra}-based cluster mass function resulting from the
\emph{Chandra Cluster Cosmology Project} was published by 
V09, for the complete redshift range of 
$0.35\!\leq\!z\!<\!0.90$ spanned by the clusters in the \emph{cosmological 
subsample}, as well as divided into three redshift bins.
Building on this mass function, \citet{2009ApJ...692.1060V} constrained
cosmological parameters, in particular $w_{\mathrm{DE}}$.

Determining accurate weak lensing masses for the distant clusters in the 
\emph{400d} cosmological subsample opens the way to observationally test the
assumptions \citet{2009ApJ...692.1033V,2009ApJ...692.1060V} make for the 
scaling relations and their evolution.
Put briefly, the WL follow-up of the \survey ~cosmological sample
clusters provides us with a control experiment for the resulting X-ray mass 
function. 
With $36$ clusters, the \emph{400d} WL sample ranks among the largest
\emph{complete} high-$z$ WL samples. 

In Paper~I, we presented the results of our feasibility study, performing a 
detailed lensing and multi-method analysis of 
CL\,0030+2618. In particular, we showed the \megacam ~instrument to be well 
suited for WL studies.
As the next step of the project, we investigate seven further clusters from
our sample, all of which were also observed with \megacam ~at MMT.
The resulting WL mass determination and the status after $8$ out of $36$
clusters have been analysed are the subjects of this paper. 

We consistently assume a $\Lambda$CDM cosmology specified by the
dimensionless Hubble parameter $h\!=\!0.72$ and matter and Dark Energy density
parameters of $\Omega_{\mathrm{m}}\!=\!0.30$ and $\Omega_{\Lambda}\!=\!0.70$. 

This paper is organised as follows: After giving a short overview on our 
data set and its reduction in Sect.~\ref{sec:sec2}, we give salient details of 
the WL analysis methods we used in Sect.~\ref{sec:sec3}. 
In Sect.~\ref{sec:sec4}, we take a closer look at two clusters which show a
more complicated shear morphology and devise a two-cluster shear model.
Comparing our MMT results to a CFHT weak lensing analysis of one of our
clusters, we once more prove MMT weak lensing to be reliable and provide an
external calibration (Sect.~\ref{sec:cfhtsec}). In Sect.~\ref{sec:accuracy}, 
we provide details of the error analysis for our main 
results, the cluster masses, which are then discussed in Sect.~\ref{sec:disc}. 
Section~\ref{sec:summary} presents our summary and conclusion.

\section{Methodology} \label{sec:sec2} 

\subsection{MMT/\megacam ~data for the \emph{400d} WL survey} \label{sec:megadata}

\begin{table*}
 \centering
 \caption{Specifications of the data sets for all eight clusters analysed so 
  far. For each cluster and filter, the dates of observation, total \megacam 
  ~exposure time $T_{\mathrm{exp}}^{\mathrm{ini}}$, usable final exposure time 
  $T_{\mathrm{exp}}^{\mathrm{fin}}$, seeing, and $5\sigma$ limiting magnitude
  (Eq.~2 in Paper~I) for the final image stack are given.
  The last column refers to a direct (D) or indirect (I)
  photometric calibration (PhC, Sect.~\ref{sec:datared}). 
  CL\,0030+2618 is the cluster analysed in Paper I.}
 \label{tab:data}
 \begin{center} 
 \begin{tabular}{cccccccc} \hline\hline
 Cluster & Filter & Observation Dates & $T_{\mathrm{exp}}^{\mathrm{ini}} [s]$ & 
 $T_{\mathrm{exp}}^{\mathrm{fin}} [s]$ & Seeing & $m_{\mathrm{lim}}$ & PhC\\ \hline
 CL\,0030+2618 & $r'$ & 2004-10-06/07 & $15300$ & $6600$ & $0\farcs82$ & $25.9$ & I\\
 & $g'$ & 2005-10-30/31,11-01 & $9150$ & $7950$ & $0\farcs87$ & $26.8$ & D\\
 & $i'$ & 2005-10-31 & $6000$ & $5700$ & $1\farcs03$ & $25.1$ & D\\ \hline
 CL\,0159+0030 & $r'$ & 2005-10-30/31,11-01 & $9900$ & $3600$ & $0\farcs85$ & $25.7$ & D \\ 
 & $g'$ & 2005-11-01, & $6000$ & $4800$ & $1\farcs05$ & $27.7$ & D\\
 & $i'$ & 2005-10-31,11-01 & $8100$ & $5700$ & $1\farcs14$ & $25.0$ & D\\ \hline
 CL\,0230+1836 & $r'$ & 2004-10-06/07; 2005-11-08 & $9600$ & $2700$ & $0\farcs68$ & $25.1$ & I \\
 & $g'$ & 2005-11-08 & $6000$ & $4200$ & $0\farcs80$ & $27.2$ & I\\
 & $i'$ & 2005-10-31,11-01/08 & $9600$ & $3600$ & $0\farcs98$ & $24.7$ & D\\ \hline
 CL\,0809+2811 & $r'$ & 2005-11-08; 2008-01-09& $9300$ & $3000$ & $0\farcs72$ & $25.4$ & D \\
 & $g'$ & 2005-10-31/11-08 & $6000$ & $3600$ & $1\farcs04$ & $26.3$ & D\\
 & $i'$ & 2005-10-31/11-01 & $7500$ & $5700$ & $0\farcs82$ & $26.1$ & D\\ \hline
 CL\,1357+6232 & $r'$ & 2005-06-07 & $7200$ & $2700$ & $0\farcs90$ & $25.4$ & D \\ \hline
 CL\,1416+4446 & $r'$ & 2005-06-08 & $7500$ & $4200$ & $0\farcs81$ & $25.8$ & D \\ \hline
 CL\,1641+4001 & $r'$ & 2005-06-07 & $8100$ & $6900$ & $0\farcs91$ & $26.0$ & D \\ \hline
 CL\,1701+6414 & $r'$ & 2005-06-08 & $7500$ & $6000$ & $0\farcs89$ & $25.8$ & D \\ \hline\hline
 \end{tabular}
 \end{center}
\end{table*}
 
Table~\ref{tab:data} summarises the observations of the eight 
$\delta\!>\!0\degr$ galaxy clusters with right ascensions\footnote{Their 
 J2000 coordinates are given by the designations in Table~\ref{tab:data}.}
$0^{\mathrm{h}}\!<\!\alpha\!<\!8^{\mathrm{h}}30^{\mathrm{m}}$ and 
$13^{\mathrm{h}}30^{\mathrm{m}}\!<\!\alpha\!<\!24^{\mathrm{h}}$ for which MMT/\megacam
~observations in the lensing ($r'$-) band have been completed.
As CL\,0030+2618 was studied in detail in Paper~I, this work focusses on the
remaining seven clusters.
Following the observation strategy described in Paper~I, with nominal exposures 
of $T^{\mathrm{nom}}\!=\!(7500\,\mathrm{s},6000\,\mathrm{s},4500\,\mathrm{s})$
in $(g'r'i')$, these seven clusters 
were observed in the four out of five MMT observing runs performed for the
\emph{400d} WL survey in which weather conditions permitted usable observations
during at least parts of the scheduled time.

\megacam ~\citep{2000fdso.conf...11M}, then located at the Fred Lawrence Whipple
Observatory's $6.5$~m MMT telescope, is a high-resolution 
($0.08\arcsec\,\mbox{px}^{-1}$), wide-field ($\sim\!24\arcmin\times 24\arcmin$
field-of-view) camera, consisting of a $4\times 9$ CCD mosaic.

The four ``winter'' clusters CL\,0030+2618, CL\,0159+0030, CL\,0230+1836, and
CL\,0809+2811 have completed observations in the $g'r'i'$ filters, 
while due to scheduling constraints, only the $r'$-imaging could be completed 
for the ``summer'' clusters CL\,1357+6232, CL\,1416+4446, CL\,1641+4001, and 
CL\,1701+6414. 
Therefore, a different strategy has to be adopted for parts of
the data reduction (Sect.~\ref{sec:ksb}) and the background 
source selection (Sect.~\ref{sec:bg7}) for these \emph{single-band} clusters
compared to \emph{three-band} clusters.

As indicated in Table~\ref{tab:data}, some clusters were observed in the 
same filter in more than one observing run. 
Using the data reduction described in Sect.~\ref{sec:datared}, we produced 
\emph{coadded} (stacked) images, for which net exposure times 
$T_{\mathrm{exp}}^{\mathrm{fin}}$, seeing, $5\sigma$ limiting magnitudes, and 
photometric calibration method (Sect.~\ref{sec:datared}) are given in the four
last columns of Table~\ref{tab:data}.

The most striking fact to note 
are the drastic reductions from the initial raw data exposure times 
$T_{\mathrm{exp}}^{\mathrm{ini}}$ to the $T_{\mathrm{exp}}^{\mathrm{fin}}$ used 
in the coadded images. In a number of cases, the required seeing in the coadded
image of $\lesssim\!1\arcsec$ in the lensing band and $\lesssim\!1\farcs2$ in 
the other bands could only be achieved by removing images such that 
$T_{\mathrm{exp}}^{\mathrm{fin}}\!<\!T^{\mathrm{nom}}$.
As this inevitably reduces the limiting
magnitude (Eq.~2 in Paper~I), the final stacks represent a compromise
between seeing and depth, aiming at an optimal WL signal.
Similarly, compromises had to be made between maintaining a low level of
anisotropy (Sect.~\ref{sec:ksb}) in the point spread function (PSF) and 
limiting magnitude. 
The ramifications of the heterogeneous data quality and the -- in some cases --
shallow exposure times, for which the good overall $r'$-band seeing could be 
obtained, will be addressed at several occasions in this article.

\subsection{Data reduction and calibration} \label{sec:datared}

The data reduction for the \emph{400d} WL survey has been described in 
detail in Paper~I. Therefore, we give only a brief recapitulation here and
refer the interested reader to Paper~I.

\begin{figure*}
 \includegraphics[width=5.8cm]{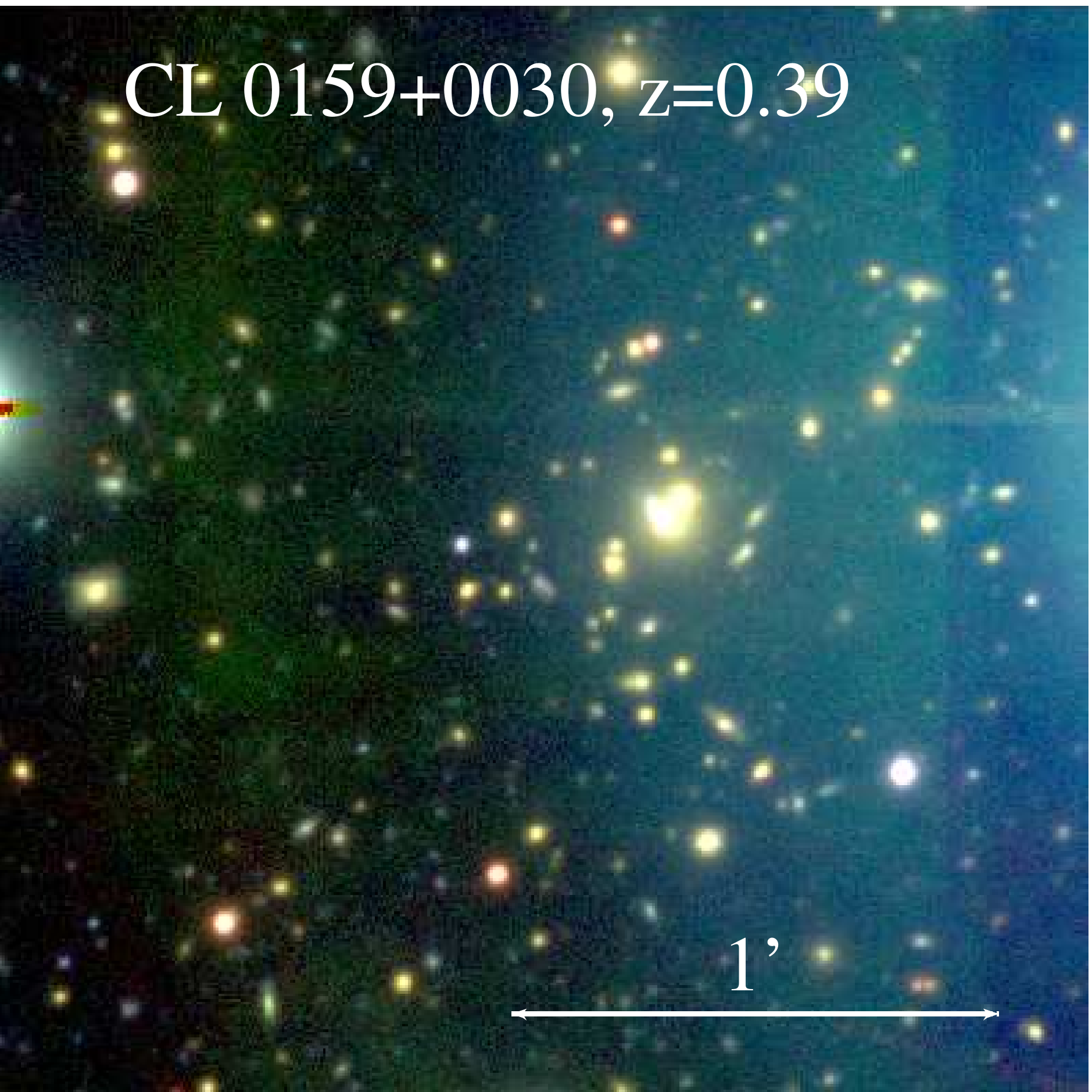}
 \includegraphics[width=5.8cm]{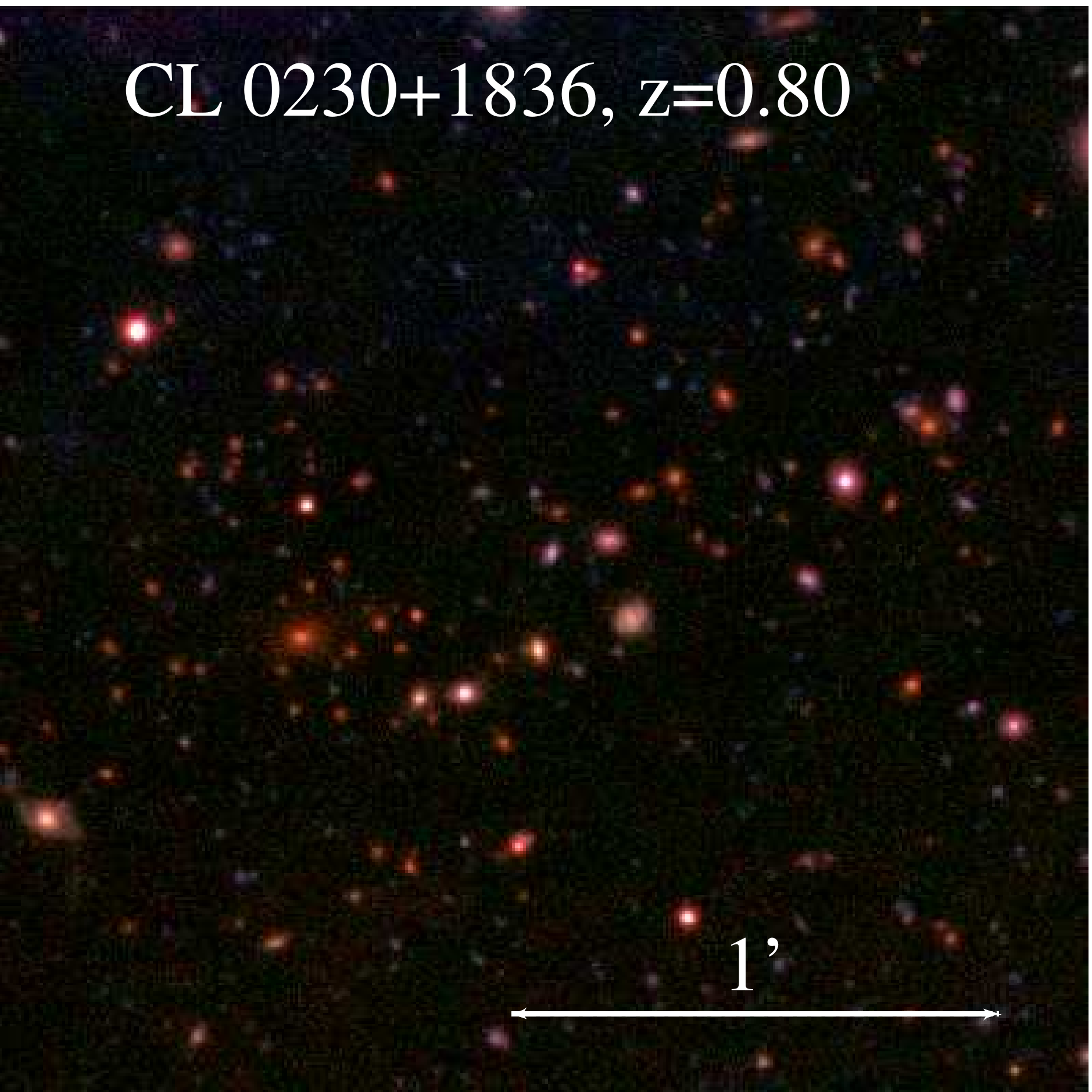}
 \includegraphics[width=5.8cm]{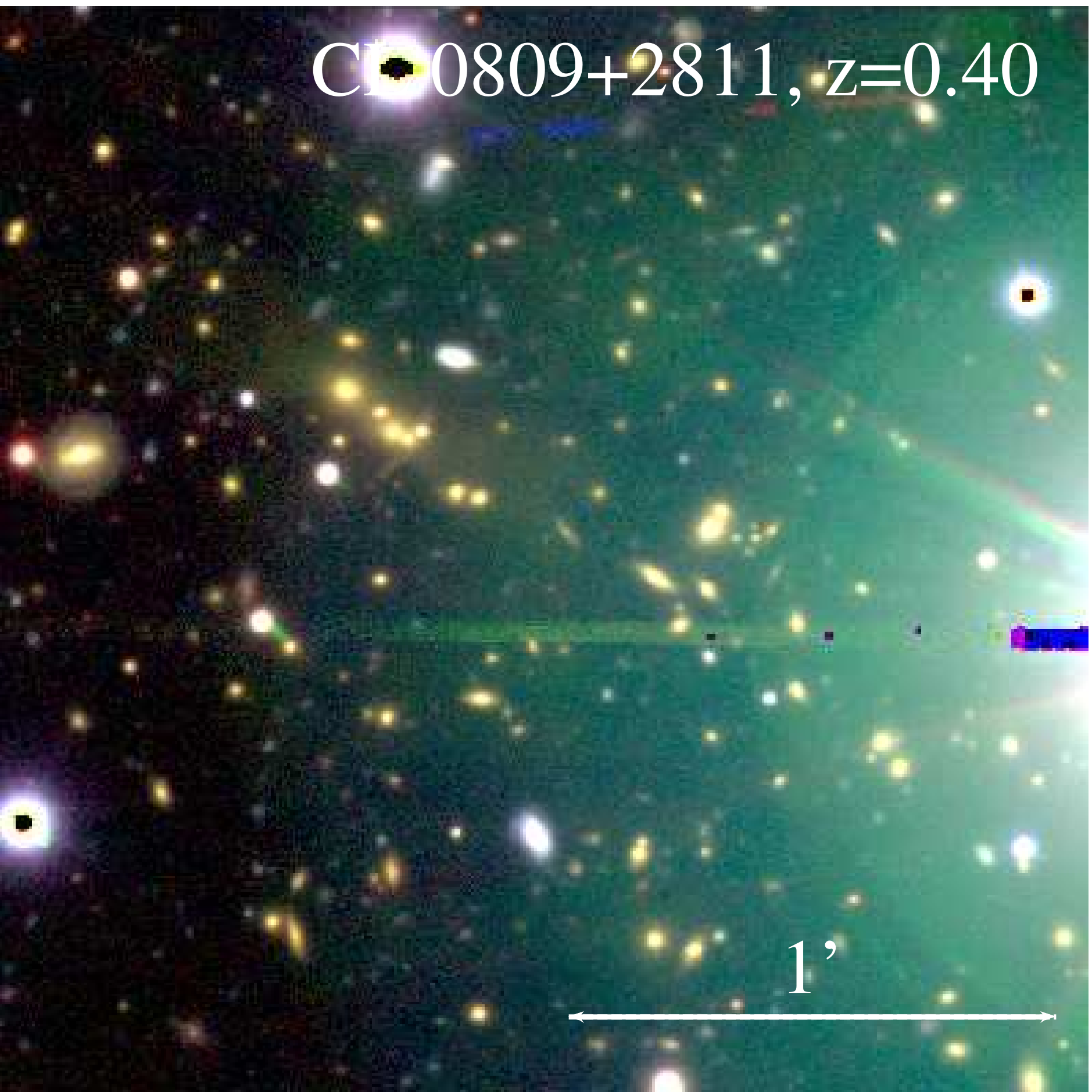}
 \includegraphics[width=5.8cm]{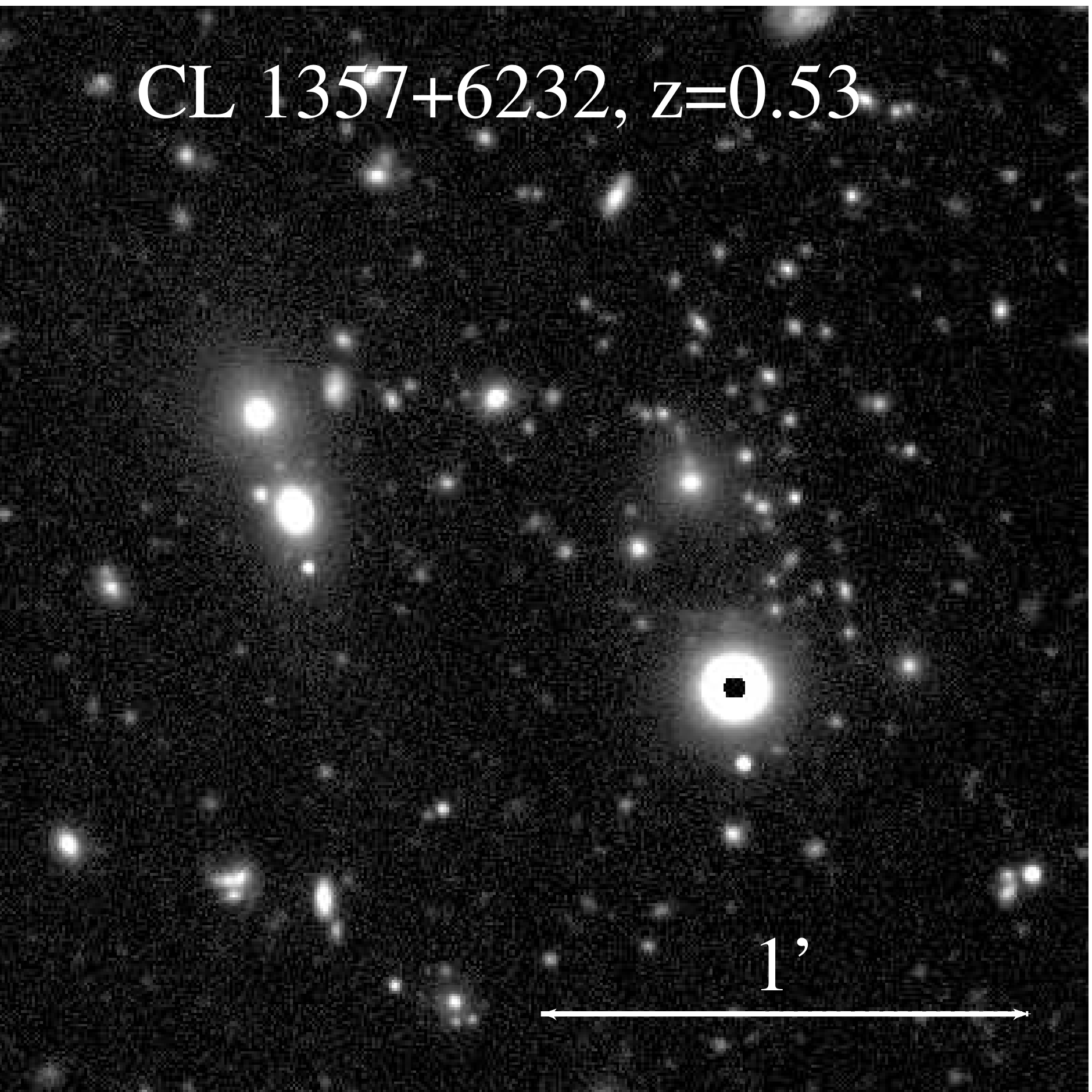}
 \includegraphics[width=5.8cm]{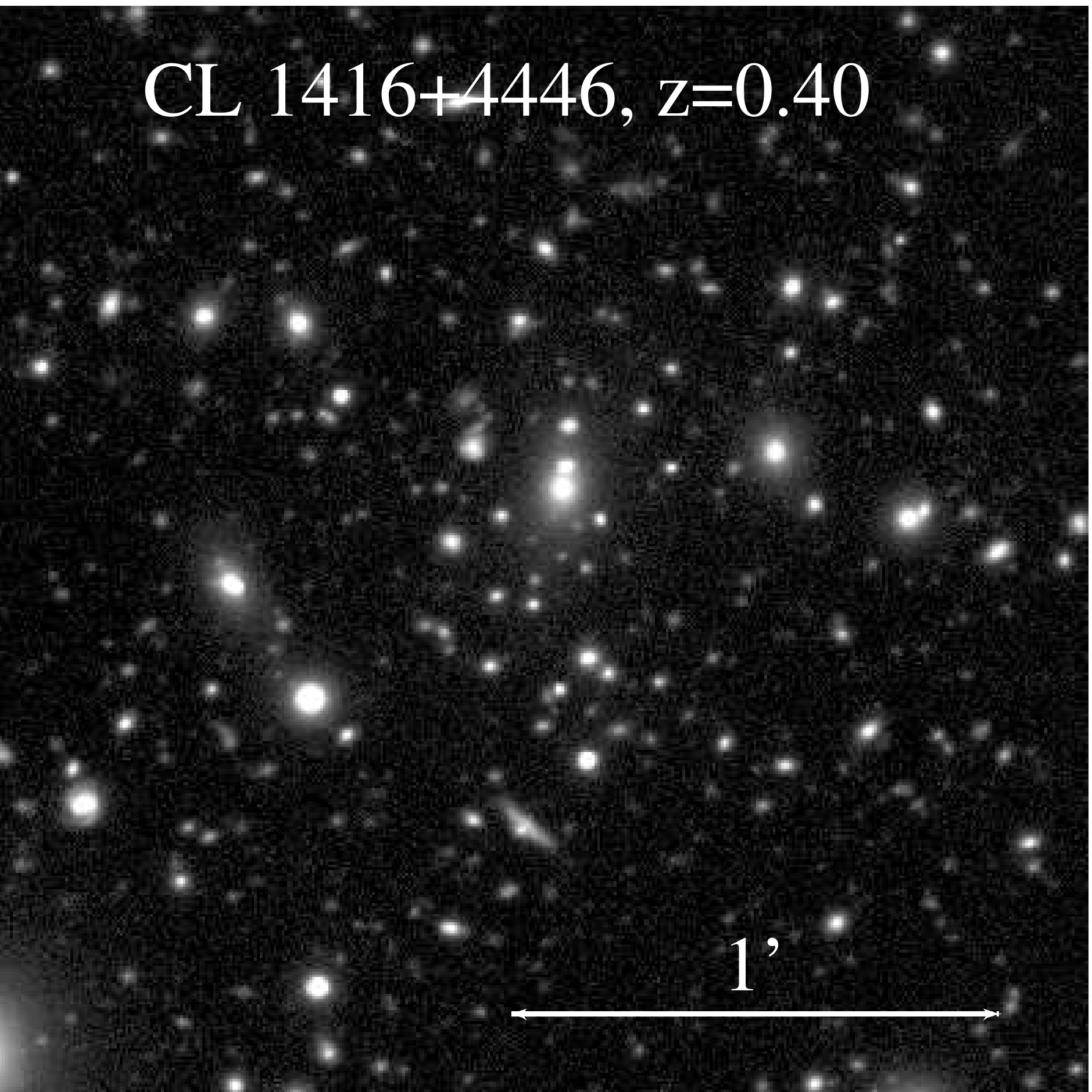}
 \includegraphics[width=5.8cm]{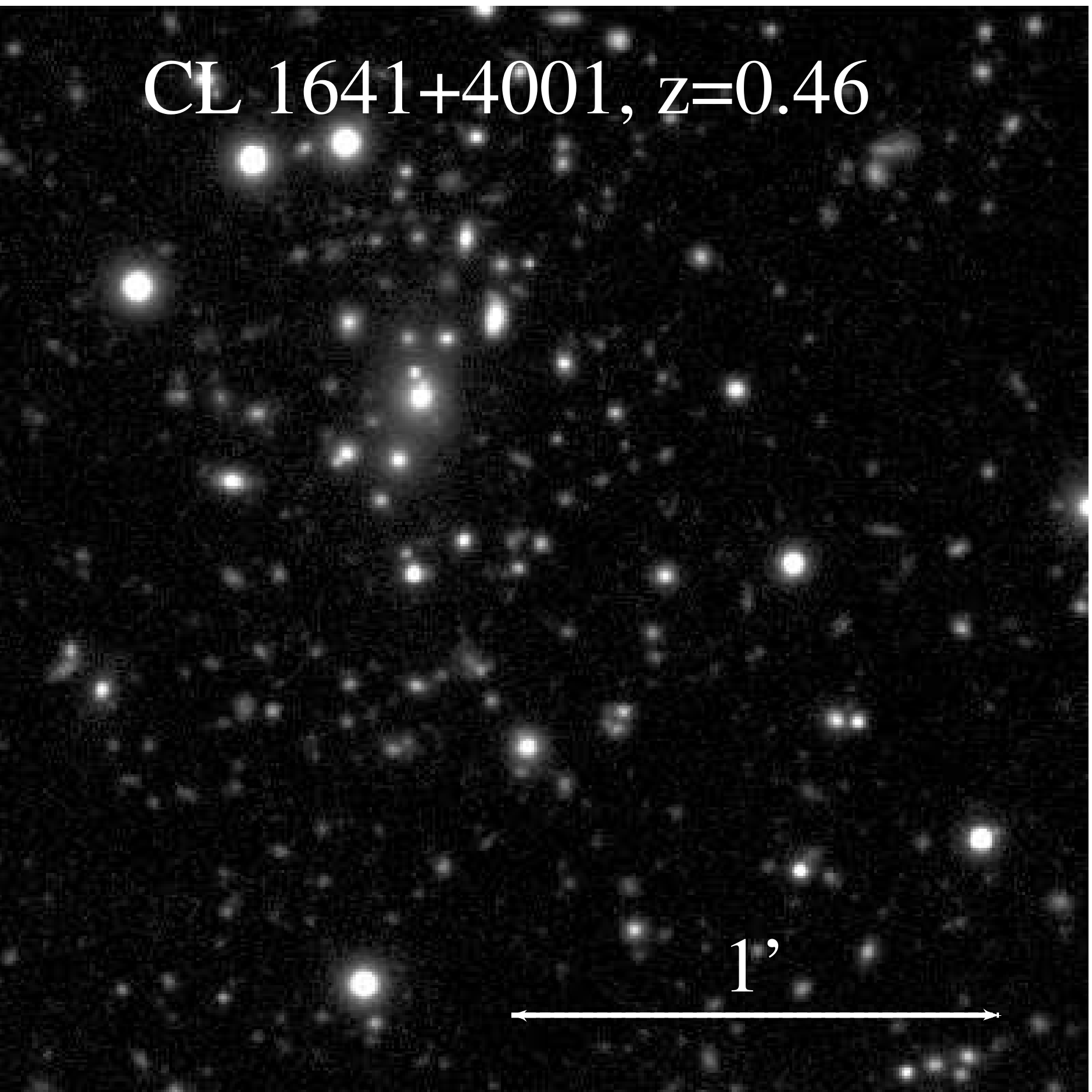}
 \includegraphics[width=5.8cm]{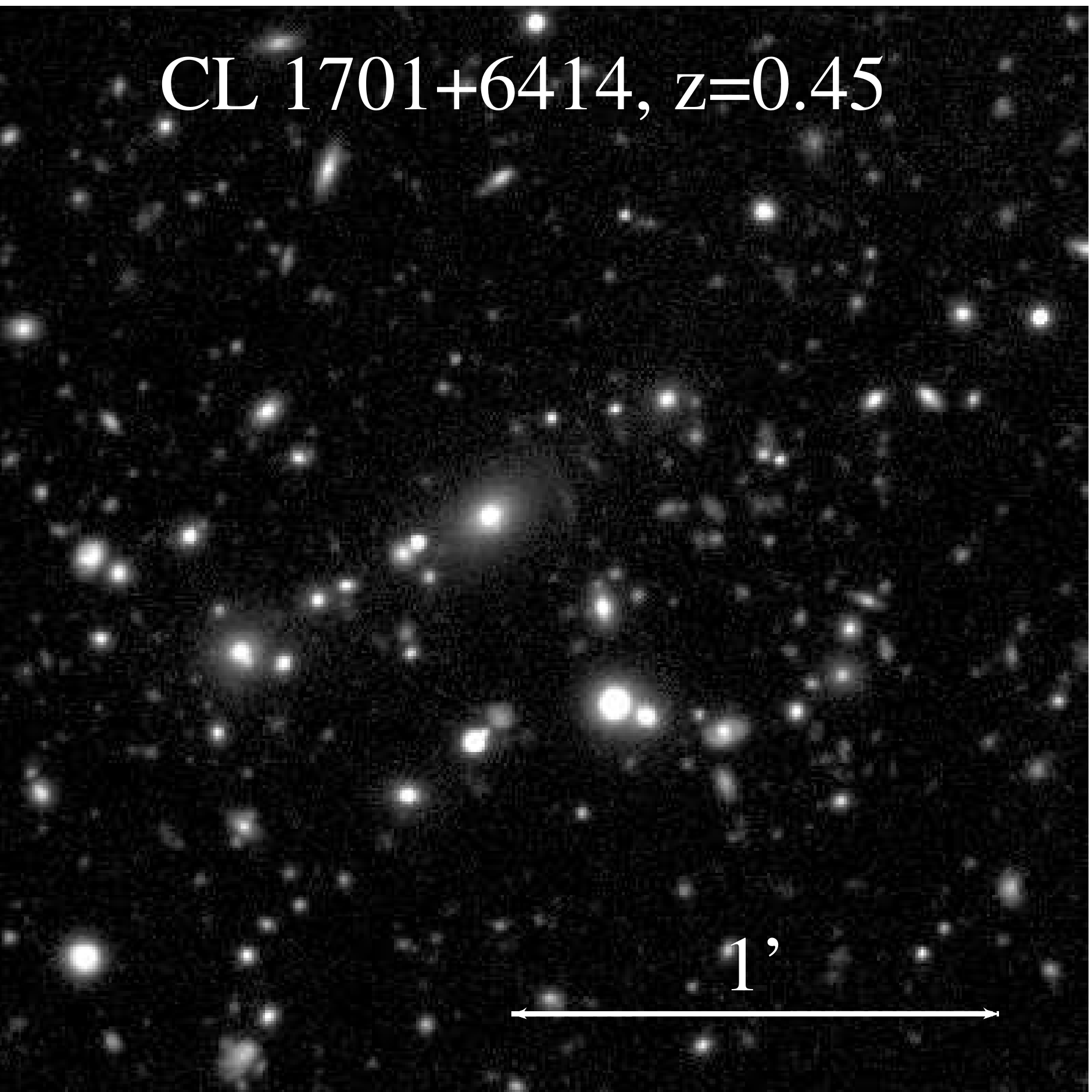}
 \hspace{3.3mm}
 \includegraphics[width=5.8cm]{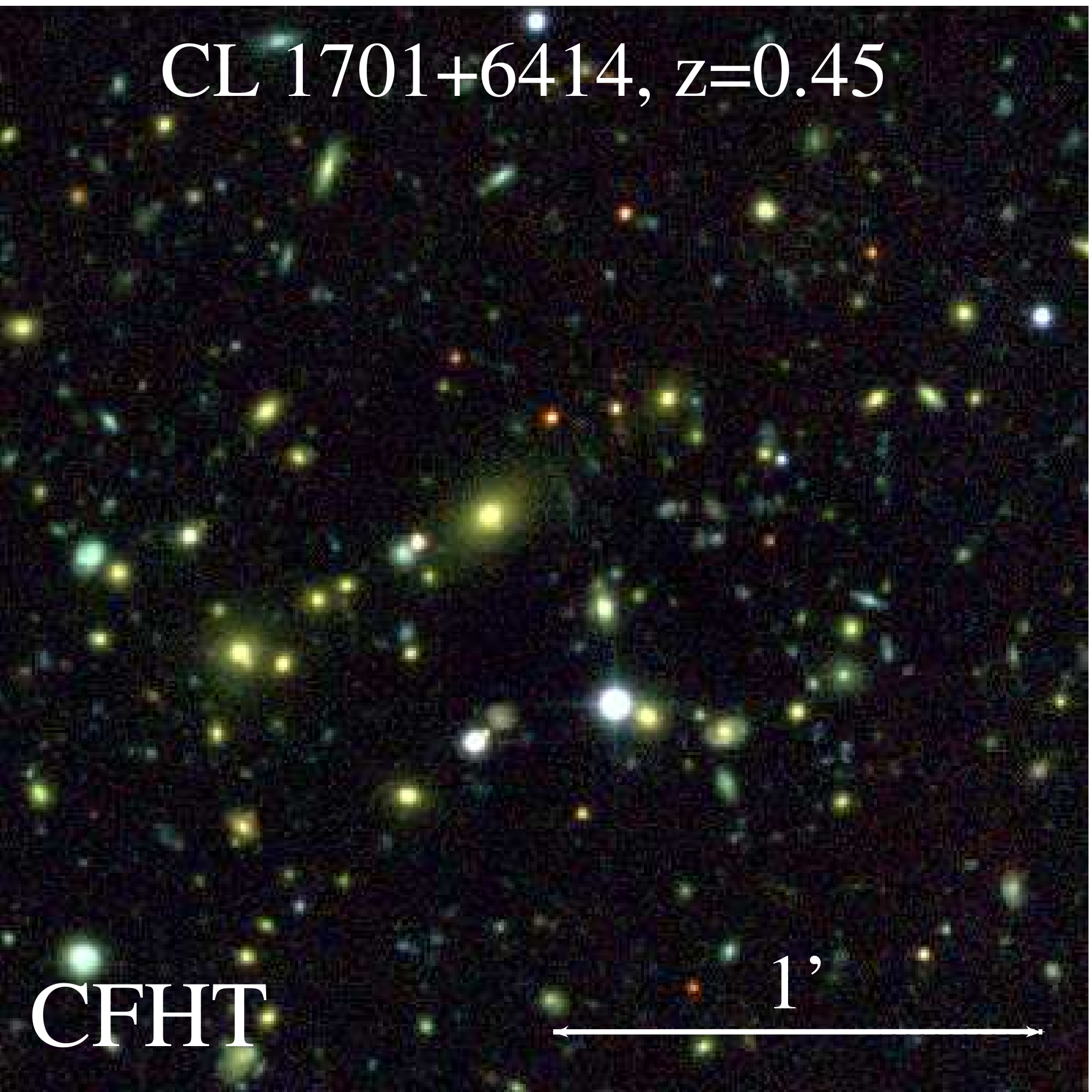}
 \hspace{3.3mm}
 \includegraphics[width=5.8cm]{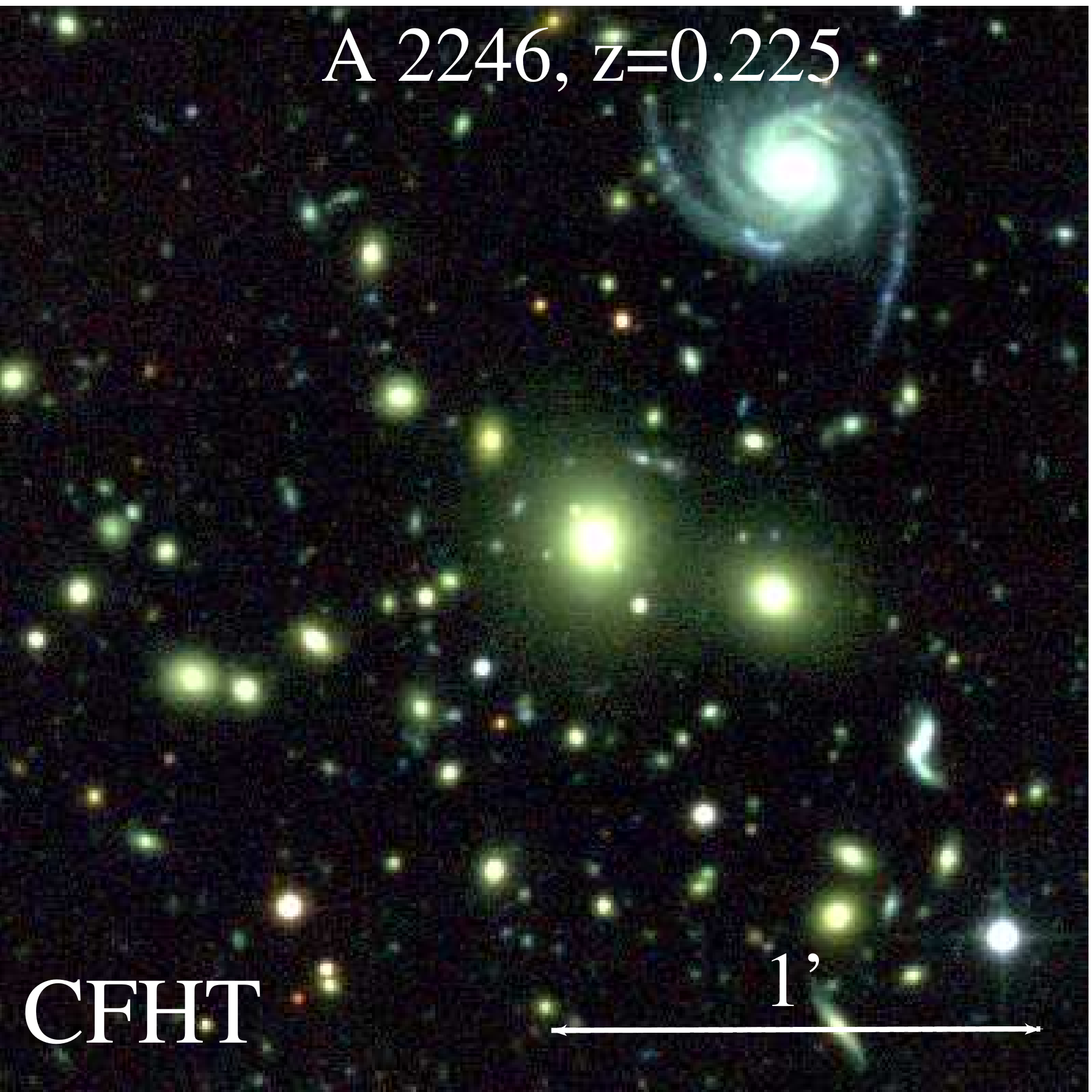}
 \caption{Clusters discussed in this paper. We show pseudo-colour images
 for the cases where colour is available, using the MMT $g'r'i'$ bands. For the
 CL\,1701+6414 field, we also show pseudo-colour images using the CFHT $g'r'i'$
 bands, both for CL\,1701+6414 and A~2246. We choose the \textsc{rosat}
 cluster coordinates as centre of the images. Note the variable background due
 to bright stars near CL\,0159+0030 and CL\,0809+2811.}
 \label{fig:zooms}
\end{figure*}

The first stage, including all tasks of elementary data reduction (de-biasing,
flatfielding, de-fringing, construction of weight images, astrometry, relative
photometry, and coaddition) are performed using the \texttt{THELI} pipeline for
optical data reduction introduced by \citet{2005AN....326..432E} and adapted to
MMT/\megacam ~in Paper~I.
Generally, we achieve a high level of homogeneity in the noise backgrounds of
coadded images, with our pipeline effectively correcting the position-dependent
transmissivity of \megacam ~filters (Appendix~\ref{sec:augias}).
The additive stray-light from very bright stars 
is not removed by \texttt{THELI}, but regions in the image in which source 
counts deviate significantly from the mean are masked as unreliable using the
algorithm described by \citet{2007A&A...470..821D}. The mask images we produce
in the second stage of data reduction for each coadded image also contain masks
for saturated stars (cf.\ Paper I) and a few manually inserted masks for, e.g.,
asteroid trails\footnote{The more common satellite streaks are masked already
during the basic data reduction, prior to coaddition.}.
The first seven panels of Fig.~\ref{fig:zooms} present the central regions 
of our clusters as observed with MMT/\textsc{megacam}. For the three-band
clusters, we prepared pseudo-colour images using the $g'r'i'$ coadded images.

Applying the method of \citet{2006A&A...452.1121H}, we computed absolute
photometric zeropoints for our coadded images. As photometric reference for
this calibration, the Sloan Digital Sky Survey Data Release Six
\citep[SDSS DR6,][]{2008ApJS..175..297A} was employed, with which six of our
clusters overlap. 
This direct calibration also yields zeropoints for fields outside the SDSS 
footprint observed in the same filter in the same photometric night, as for the
$i'$-band of CL\,0230+1836 (Table~\ref{tab:data}).
The remaining observations were done in nights in which no cluster with SDSS
overlap was observed under photometric conditions. To these data, labelled with
``I'' in Table~\ref{tab:data}, we applied an indirect calibration described in
Paper~I, basically a rudimentary but effective stellar locus regression
\citep{2009AJ....138..110H}.
Details concerning the results and accuracy of the photometric calibration can
be found in Appendix~\ref{sec:augias}.

\subsection{From images to shape catalogues}  \label{sec:ksb}

\begin{table*} 
 \caption{Definitions of  the galaxy shape and lensing catalogues.
  The parameters $\vartheta^{*}_{\mathrm{min}}$, $\vartheta^{*}_{\mathrm{max}}$, 
  $r'^{*}_{\mathrm{min}}$, and $r'^{*}_{\mathrm{max}}$ delineate the stellar locus. 
  The galaxy shape catalogue considers sources 
  $\vartheta\!\!>\!\vartheta^{*}_{\mathrm{max}}$ for 
  $r'^{*}_{\mathrm{min}}\!\!<\!r'_{\mathrm{AUTO}}\!\!<\!r'^{*}_{\mathrm{max}}$ and 
  $\vartheta\!\!>\!\vartheta^{\mathrm{ana}}_{\mathrm{min}}$ for 
  $r'_{\mathrm{AUTO}}\!\!\geq\!r'^{*}_{\mathrm{max}}$ (Cols. (1) to (5)). 
  The number densities $n_{\mathrm{KSB}}$ and $n_{\mathrm{gal}}$ of sources in the
  KSB and galaxy shape catalogues are shown in columns (6) and (7); followed by
  the degree $d_{\mathrm{ani}}$ of the polynomial for PSF anisotropy correction.
  Columns (9) to (13) present the photometric cuts $m_{\mathrm{bright}}$ and
  $m_{\mathrm{faint}}$, defining the lensing catalogue with a source density
  $n_{\mathrm{lc}}$. Using a filter scale of $\theta_{\mathrm{out}}$ in 
  Eq.~(\ref{eq:s-stat}), we find a maximum shear peak significance 
  $S_{\mathrm{\max}}$. 
  CL\,0030+2618 (Paper I) is included for completeness.}
 \begin{center}
 \begin{tabular}{c|ccccc|ccc||ccccc} \hline\hline
 Cluster field & $\vartheta^{*}_{\mathrm{min}}$ & 
 $\vartheta^{\mathrm{ana}}_{\mathrm{min}}$ & $\vartheta^{*}_{\mathrm{max}}$ &  
 $r'^{*}_{\mathrm{min}}$ & $r'^{*}_{\mathrm{max}}$ & $n_{\mathrm{KSB}}$ & 
 $n_{\mathrm{gal}}$ & $d_{\mathrm{ani}}$ & 
 $m_{\mathrm{bright}}$ & $m_{\mathrm{faint}}$ & $n_{\mathrm{lc}}$ &
 $\theta_{\mathrm{out}}^{\mathrm{opt}}$ & $S_{\!\mathrm{max}}$ \\
  & [px] & [px] & [px] & [mag] & [mag] & $[\mbox{arcmin}^{-2}]$ & $[\mbox{arcmin}^{-2}]$ & & [mag] & [mag] & $[\mbox{arcmin}^{-2}]$ & & \\
  & (1) & (2) & (3) & (4) & (5) & (6) & (7) & (8) & (9) & (10) & (11) & (12) & (13) \\\hline
 CL\,0030+2618 & $2.55$ & $2.80$ & $2.95$ & $16.75$ & $22.5$ & $44.79$ & $22.64$ & $5$ & $20.0$ & $22.5$ & $21.28$ & $15\farcm5$ & $5.84$ \\ 
 CL\,0159+0030 & $2.65$ & $2.95$ & $3.10$ & $17.5$ & $22.5$ & $37.27$ & $18.91$ & $3$ & $20.0$ & $24.6$ & $11.58$ & $10\farcm5$ & $4.12$ \\ 
 CL\,0230+1836 & $2.05$ & $2.33$ & $2.45$ & $16.75$ & $22.5$ & $29.28$ & $13.22$ & $4$ & $21.5$ & $23.0$ & $11.04$ & $8\farcm5$ & $3.64$ \\ 
 CL\,0809+2811 & $2.35$ & $2.66$ & $2.80$ & $17.0$ & $22.5$ & $33.39$ & $16.56$ & $2$ & $22.5$ & $24.8$ & $10.31$ & $13\farcm5$ & $5.39$ \\
 CL\,1357+6232 & $2.85$ & $3.18$ & $3.35$ & $17.5$ & $22.25$ & $29.52$ & $14.24$ & $5$ & -- & $18.5$ & $14.23$ & $10\farcm17$ & $4.47$ \\ 
 CL\,1416+4446 & $2.55$ & $2.80$ & $2.95$ & $16.75$ & $22.5$ & $42.50$ & $21.51$ & $5$ & -- & $18.5$ & $21.50$ & $4\farcm83$ & $4.25$ \\ 
 CL\,1641+4001 & $2.90$ & $3.135$ & $3.30$ & $16.75$ & $22.5$ & $39.95$ & $19.58$ & $3$ & -- & $22.7$ & $17.65$ & $16\farcm0^{\dagger}$ & $4.12$ \\ 
 CL\,1701+6414 & $2.60$ & $2.95$ & $3.10$ & $16.75$ & $22.5$ & $40.71$ & $20.72$ & $4$ & -- & $21.9$ & $19.47$ & $15\farcm5^{\dagger}$ & $3.75$ \\ \hline\hline
 \end{tabular}  \label{tab:shapecats}
 \end{center}
 \begin{minipage}{170mm}
  \smallskip
  $\dagger$ No significant decrease of $S(\theta_{\mathrm{out}})$ at the shear 
  peak was found out to the largest probed value 
  $\theta_{\mathrm{out}}\!=\!16\arcmin$. 
  See Sect.~\ref{sec:sec4} for more details on the
  CL\,1701+6414 and CL\,1641+4001 cases.
 \end{minipage}
\end{table*}
A detailed description of how we distill from a coadded image a 
\emph{galaxy shape catalogue}, containing positions, ellipticity measurements, 
and photometric data for sources that can be considered galaxies can be found 
in Paper~I. 
For the single-band clusters, we use straight-forward calls to
\texttt{SExtractor} \citep{1996A&AS..117..393B}.
For the three-band clusters, the convolution of images to the seeing in
the poorest band and calls to \texttt{SExtractor} in
double-detection mode, using the (unconvolved) $r'$-band image as the 
``detection image'', as described in Paper~I are being performed.

We apply the ``TS'' shear measurement pipeline 
\citep{2006MNRAS.368.1323H,2007A&A...468..823S,2009A&A...504..689H}, 
an implementation of the KSB+ algorithm 
\citep{1995ApJ...449..460K,2001A&A...366..717E}, which determines moments of
the brightness distribution for each source and corrects for the convolution
with an anisotropic PSF.
The PSF anisotropy is traced by measuring the brightness distribution of
sources identified as stars in a plot of their magnitude $r'_{\mathrm{AUTO}}$
against the half-light radius $\vartheta$. The values we used to define the
boundaries of the stellar locus are given in Table~\ref{tab:shapecats}.
Only sources in the \emph{KSB catalogue}, consisting of detections with a viable
measurement of $\vartheta$, are further considered.

Consistent with our Paper~I findings, MMT/\megacam ~exhibits a smooth,
albeit variable pattern of PSF anisotropy which can be modelled by a low-order
($2\!\!\leq\!\!d_{\mathrm{ani}}\!\!\leq\!\!5$, see Table~\ref{tab:shapecats}) 
polynomial in image coordinates
such that the residual PSF anisotropy has a practically vanishing mean value
and a dispersion $0.005\!\leq\!\sigma(e^{\mathrm{ani}})\!\leq\!0.010$ in the 
$r'$-band image stacks. 
In terms of the uncorrected PSF anisotropy, however, there are considerable 
differences in the input images for different cluster fields. Excessive PSF
anisotropy observed in several input frames -- which thus had to be
removed from the coadded images -- can be attributed to either tracking or
focussing issues of the telescope. In most fields, no extreme outliers
were present or could be easily identified. Only frames with average 
PSF ellipticity $|\langle e\rangle|\!<\!0.06$ entered the coaddition. No clear
distinction leaving a sufficient number of low-anisotropy frames was possible
for the CL\,1641+4001 and CL\,1701+6414 fields. In these cases, all frames
with $|\langle e\rangle|\!<\!0.10$ were used for coaddition. 

We classify as galaxies all sources fainter than the brightest unsaturated
point sources ($r'_{\mathrm{AUTO}}>r'^{*}_{\mathrm{min}}$) and more extended than the 
PSF ($\vartheta\!>\!\vartheta^{*}_{\mathrm{max}}$). Because even poorly resolved
galaxies carry a lensing signal, we add sources 
$r'_{\mathrm{AUTO}}>r'^{*}_{\mathrm{max}}$ and 
$\vartheta\!>\!\vartheta^{\mathrm{ana}}_{\mathrm{min}}$ with 
$\vartheta^{\mathrm{ana}}_{\mathrm{min}}\!\approx\!0.95 \vartheta^{*}_{\mathrm{max}}$ 
to the \emph{galaxy shape catalogue} (cf.\ Fig.~4 of Paper I). The parameters defining
this catalogue for each field are tabulated in Table~\ref{tab:shapecats},
together with its number density $n_{\mathrm{gal}}\!\approx\!n_{\mathrm{KSB}}/2$.

\subsection{Background Selection} \label{sec:bg7}

Cluster weak lensing studies rely on carefully selected catalogues of background
galaxies, the carriers of the lensing signal. While falling short of yielding a
reliable photometric redshift (photo-$z$) estimate for each individual galaxy,
three-colour imaging makes possible a selection of foreground, cluster, and
background sources based on their distribution in colour-colour-magnitude space
\citep[cf.][Klein et al., in prep.]{2010MNRAS.405..257M}.
The method described below that we use for the three-band clusters is an
improved version of the background selection in Paper I, to which we refer for
concepts and terminology: While considering all galaxies fainter than
$m_{\mathrm{faint}}$ in the \emph{lensing catalogue}, galaxies brighter than
$m_{\mathrm{bright}}$ are rejected. In the intermediate regime 
($m_{\mathrm{bright}}\!<\!r'\!<\!m_{\mathrm{faint}}$), we include galaxies whose
$g'\!-\!r'$ versus $r'\!-\!i'$ colours are consistent with sitting in the
background of the cluster, based on the -- similarly deep --
\citet{2009ApJ...690.1236I} photo-$z$ 
catalogue (see Appendix~\ref{sec:bgdetails} for more details).

For the single-band clusters, the background selection simplifies to a magnitude
cut, meaning that lensing catalogue consists of all galaxies 
$r'\!>\!m_{\mathrm{faint}}$.
For each of these sources, our KSB+ implementation yields a PSF-corrected
ellipticity $\varepsilon\!=\!\varepsilon_{1}\!+\!\mathrm{i}\varepsilon_{2}$, 
a noisy but, in principle, unbiased estimate of
the \emph{reduced gravitational shear} $g\!=\!g_{1}\!+\!\mathrm{i}g_{2}$ 
\citep[cf.][]{2006glsw.book.....S}.
We choose the values of $m_{\mathrm{faint}}$ (and $m_{\mathrm{bright}}$ where 
applicable) such that the signal-to-noise ratio of the aperture mass estimator,
or $S$-statistics \citep{1996MNRAS.283..837S} is optimised:
\begin{equation} \label{eq:s-stat}
S(\pmb{\theta}_{\mathrm{c}};\theta_{\mathrm{out}})\!=\!
\frac{\sqrt{2}}{\sigma_{\!\varepsilon}}\frac{\sum_{j}{\varepsilon_{\mathrm{t},j}\,
Q_{j}(|\pmb{\theta}_{\!j}\!-\!\pmb{\theta}_{\mathrm{c}}|/\theta_{\mathrm{out}})}}
{\sqrt{\sum_{j}{Q_{j}^{2}(|\pmb{\theta}_{\!j}\!-\!
\pmb{\theta}_{\mathrm{c}}|/\theta_{\mathrm{out}})}}}\quad.
\end{equation}
By $\varepsilon_{\mathrm{t},j}\!=\!\mathrm{Re}[\varepsilon\exp(-2\mathrm{i}\varphi)]$, 
we denote the tangential ellipticity of the galaxy at position 
$\pmb{\theta}_{\!j}$, which with respect to the point 
$\pmb{\theta}_{\mathrm{c}}$ has a phase angle $\varphi$. 
Equation~(\ref{eq:s-stat}) considers the noise from intrinsic source ellipticity,
measured as $\sigma_{\!\varepsilon}\!=\!\langle\varepsilon_{1}^{2}\!+\!\varepsilon_{2}^{2}\rangle^{1/2}$; while 
$Q_{j}(|\pmb{\theta}_{\!j}\!-\!\pmb{\theta}_{\mathrm{c}}|/\theta_{\mathrm{out}})$
is the \citet{2007A&A...462..875S} filter function with outer radius 
$\theta_{\mathrm{out}}$, maximising $S$ for a cluster-like radial shear profile.

We evaluate Eq.~(\ref{eq:s-stat}) on a regular grid with $15\arcsec$ mesh size.
With the notable exception of CL\,1701+6414 (Sect.~\ref{sec:cl1701}), 
the signal peaks are found close to the \textsc{rosat}-determined cluster 
centres and can be easily identified with our clusters.
The adopted values of $m_{\mathrm{bright}}$, $m_{\mathrm{faint}}$, and 
$\theta_{\mathrm{out}}$, yielding an optimal 
signal-to-noise ratio
$S_{\!\mathrm{max}}$ are summarised in Table~\ref{tab:shapecats}, as well as the
number density $n_{\mathrm{lc}}$ in the lensing catalogue\footnote{In addition to
the photometric cuts, we restrict ourselves to high-quality sources defined
by $|\varepsilon|\!<\!0.8$, \texttt{SExtractor} detection significance 
$\nu\!>\!4.5$ and $\tr{(\tens{P}^{\mathrm{g}})}\!>\!0.1$ for the KSB 
pre-seeing polarisability tensor.}.
We refer to the grid cell in which $S_{\!\mathrm{max}}$ occurs as the cluster 
\emph{shear peak} and discuss the significance of our cluster detections in 
Sect.~\ref{sec:mapdisc}. 

\subsection{Shear profile modelling} \label{sec:spm}

\begin{table*}
 \caption{Additional parameters defining the ``default'' cluster models.
  Columns (1) to (5) tabulate the cluster redshift $z_{\mathrm{d}}$, the average
  $\langle\langle\beta\rangle\rangle$ and dispersion 
  $\sigma(\langle\beta\rangle)$ of the distance ratio estimated from the 
  CFHTLS \textsl{Deep} fields, the estimated fraction $\hat{f}_{\mathrm{d}}$ of 
  residual foreground galaxies in the lensing catalogue, based on the same 
  \citet{2006A&A...457..841I} photo-$z$ catalogue, and whether or not a 
  dilution correction $f_{1}(\theta)$ has been applied.
  Columns (6) and (7) present the celestial coordinates of the assumed
  cluster centre, with respect to which the fitting range
  $r_{\mathrm{min}}\!\leq\!\theta\!\leq\!r_{\mathrm{max}}$ in 
  columns (8) and (9) is defined. Columns (10) and (11) repeat the fitting
  range in terms of angular separation.
  Finally, we give the separation between \textsc{rosat}
  and lensing centres in column (12).}
 \begin{center} 
  \begin{tabular}{cccccc|ccccccc}\hline\hline
   Cluster & $z_{\mathrm{d}}$ & $\langle\langle\beta\rangle\rangle$ & $\sigma(\langle\beta\rangle)$ & $\hat{f}_{\mathrm{d}}$ & $f_{1}(\theta)$ & $\alpha_{\mathrm{c,J2000}}$ & $\delta_{\mathrm{c,J2000}}$ & $r_{\mathrm{min}}$ & $r_{\mathrm{max}}$ & $\theta_{\mathrm{min}}$ & $\theta_{\mathrm{max}}$ & $\Delta\theta$ \\
 & (1) & (2) & (3) & (4) & (5) & (6) & (7) &  (8) & (9) & (10) & (11) & (12) \\ \hline
   CL\,0030+2618 & $0.50$ & $0.348^{\dagger}$ & $0.024^{\dagger}$ & $0.152^{\dagger}$ & $\checkmark$ & $00^{\mathrm{h}}30^{\mathrm{m}}33\fs6$ & $+26\degr18\arcmin16\arcsec$ & $0.2\,\mbox{Mpc}$ & $5.0\,\mbox{Mpc}$ & $0\farcm56$ & $14\farcm04$ & $23\arcsec$  \\
   CL\,0159+0030 & $0.39$ & $0.447$ & $0.023$ & $0.087$ & $\checkmark$ & $01^{\mathrm{h}}59^{\mathrm{m}}18\fs2$ & $+00\degr30\arcmin09\arcsec$ & $0.2\,\mbox{Mpc}$ & $5.0\,\mbox{Mpc}$ & $0\farcm65$ & $16\farcm20$ & $79\arcsec^{\S}$ \\
   CL\,0230+1836 & $0.80$ & $0.168$ & $0.020$ & $0.321$ & $\checkmark$ & $02^{\mathrm{h}}30^{\mathrm{m}}26\fs6$ & $+18\degr36\arcmin22\arcsec$ & $0.2\,\mbox{Mpc}$ & $5.0\,\mbox{Mpc}$ & $0\farcm46$ & $11\farcm42$ & $20\arcsec$ \\
   CL\,0809+2811 & $0.40$ & $0.437$ & $0.023$ & $0.105$ & $\checkmark$ & $08^{\mathrm{h}}09^{\mathrm{m}}41\fs0$ & $+28\degr11\arcmin58\arcsec$ & $0.2\,\mbox{Mpc}$ & $5.0\,\mbox{Mpc}$ & $0\farcm64$ & $15\farcm95$ & $178\arcsec^{\S}$ \\
   CL\,1357+6232 & $0.53$ & $0.324$ & $0.024$ & $0.198$ & -- & $13^{\mathrm{h}}57^{\mathrm{m}}19\fs4$ & $+62\degr32\arcmin42\arcsec$ & $0.2\,\mbox{Mpc}$ & $5.0\,\mbox{Mpc}$ & $0\farcm54$ & $13\farcm62$ & $50\arcsec$ \\
   CL\,1416+4446 & $0.40$ & $0.437$ & $0.023$ & $0.136$ & -- & $14^{\mathrm{h}}16^{\mathrm{m}}28\fs1$ & $+44\degr46\arcmin38\arcsec$ & $0.2\,\mbox{Mpc}$ & $5.0\,\mbox{Mpc}$ & $0\farcm64$ & $15\farcm95$ & $19\arcsec$ \\
   CL\,1641+4001 & $0.46$ & $0.381$ & $0.024$ & $0.127$ & -- & $16^{\mathrm{h}}41^{\mathrm{m}}52\fs3$ & $+40\degr01\arcmin27\arcsec$ & $0.2\,\mbox{Mpc}$ & $5.0\,\mbox{Mpc}$ & $0\farcm59$ & $14\farcm70$ & $95\arcsec$ \\
   CL\,1701+6414 & $0.45$ & $0.381$ & $0.024$ & $0.134$ & -- & $17^{\mathrm{h}}01^{\mathrm{m}}22\fs5$ & $+64\degr14\arcmin08\arcsec$ & $0.2\,\mbox{Mpc}$ & $5.0\,\mbox{Mpc}$ & $0\farcm60$ & $14\farcm88$ & $66\arcsec$ \\ \hline\hline
  \end{tabular} \label{tab:defmod}
 \end{center}
 \begin{minipage}{170mm}
  \smallskip
  $\dagger$ We use an improved fit for the redshift distribution and corrected
            $\hat{f}_{\mathrm{d}}$ for CL\,0030+2618 with respect to Paper~I.\\
  $\S$ Lensing centre is within a large masked area, which probably diminished the accuracy.
 \end{minipage} 
\end{table*}

Pursuing the approach adopted in Paper~I, we model the tangential
ellipticity profile $\varepsilon_{\mathrm{t}}(\theta)$ of our clusters with the 
reduced shear profile $g(\theta;r_{200},c_{\mathrm{NFW}})$
\citep{1996A&A...313..697B,2000ApJ...534...34W} corresponding to the
\citet[NFW]{1995MNRAS.275..720N,1996ApJ...462..563N,1997ApJ...490..493N} 
density profile. From the estimate of the radius $r_{200}$ -- defined such that 
the density of the enclosed matter exceeds the critical density 
$\rho_{\mathrm{c}}(z_{\mathrm{d}})$ at the cluster redshift  $z_{\mathrm{d}}$ 
by a factor of $\Delta\!=\!200$ -- we infer the cluster mass $M_{200}$ via
\begin{equation} \label{eq:mdelta}
M_{\Delta}\!=\!\Delta \frac{4\pi}{3} \rho_{\mathrm{c}}(z_{\mathrm{d}}) r_{\Delta}^{3}\quad.
\end{equation}
The best matching cluster mass profile parameters $r_{200}$ and 
$c_{\mathrm{NFW}}$ minimise the merit function
\begin{equation} \label{eq:merit}
\chi^{2}\!=\!\sum_{i=1}^{N}{\frac{\left|g_{i}(\theta_{i};r_{200},c_{\mathrm{NFW}})\!-\!
\tilde{\varepsilon}_{\mathrm{t},i}(\theta_{i})\right|^{2}}
{\tilde{\sigma}_{\!i}^{2}(\theta_{i})\left(1\!-\!\left|
g_{i}(\theta_{i};r_{200},c_{\mathrm{NFW}})\right|^{2}\right)^{2}}}\quad,
\end{equation}
which we evaluate on a regular grid in $r_{200}$ and $c_{\mathrm{NFW}}$.
By $\tilde{\varepsilon}_{\mathrm{t},i}$, we denote the tangential component of the
scaled ellipticity $\tilde{\varepsilon}_{i}=f_{0}f_{1}(\theta_{i})\varepsilon_{i}$
for the $i$-th galaxy, including a global shear calibration factor 
$f_{0}\!=\!1.08$ \citep[see Paper~I and][]{2009A&A...504..689H} as well as a 
separation-dependent correction $f_{1}(\theta)$ for the shear dilution by 
cluster members (detailed below).
Accordingly, the error scales as 
$\tilde{\sigma}_{\!i}(\theta_{i})\!=\!f_{0}f_{1}(\theta_{i})\sigma_{\mathrm{\!\varepsilon}}/\!\sqrt{2}$
with $\sigma_{\!\mathrm{\varepsilon}}$ from Sect.~\ref{sec:bg7}.
The index $i$ runs over all lensing catalogue galaxies with separations within
the fitting range $\theta_{\mathrm{min}}\!\leq\!\theta\!\leq\!\theta_{\mathrm{max}}$
from the assumed \textsc{rosat} cluster centre, presented in Table~\ref{tab:defmod}.
We choose separations $\theta_{\mathrm{min}}$ and $\theta_{\mathrm{max}}$
corresponding to distances of $r_{\mathrm{min}}\!=\!0.2\,\mbox{Mpc}$ and
 $r_{\mathrm{max}}\!=\!5.0\,\mbox{Mpc}$ at the respective cluster redshift.
The denominator of Eq.~(\ref{eq:merit}) accounts for the dependence of the
noise on $g_{i}(\theta_{i})$ itself \citep{2000A&A...353...41S}.

\paragraph{Source redshift distributions}
The reduced shear $g_{i}(\theta_{i})$  exerted by a lens on the image of a 
background source further depends on the ratio of angular diameter distances 
between deflector and source $D_{\mathrm{ds}}$ and source and observer
$D_{\mathrm{s}}$. 
For each of our fields, we estimate a catalogue-average 
$\langle\beta\rangle\!=\!\langle D_{\mathrm{ds}}/D_{\mathrm{s}}\rangle_{i=1\ldots N}$ 
using the \citet{2006A&A...457..841I} photo-$z$ catalogue, drawn from the CFHTLS
\textsl{Deep} fields with similar source number counts as a function of 
magnitude $r'$ as our MMT observations. 
Applying the same photometric cuts as to the MMT data to the catalogues of
reliable photo-$z$ sources (cf.\ Paper I), we thus obtain proxy
redshift distributions for
our cluster observations. We repeat the fit of a \citet{2001A&A...374..757V} 
redshift distribution and subsequent calculation of $\langle\beta\rangle$ as 
described in Paper~I -- but to an improved accuracy -- for all combinations of 
MMT and CFHT \textsl{Deep} fields.
As an input to Eq.~(\ref{eq:merit}), we use the mean 
$\langle\langle\beta\rangle\rangle_{k=1\ldots 4}$
(Table~\ref{tab:defmod}) measured for the \textsl{Deep} fields and consider its
dispersion $\sigma(\langle\beta\rangle)$ in the error analysis 
(Sect.~\ref{sec:err7}).

We further employ the \citet{2006A&A...457..841I} catalogue to test the efficacy
of the background selection. Applying the respective background selection to the
\textsl{Deep~1} photo-$z$ catalogue, we determine the fraction 
$\hat{f}_{\mathrm{d}}$ of residual foreground galaxies in the lensing catalogues
(Table~\ref{tab:defmod}, cf.\ Sect.~\ref{sec:err7}).

\paragraph{Dilution by cluster members}
Although the selection of lensing catalogue galaxies is designed to include
preferentially background galaxies, we detect an increase in the fraction 
$f_{\mathrm{rsc}}(\theta)$ of galaxies whose $g'\!-\!i'$ colours are consistent 
with the red sequences at $z_{\mathrm{d}}$ towards the centres of our three-band 
clusters.
Tentative red sequence galaxies are defined using an interval in $g'\!-\!i'$
empirically found in the $g'\!-\!i'$ versus $i'$ colour-magnitude diagram,
around the expected colour of a \citet{1980ApJS...43..393C} early-type galaxy
calculated with the \citet{2000A&A...363..476B} photo-$z$ code.
To correct for the dilution effect of these likely unlensed sources in the
shear catalogues, the corrective factor
$f_{1}(\theta)\!=\!1+\Sigma(\theta)/[\Sigma(\theta)+B]$ is introduced.
The NFW surface mass profile $\Sigma(\theta)$ and background term $B$ are
determined by a fit to $f_{\mathrm{rsc}}(\theta)$.
We apply this correction only to the three-band clusters for which the
$g'\!-\!i'$ information is available (see Table~\ref{tab:defmod}).
Because we have $f_{1}(\theta)$ measured for only four clusters, three of which
suffer from large masks in the crucial central regions, we decide against using
an averaged $f_{1}(\theta)$ for the single-band clusters at this stage of
the survey.

\subsection{Surface mass maps}

While we use the tangential shear profile to determine cluster masses,
we are interested as well in the (projected) mass distributions of our clusters
in order to distinguish possibly merging systems of disturbed morphology from
relaxed clusters.
The non-local relation between shear and convergence 
$\kappa\!=\!\Sigma/\Sigma_{\mathrm{crit}}$ can be inverted, as shown by 
\citet{1993ApJ...404..441K}. We perform mass reconstructions using the 
\citet{1996A&A...305..383S,2001A&A...374..740S} 
finite-field inversion algorithm. 
Concerning the mass sheet degeneracy \citep[cf.][]{2006glsw.book.....S}, the 
mean $\kappa$ along the edge of the field-of-view is assumed to vanish.

The dimensionless surface mass $\tilde{\kappa}\!\propto\!\kappa$, with an
arbitrary normalisation, is calculated on a regular grid.
Because each cluster field has to be divided into an integral number of grid
cells, the mesh size cannot be fixed to the same constant for all clusters,
but varies slightly, with a mean of $40\farcs93$ and a standard deviation of
$0\farcs36$. For all clusters and grid points, the algorithm accounts for
lensing catalogue galaxies within a radius of $\theta_{\mathrm{s}}\!=\!2\arcmin$.
The input shear field is smoothed with a truncated Gaussian filter of  
$0.555\,\theta_{\mathrm{s}}$ full-width half-maximum, which drops to zero at 
$\theta_{\mathrm{s}}$.

\section{Results for normal clusters} \label{sec:sec3}

\begin{figure*}
\vspace{-0.5cm}
 \includegraphics[width=16cm]{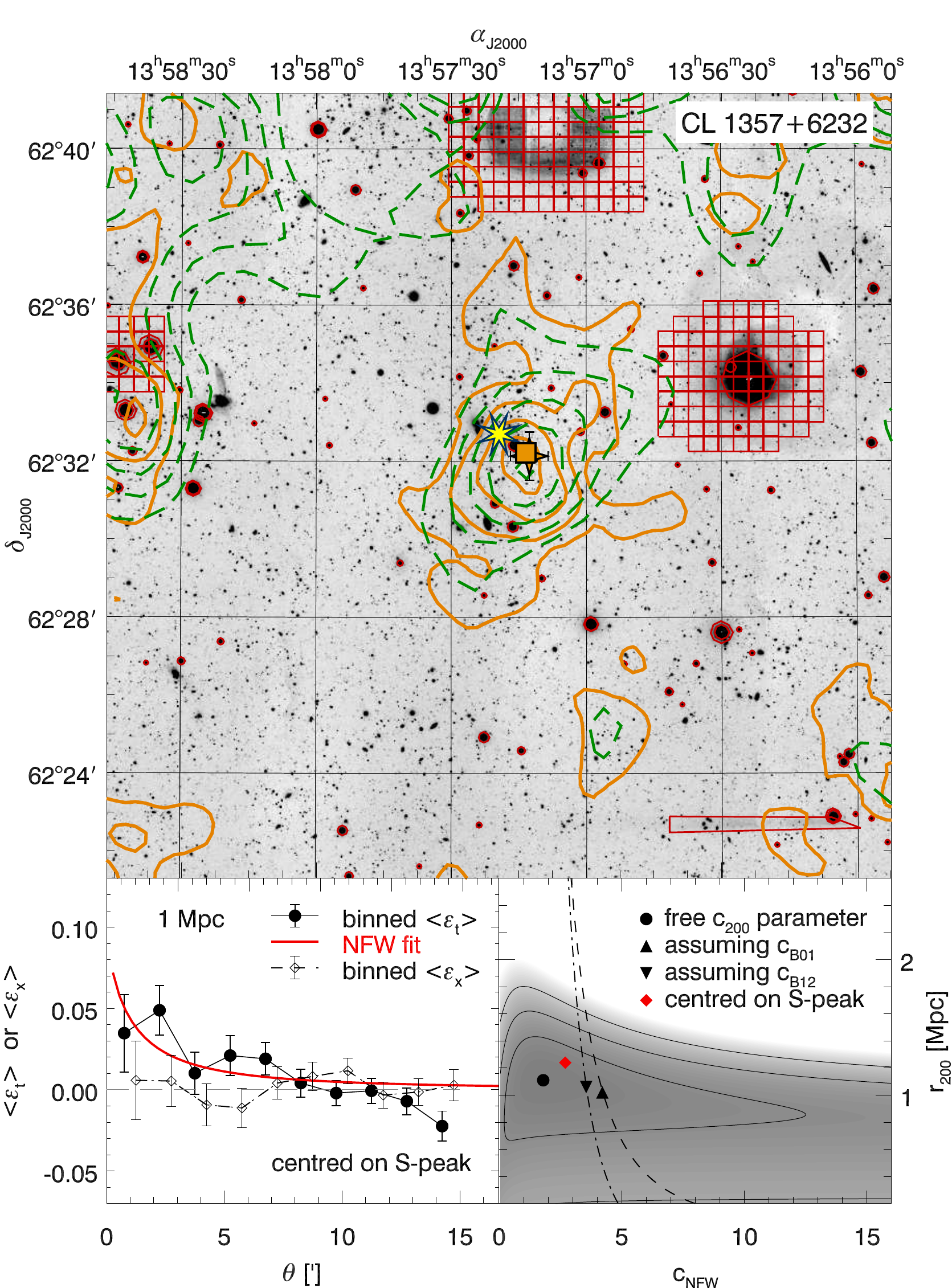}
 \caption{Lensing results for CL\,1357+6232. \emph{Upper panel:} \megacam 
  ~$r'$-band image (cut-out of $\sim\!20\arcmin$ side length), overlaid with 
  $S$-statistics (orange solid) and $\tilde{\kappa}$ (green dashed) contours.
  Contours start at $S\!=\!1.0$ and $\tilde{\kappa}\!=\!0.01$, with increments 
  of $\Delta S\!=\!1.0$ and $\Delta\tilde{\kappa}\!=\!0.01$, respectively.
  The eight-pointed star symbol marks the \textsc{rosat} cluster centre, while
  the filled square shows the shear peak grid cell. A star symbol with error 
  bars denotes the lensing centre from bootstrapping.
  Regions inside red polygons are masked out of the analysis.
 \emph{Lower left panel:} Profiles of the binned tangential 
 ($\langle\varepsilon_{\mathrm{t}}\rangle$, filled circles) and binned cross 
 ($\langle\varepsilon_{\times}\rangle$, open diamonds) ellipticities. 
 Error bars give the bin dispersions. The thick blue curve denotes the best-fit
 NFW model (Eq.~\ref{eq:merit}). Here, the $S$-peak was assumed as centre.
 \emph{Lower right panel:} $\Delta\chi^{2}(r_{200},c_{\mathrm{NFW}})$ with respect 
 to its minimum (filled circle), shown as grey shading and contours indicating 
 $1\sigma$, $2\sigma$, and $3\sigma$ confidence. An upward (downward) triangle 
 on a dashed (dash-dotted) curves mark the best-fit values assuming the B01 and
 B12 mass--$c_{200}$--relations. A diamond marks cluster parameters assuming
 the $S$-peak as centre.}
 \label{fig:cl1357}
\end{figure*}

\begin{table*}
 \caption{
 Synopsis of cluster parameters and resulting weak lensing masses:
 Columns (1) to (3) give $r_{\mathrm{200}}$ and $c_{\mathrm{NFW}}$ from
 the default cluster models, and the corresponding mass, 
 $M_{200}^{\mathrm{wl}}(r_{200})$. Fixing the concentration to the value
 $c_{\mathrm{B01}}$ expected from the \citet{2001MNRAS.321..559B} 
 mass--concentration relation, we obtain the best-fit radius $r_{\mathrm{200,B01}}$
 and mass $M_{\mathrm{200,B01}}^{\mathrm{wl}}(r_{\mathrm{200,B01}})$, in Columns (4) to (6).
 Columns (7) to (9) contain the respective quantities calculated 
 assuming the \citet{2011arXiv1112.5479B} mass-concentration relation.}
 \begin{center}
 \begin{tabular}{c|ccc|ccc|ccc} \hline\hline
Cluster & $r_{\mathrm{200}}$ & $c_{\mathrm{NFW}}$ & $M_{200}^{\mathrm{wl}}(r_{200})$ & $r_{\mathrm{200,B01}}$ & $c_{\mathrm{B01,NFW}}$ & $M_{\mathrm{200,B01}}^{\mathrm{wl}}(r_{\mathrm{200,B01}})$ & $r_{\mathrm{200,B12}}$ & $c_{\mathrm{B12,NFW}}$ & $M_{\mathrm{200,B12}}^{\mathrm{wl}}(r_{\mathrm{200,B12}})$\\ 
 & [Mpc] & & [$10^{14}\,\mbox{M}_{\sun}$] & [Mpc] & & [$10^{14}\,\mbox{M}_{\sun}$] & [Mpc] & & [$10^{14}\,\mbox{M}_{\sun}$]  \\
 & (1) & (2) & (3) & (4) & (5) & (6) & (7) & (8) & (9) \\ \hline 
 CL\,0030+2618 & $1.52_{-0.16}^{+0.14}$ & $1.7_{-0.8}^{+1.3}$ & $7.25_{-2.06}^{+2.19}$ & $1.39_{-0.14}^{+0.13}$ & $3.9_{-1.3}^{+2.0}$ & $5.50_{-1.50}^{+1.65}$ & $1.45_{-0.15}^{+0.13}$ &  $3.4_{-1.1}^{+1.7}$ & $6.23_{-1.76}^{+1.87}$ \\
 CL\,0159+0030 & $1.37_{-0.22}^{+0.18}$ & $>\!20$ & $4.67_{-1.91}^{+2.09}$ & $1.34_{-0.25}^{+0.21}$ & $4.3_{-1.5}^{+2.2}$ & $4.39_{-2.03}^{+2.44}$  & $1.31_{-0.27}^{+0.22}$ & $3.6_{-1.2}^{+1.9}$ & $4.10_{-1.76}^{+2.46}$ \\
 CL\,0230+1836 & $1.54_{-0.32}^{+0.28}$ & $2.8_{-1.6}^{+3.7}$ & $10.78_{-5.42}^{+7.01}$ & $1.49_{-0.31}^{+0.26}$ & $3.0_{-1.0}^{+1.5}$ & $9.68_{-4.83}^{+6.02}$ & $1.51_{-0.32}^{+0.27}$ & $2.9_{-1.0}^{+1.5}$ & $10.23_{-5.18}^{+6.59}$ \\
 CL\,0809+2811 & $1.75_{-0.28}^{+0.23}$ & $1.1_{-0.8}^{+1.9}$ &  $9.84_{-4.01}^{+4.41}$ & $1.69_{-0.21}^{+0.18}$ & $3.9_{-1.3}^{+2.0}$ & $8.87_{-2.93}^{+3.15}$ & $1.71_{-0.22}^{+0.19}$ & $3.4_{-1.2}^{+1.7}$ & $9.24_{-3.14}^{+3.46}$ \\
 CL\,1357+6232 & $1.11_{-0.25}^{+0.21}$ & $1.8_{-1.2}^{+3.3}$ & $2.92_{-1.56}^{+1.99}$ & $1.02_{-0.23}^{+0.18}$ & $4.2_{-1.4}^{+2.2}$ & $2.26_{-1.20}^{+1.46}$ & $1.06_{-0.24}^{+0.20}$ & $3.6_{-1.2}^{+1.8}$ & $2.52_{-1.34}^{+1.74}$ \\
 CL\,1416+4446 & $0.98_{-0.18}^{+0.15}$ & $5.5_{-3.3}^{+13.9}$ & $1.73_{-0.79}^{+0.92}$ & $0.97_{-0.17}^{+0.14}$ & $4.8_{-1.6}^{+2.5}$ & $1.69_{-0.74}^{+0.83}$ & $0.98_{-0.18}^{+0.16}$ & $3.9_{-1.3}^{+2.0}$ & $1.71_{-0.79}^{+0.95}$ \\
 CL\,1641+4001 & $1.06_{-0.26}^{+0.30}$ & $0.1_{-0.1}^{+0.3}$ & $2.34_{-1.34}^{+2.61}$ & $0.86_{-0.36}^{+0.22}$ & $4.8_{-1.6}^{+2.5}$ & $1.27_{-1.02}^{+1.26}$ & $1.01_{-0.28}^{+0.21}$ & $3.7_{-1.3}^{+1.9}$ & $2.05_{-1.26}^{+1.51}$ \\
 CL\,1701+6414$\dagger$ & $0.94_{-0.29}^{+0.32}$ & $0.1_{-0.1}^{+1.1}$ & $1.62_{-1.08}^{+2.28}$ & $0.95_{-0.19}^{+0.16}$ & $4.6_{-1.6}^{+2.4}$ & $1.69_{-0.83}^{+1.02}$  & $1.01_{-0.20}^{+0.17}$ & $3.8_{-1.3}^{+1.9}$ & $2.03_{-0.98}^{+1.17}$ \\\hline\hline
 \end{tabular}
  \label{tab:massescb}
 \end{center}
 \begin{minipage}{180mm}
  \smallskip
  $\dagger$ Fixing the radius and concentration of A~2246 to $r_{\mathrm{s,200}}\!=\!0.90\,\mbox{Mpc}$ and $c_{\mathrm{NFW}}\!=\!20$ and hence using $\Delta\chi^{2}\!=\!1$ for the $1\sigma$ error margins.
 \end{minipage}
\end{table*}
In this Section, we present the outcome of the WL modelling, by showing a 
comprehensive figure combining the lensing signal maps, shear profile, and NFW
modelling for each cluster. CL\,1357+6232 (Fig.~\ref{fig:cl1357}) serves as our
example; for more details on the other clusters, we refer to 
Figs.~\ref{fig:cl0159} to Figs.~\ref{fig:cl1416} in Appendix~\ref{sec:indivcl}.
Two clusters, CL\,1701+6414 (Fig.~\ref{fig:cl1701}) and 
CL\,1641+4001 (Fig.~\ref{fig:cl1641}), exhibit multiple shear peaks and shear 
profiles that are very flat but positive over a large radial range.
The more involved modelling of these ``special cases'' -- as opposed to the 
``normal clusters'' -- is described in Sect.~\ref{sec:sec4}.

In the upper panel of Fig.~\ref{fig:cl1357}, we present the $S$-statistics 
(solid orange) and $\tilde{\kappa}$-contours (green dashed) for CL\,1357+6232, 
overlaid on a cut-out of the \megacam ~$r'$-band image with $\sim\!\!20\arcmin$
side length.
Masked areas can be identified from the red polygons (mostly squares). 
The \textsc{rosat} centre is given by a yellow, eight-pointed star symbol. 
A filled orange square denotes the shear peak grid cell (Sect.~\ref{sec:bg7}), 
while a star symbol with error bars shows the WL centre from bootstrapping 
(Sect.~\ref{sec:wlcc}).

The lower left panel of Fig.~\ref{fig:cl1357} shows the binned shear profile
$\langle\varepsilon_{\mathrm{t}}(\theta)\rangle$ as filled circles with error bars
giving the dispersion of the measured values. Open diamonds give
the cross component $\langle\varepsilon_{\times}(\theta)\rangle$ which is
\emph{on average} expected to be consistent with zero for cluster lenses.
The red solid line denotes the best-fit NFW model.
Finally, the lower right panel of Fig.~\ref{fig:cl1357} presents
$\Delta\chi^{2}(r_{200},c_{\mathrm{NFW}})\!=\!\chi^{2}\!-\!\min{(\chi^{2})}.$
The minimum is indicated by filled circle; contour lines enclose the 
$99.73$\%, $95.4$\%, and $68.3$\% confidence regions, 
(i.e.\ $\Delta\chi^{2}\!=\!2.30$, $6.17$, and $11.30$). 
An upward triangle marks the minimum of $\Delta\chi^{2}$ when restricting 
$c_{\mathrm{NFW}}$ to its \citet[dashed line]{2001MNRAS.321..559B} value.

\subsection{Cluster detection and lensing morphology}

We successfully detect all observed \emph{400d} clusters using the 
$S$-statistics with at least $3.5\sigma$ significance and are able to derive 
a weak lensing mass estimate for each cluster. 
Table~\ref{tab:shapecats} summarises the maximum detection levels 
$S$ and the optimal filter scales $\theta_{\mathrm{out}}^{\mathrm{opt}}$. 
The most significant detection is CL\,0030+2618 at $z\!=\!0.50$ with 
$S\!=\!5.84$ (Paper I); the formally least significant detection is 
CL\,0230+1836 at $z\!=\!0.80$ with $S\!=\!3.64$. The $S\!=\!3.75$ measured for 
CL\,1701+6414 has a contribution from the nearby cluster A~2246 at 
$\theta\!\approx\!270\arcsec$ separation (Sect.~\ref{sec:cl1701}), rendering 
it the least secure detection: For $\theta_{\mathrm{out}}\!=\!220\arcsec$, we 
detect CL\,1701+6414 at the $2.5\sigma$ level.
By detecting CL\,0230 +1836, we demonstrate the feasibility of
\megacam ~WL studies out to the highest redshifts accessible for current
ground-based weak lensing.

In general, we find a very good agreement between the signal morphologies, 
of the $S$-statistics and mass reconstruction, i.e.\ we detect the same
structures at comparable relative signal strength. This result reaffirms
that our detections are not caused by artifacts in the (independent) analysis
methods. 

\subsection{WL cluster centres} \label{sec:wlcc}

We define a ``default'' model for the NFW modelling of each cluster, determined
by the parameters in Table~\ref{tab:defmod}, i.e.\ the cluster centre, fitting 
range, $\langle\langle\beta\rangle\rangle$, and dilution correction.
We acknowledge that a careful and consistent treatment of cluster centres is
important to prevent masses from being biased. 
In the default model, we use the lensing-independent \textsc{rosat} 
X-cluster centres. For comparison, we also consider cluster centres
based on the $S$-map, which provide us with a high 
signal-to-noise shear profile. The shear peaks (most significant cell in the
$S$-map) are thoroughly studied with respect to the background selection
parameters and their interpretation as significances (Sect.~\ref{sec:mapdisc}).

The $S$-peak of CL 0159+0030 is located conspicuously close to the edge of an
extended shear plateau which is likely caused by a large masked area
\footnote{
Due to the filtering with large scales $\theta_{\mathrm{out}}$, we measure a
signal also in masked areas. Naturally, the correlation between neighbouring
grid cells is even higher than in unmasked regions.} around a bright star 
($V\!=\!8.3$, Figs.~\ref{fig:zooms} and \ref{fig:cl0159}). 
Similarly bright stars are present also close to CL\,0230+1836 and 
CL\,0809+2811 (Figs.~\ref{fig:cl0230} and \ref{fig:cl0809}).
In the latter case, where the $S$-peak lies within the masked area, we discuss
the effect of masking in Sect.~\ref{sec:mmask}.

As noise can boost $S$ in a grid cell compared to its neighbours, we perform a
bootstrap resampling of the $S$-map (cf.\ Paper~I) in two cases, CL\,1357+6232
and CL\,1416+4446. Averaging over $10^{5}$ realisations, for which we draw
$N_{\mathrm{lc}}$ galaxies with repetitions from the lensing catalogue, we
determine a lensing centre.
We find the bootstrap lensing centres to be in good agreement with the shear
peaks of CL\,1357+6232 and CL\,1416+4446, well within the standard deviation
of the bootstrap samples.
In Sect.~\ref{sec:centdisc}, the implications of the choice of cluster centres 
for the mass estimates are discussed.

\subsection{Shear profiles and NFW modelling}

Five of our clusters can be classified as ``normal'', characterised by centrally
increasing $\langle\varepsilon_{\mathrm{t}}\rangle(\theta)$ profiles, in good 
agreement with the NFW models. 
As expected, their $\langle\varepsilon_{\times}\rangle(\theta)$ profiles are
consistent with zero, with fluctuations that can be explained by shape noise.
The two other clusters, CL\,1641+4001 and CL\,1701+6414 show a more complicated
morphology in their $S$-maps (Sect.~\ref{sec:sec4}).

Table~\ref{tab:massescb} provides the cluster parameters resulting from the
NFW modelling. Uncertainties in $r_{200}$ and $c_{\mathrm{NFW}}$ are calculated
from $\Delta\chi^{2}$ corresponding to
a $68.3$\% confidence limit for one interesting parameter ($\Delta\chi^{2}\!=\!1$).
Cluster masses $M_{200}^{\mathrm{wl}}(r_{200})$ are inferred via Eq.~(\ref{eq:mdelta}).

\subsection{Mass--Concentration Relations}

Weak lensing hardly constrains cluster concentration parameters, because the
dependence on $c_{\mathrm{NFW}}$ is highest in the cluster centre where few lensed
sources are observed. This is reflected also in our results, with huge 
uncertainties measured for $c_{\mathrm{NFW}}$ in several objects. 
Hence, we perform two additional measurements, 
in which we fix the value of $c_{\mathrm{NFW}}$. 

The first mass--concentration relation we assume is the one found by
\citet[B01]{2001MNRAS.321..559B} for simulated clusters:
\begin{equation} 
c_{\mathrm{B01}}=\frac{c_{\mathrm{B01},0}}{1+z}\left(\frac{M_{\mathrm{vir}}}{M_{\ast}}\right)^{\alpha_{\mathrm{B01}}}
\label{eq:cm}
\end{equation}
with $c_{\mathrm{B01},0}\!=\!9.0$, $\alpha_{\mathrm{B01}}\!=\!-0.13$, and
$M_{\ast}\!=\!1.3\times 10^{13}\,h^{-1}\mathrm{M}_{\sun}$.
In their simulations, B01 observe a scatter of 
$\sigma(\log{c_{\mathrm{vir}}})\!=\!0.18$ for a fixed $M_{\mathrm{vir}}$.

For our purposes,
we insert $M_{200}^{\mathrm{wl}}(r_{200})$ for $M_{\mathrm{vir}}$ in Eq.~(\ref{eq:cm}).
Due to the weak dependence of $c_{\mathrm{B01}}$ on $M_{\mathrm{vir}}$ this results
only in a very small underestimate of $c_{\mathrm{B01}}$.
For two of the total of eight clusters we analysed, $c_{\mathrm{B01}}$ is
very close to the $c_{\mathrm{NFW}}$ obtained by lensing, while for others it
differs strongly (see Table~\ref{tab:massescb}).

Assuming the B01 mass--concentration relation, we apply a Gaussian prior
$p_{\mathrm{c}}(r_{200},c_{200})$ with standard deviation 
$\sigma(\log{c_{\mathrm{200}}})\!=\!0.18$ to the tabulated values of 
$\Delta\chi^{2}(r_{200},c_{\mathrm{NFW}})$ for each of our clusters, 
and marginalise over the $c_{200}$ dimension.
The radii $r_{200,\mathrm{B01}}$ and the corresponding masses 
$M_{200,\mathrm{B01}}^{\mathrm{wl}}(r_{200,\mathrm{B01}})$, found from the minimum of
$\sum_{j}{p_{\mathrm{c}}(r_{200},c_{200,j})\Delta\chi^{2}(r_{200},c_{200,j})}$
are listed in Table~\ref{tab:massescb}.

We notice that the simulations from which the B01 relation was measured assume
$\sigma_{8}\!=\!1.0$ to fix the normalisation of the matter power spectrum.
This value is inconsistent with more recent measurements of cosmological 
parameters \citep[e.g.][]{2011ApJS..192...16L,2012arXiv1202.2889B}.
Hence, we consider a second mass--concentration relation, based on a recent suite
of simulations employing $\sigma_{8}\!=\!0.8$ as favoured by currents models:
\citet[B12]{2011arXiv1112.5479B} study Dark Matter haloes of massive clusters
and find that the concentration parameter can be modelled with a single power
law when expressed in terms of the \emph{peak height parameter} $\nu$ from
linear collapse theory.\footnote{We use the fitting formula for $\nu(M,z)$ from
Table~2 of \citet{2011arXiv1112.5479B}.} Their simulated clusters are best 
represented by:
\begin{equation}
c_{200,\mathrm{B12}}(\nu)\!=\!D(z)^{0.5}\times 5.9\nu^{-0.35}\quad,
\label{eq:bhh12}
\end{equation}
with a variance of $\sigma_{\mathrm{c}}\!=\!0.33c_{200}$. We compute the growth 
factor $D(z)$ for a flat Universe with a cosmological constant.
In complete analogy to the B01 relation, we compute cluster radii
$r_{200,\mathrm{B12}}$ and masses $M_{200,\mathrm{B12}}^{\mathrm{wl}}(r_{200,\mathrm{B12}})$
for each cluster, given the $c_{200}$-$M_{200}$--relation resulting from
Eq.~(\ref{eq:bhh12}). The results are presented in Table~\ref{tab:massescb}.

\section{Special Cases} \label{sec:sec4}

\subsection{CL\,1701+6414} \label{sec:cl1701}

\begin{figure*}
\vspace{-0.5cm}
\centering
\includegraphics[width=17cm]{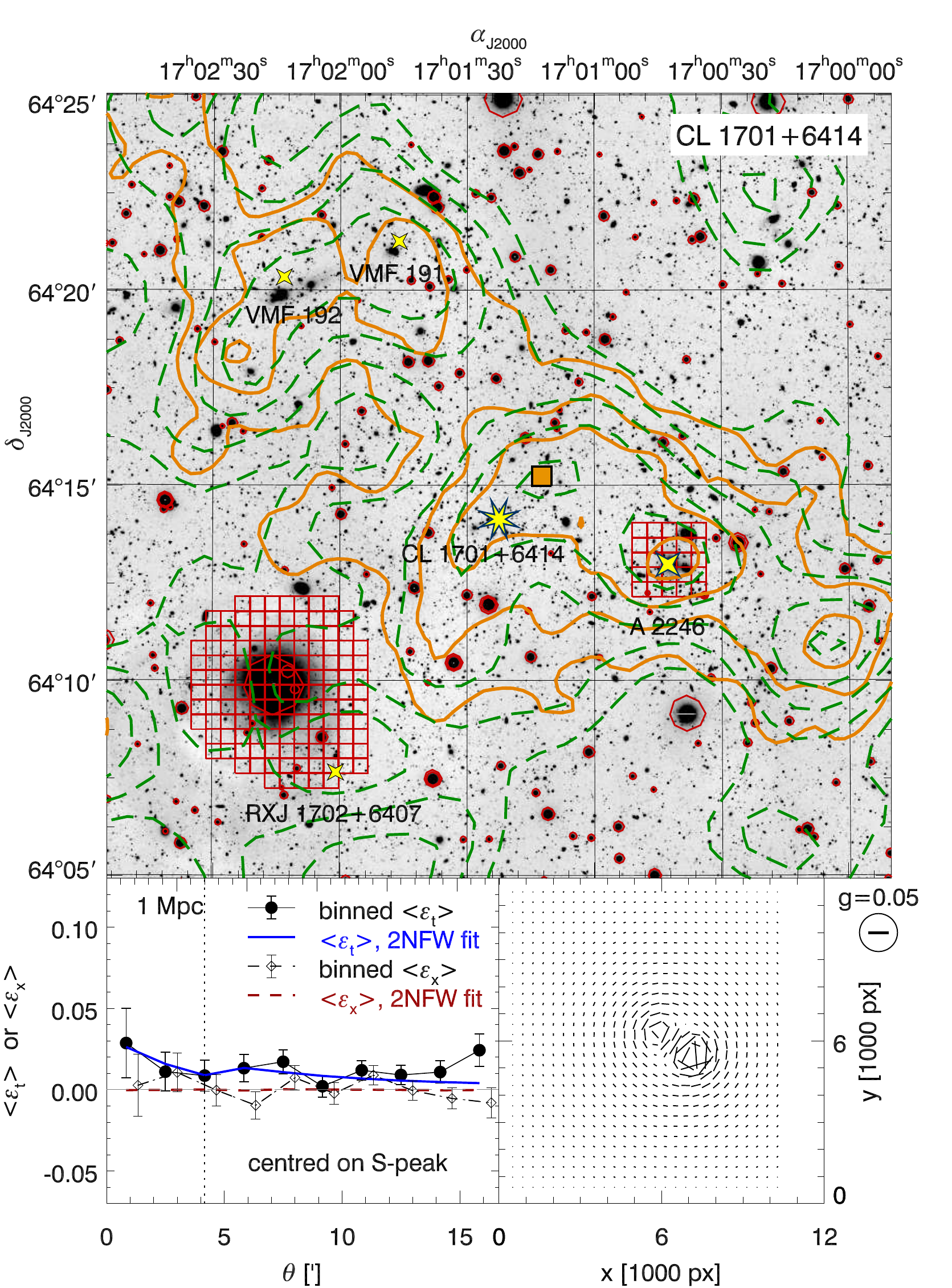}
\caption{Shear signal in the CL\,1701+6414 field and its best-fit model with
 two NFW components accounting for CL\,1701+6414 and A~2246. 
 \textit{Upper plot:} The layout follows Fig.~\ref{fig:cl1357}. 
 The \textsc{rosat} position of A~2246 is marked by a big four-pointed star 
 symbol. Smaller star symbols denote positions of further X-ray clusters.
 \textit{Lower left plot:} The layout follows Fig.~\ref{fig:cl1357}. The solid
 blue and dashed red lines give the \emph{mean} tangential and cross shear
 components, averaged in bins around the CL\,1701+6414 shear peak, as expected 
 from the two-cluster model. The separation of the two main clusters is
 indicated by a vertical dotted line.
 \textit{Lower right plot: } The orientations and amplitudes of the shear, as 
 expected from the best-fit two-cluster model, calculated on a regular grid.} 
\label{fig:cl1701}
\end{figure*}
\begin{figure*}[t]
\sidecaption
\centering
\includegraphics[width=13.5cm]{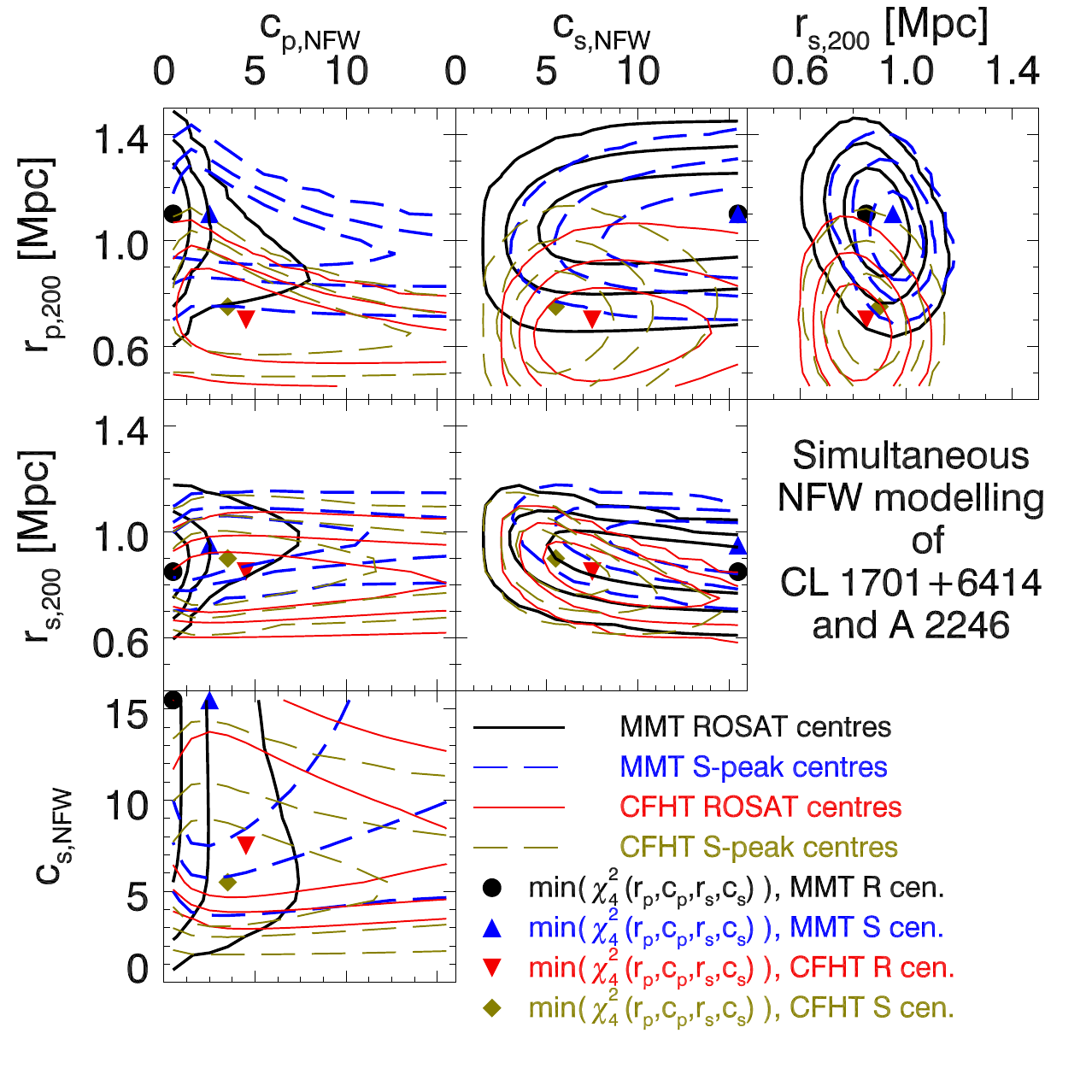}
\caption{Simultaneous NFW modelling of CL\,1701\-+6414 and A~2246. Each panel 
shows the dependencies between two of the four parameters, with the other two
marginalised. Solid confidence contours ($1\sigma$, $2\sigma$, $3\sigma$) 
denote the default case, using the \textsc{rosat} centres;
dashed contours denote models centred on the $S$-peaks. 
The respective parameters minimising 
$\chi_{4}^{2}$ are indicated by a filled circle and a upward triangle.
Sets of thin contours denote the confidence contours
and parameters minimising $\chi_{4}^{2}$ obtained from the analogous analysis of
the CFHT lensing catalogue. The best-fit values for the CFHT data 
are marked by downward- and upward-pointing triangles for the
 \texttt{ROSAT} and $S$-peak centres, respectively.}
\label{fig:l4d}
\end{figure*}

\begin{figure*}
 \includegraphics[width=17cm]{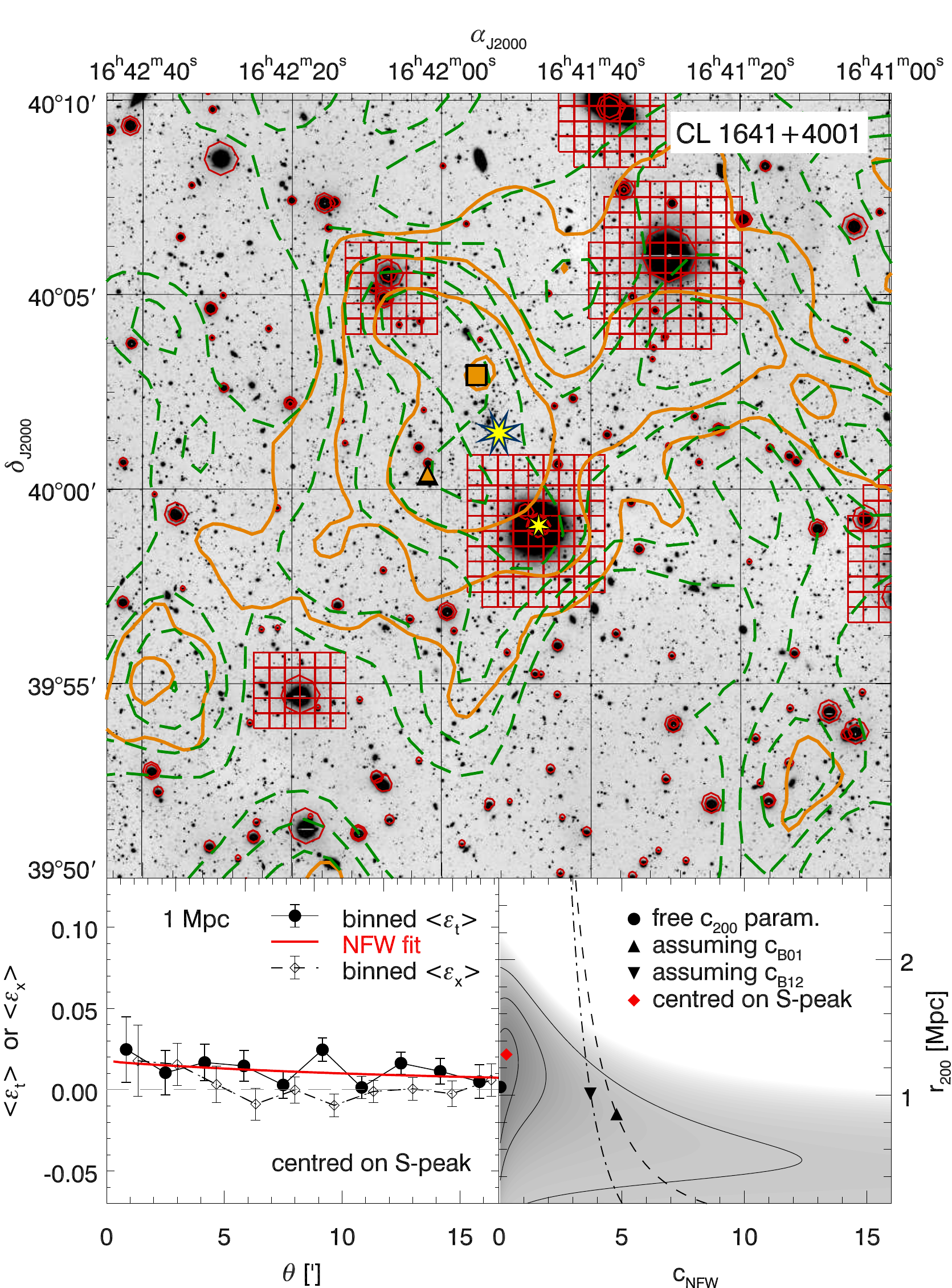}
 \caption{Same as Fig.~\ref{fig:cl1357}, but for CL\,1641+4001. In the map,
  a triangle denotes the secondary shear peak, while a small star symbol marks
  the position of the \citet{2007MNRAS.379..867V} cluster candidate.
  Note that no peak in the complex pattern of shear peaks correlates 
  with its position.}
 \label{fig:cl1641}
\end{figure*}

\subsubsection{X-ray clusters and shear peaks}
A weak lensing analysis of CL\,1701+6414 has to deal with shear by multiple 
structures. The strongest shear peak ($S\!=\!4.3\sigma$) in 
Fig.~\ref{fig:cl1701} coincides with the most prominent cluster in the field
amongst optical galaxies\footnote{Due to the high concentration of galaxies, 
the region is masked.}, Abell 2246 (big four-pointed star symbol in 
Fig.~\ref{fig:cl1701}), $4\farcm2$ to the west of CL\,1701+6414.
With a redshift of $z\!=\!0.225$ 
\citep{1998ApJ...502..558V,2007ApJS..172..561B}, A~2246 is part of the
\emph{400d} parent sample, but not of the distant cosmological sample. 
CL\,1701+6414, for whose detection in the $S$-statistics the lensing catalogue 
was optimised, is detected at the $3.7\sigma$ level.
The \textsc{rosat} catalogue of \citet{1998ApJ...502..558V} lists two 
further clusters in the field, VMF~191 at $z\!\!=\!\!0.220$ and 
VMF~192 at $z\!\!=\!\!0.224$ (small star symbols in Fig.~\ref{fig:cl1701}), 
which we identify with $S$-peaks of $2.9\sigma$ and $2.7\sigma$, respectively. 
Another $3.1\sigma$ peak lies  close-by. A zone of positive shear signal extends
over $>\!20\arcmin$, from the north-east of VMF~192 to a $3.6\sigma$ shear peak
south-west of A~2246, which does not correspond to a known cluster.
Noticing the very similar redshifts of A~2246, VMF~191, and VMF~192, we likely 
are observing a physical filament at $z\!=\!0.22$, through whose centre we see 
CL\,1701+6414 in projection. 
Luckily, a likely strong lensing arc, $10\arcsec$ to the west of the BCG of 
CL\,1701+6414 \citep[$z\!=\!0.44\!\pm\!0.01$,][]{1997A&A...326..489R} gives
direct evidence that CL\,1701+6414 acts as a gravitational lens.
We find no significant WL signal near the \textsc{rosat} source RX\,J1702+6407
\citep[cf.\ Appendix~\ref{sec:detcl1701}]{2002ApJ...569..689D}.  

\subsubsection{Two-cluster modelling of MMT data}
Plotting the binned tangential shear around the lensing centre (lower left panel
of Fig.~\ref{fig:cl1701}), we find a flat profile whose average 
$\langle\varepsilon_{\mathrm{t}}(\theta)\rangle\!>\!0$ is consistent with the 
extended shear signal in the $S$-map. The cross-component 
$\langle\varepsilon_{\times}(\theta)\rangle$ is consistent with zero.  
Acknowledging the prominent signal related to A~2246, we 
model the shear of CL\,1701+6414 and A~2246, simultaneously, using an 
NFW shear profile for each deflector. 

We assume both the shear $g_{\mathrm{p}}$ of the primary and $g_{\mathrm{s}}$ of the
secondary component to be small. In this limit, the shear components 
originating from both lenses become additive: 
\begin{equation}
g_{\mathrm{add},\alpha}(\pmb{\theta})\!=\!
g_{\mathrm{p},\alpha}(\pmb{\theta}; r_{\mathrm{p},200}, c_{\mathrm{p,NFW}})+
g_{\mathrm{s},\alpha}(\pmb{\theta}; r_{\mathrm{s},200}, c_{\mathrm{s,NFW}})\quad,
\end{equation}
with $\alpha\!=\!1,2$. Here, $r_{\mathrm{p},200}$, $r_{\mathrm{s},200}$, 
$c_{\mathrm{p,NFW}}$, and $c_{\mathrm{s,NFW}}$ are the radii and concentration 
parameters of the primary and secondary component. 
Note that $g_{\mathrm{add},\alpha}(\pmb{\theta})$ explicitly depends on the 
two-dimensional coordinate vector $\pmb{\theta}$: the shear field of two 
clusters no longer has radial, but only axial symmetry. This is illustrated in
the lower right panel of Fig.~\ref{fig:cl1701}, showing the shear fit expected
from the best-fit two-cluster model for the CL\,1701+6414 lensing catalogue, 
evaluated on a regular grid.
We consider a modification of the merit function given by Eq.~(\ref{eq:merit}):
\begin{equation} \label{eq:likeli4}
\chi^{2}_{\mathrm{4}}\!=\!\sum_{i=1}^{N_{\mathrm{gal}}}{\frac{\left|g_{\mathrm{add},i}
(r_{\mathrm{p,200}},c_{\mathrm{p,NFW}},r_{\mathrm{s,200}},c_{\mathrm{s,NFW}})\!-\!
\varepsilon_{i}\right|^{2}}{\sigma_{\mathrm{fit}}^{2}\left(1\!-\!\left|g_{\mathrm{add},i}
(r_{\mathrm{p,200}},c_{\mathrm{p,NFW}},r_{\mathrm{s,200}},c_{\mathrm{s,NFW}})\right|^{2}
\right)^{2}}}\quad.
\end{equation}
The symbol $\chi^{2}_{\mathrm{4}}$ highlights the dependence on four parameters, 
the radii and concentrations of the two clusters.
Note that $\chi^{2}_{\mathrm{4}}$ models the measured $\varepsilon_{i}$ directly, 
without recourse to a definition of the tangential component.

We assumed $\langle\langle\beta\rangle\rangle\!=\!0.381$ for 
CL\,1701+6414 (Table~\ref{tab:defmod}) and 
$\langle\langle\beta\rangle\rangle\!=\!0.640$ for A~2246, calculated
the same way as for the other clusters.
The \emph{average} tangential and cross-component of the shear expected 
in concentric annuli around the centre of CL\,1701+6414 are presented in the
lower left panel of Fig.~\ref{fig:cl1701}. 
A vertical dotted line denotes the separation of CL\,1701+6414 and A~2246.
We find a good agreement to the measured shear
and note that due to the lack of radial symmetry the dispersion of 
the \emph{model} values in the annuli is of the same order as the measurement 
errors. Although the cross-component can be large at some points in the image 
plane, $\langle g_{\times}\rangle$ cancels out nearly completely when averaging 
over the annuli.

Figure~\ref{fig:l4d} presents the confidence contours and parameters minimising
Eq.~(\ref{eq:likeli4}) for the default model (filled circle and solid contours).
The panels of Fig.~\ref{fig:l4d} show all combinations of two fit parameters, 
where we marginalised over the two remaining ones. 
Owing to the $4$-dimensional parameter space, we tested a coarse grid of
points to avoid excessive computing time. 
The picture emerges that $r_{\mathrm{p,200}}$ and $r_{\mathrm{s,200}}$ are relatively
independent of each other (top right panel). 
Hence, the presence of the respective other cluster does not seem to affect the
accuracy with which we can determine the masses of the two clusters strongly. 
The data favour the smallest tested value, $c_{\mathrm{p,NFW}}\!=\!0.5$ for the 
concentration of CL\,1701+6414, and the largest one, 
$c_{\mathrm{s,NFW}}\!=\!15.5$, for A~2246. 
Using shear peak cluster centres (dashed contours and
upward pointing triangle in
Figure~\ref{fig:l4d}), $c_{\mathrm{p,NFW}}$ is also very low, but
the uncertainties are large. 
The poor constraint on $c_{\mathrm{s,NFW}}$ might be partly due to the
masking of the centre of A~2246 or shear contribution by the BCG.

Given the absence of a strong covariance between the parameters of
A~2264 and CL\,1701+6414, we fix the parameters of the foreground cluster
to $r_{\mathrm{s,200}}\!=\!0.90\,\mbox{Mpc}$ and $c_{\mathrm{s,NFW}}\!=\!20$
and repeat the analysis with our usual, finer parameter grid. 
The best model is found for
$r_{\mathrm{p},200}^{\mathrm{min}}\!=\!0.94_{-0.29}^{+0.32}\,\mbox{Mpc}$ 
$c_{\mathrm{p,NFW}}^{\mathrm{min}}\!=\!0.1_{-0.1}^{+1.1}$, 
confirming the results from Fig.~\ref{fig:l4d}.
We note that we find a low $c_{\mathrm{p,NFW}}$, although our model
explicitly accounts for the extra shear by A~2246.
Using the default model, we compute
masses of $1.6_{-1.1}^{+2.3}\!\times\!10^{14}\,\mbox{M}_{\sun}$ 
for CL\,1701+6414 and 
$1.1_{-0.3}^{+0.4}\!\times\!10^{14}\,\mbox{M}_{\sun}$ for A~2246,
based on $r_{\mathrm{s},200}^{\mathrm{min}}\!=\!0.9\pm0.1\,\mbox{Mpc}$.

\subsubsection{Comparison to CFHT data} \label{sec:cfhtpre}

In addition, Fig.~\ref{fig:l4d} shows confidence contours obtained from a WL
analysis of CFHT observations of the CL\,1701+6414 field ($r'$-band, 
$\approx\!7200\,\mbox{s}$), which we discuss in greater detail in
Sect.~\ref{sec:cfhtsec}.
We repeated the two-cluster modelling using Eq.~(\ref{eq:likeli4})  
following the same data reduction and shear measurement pipelines. 
Besides  $m_{\mathrm{faint}}\!=\!20.2$ and the PSF-dependent galaxy selection, 
parameters are kept at their MMT values. 

The resulting cluster parameters minimising $\chi_{4}^{2}$ (red downward
triangles for \textsc{rosat} and diamonds for $S$-peak centres) 
and the corresponding thin confidence contours in
Fig.~\ref{fig:l4d} show agreement with the
MMT data within the $2\sigma$ margins or better.
With $r_{\mathrm{p},200}^{\mathrm{min}}\!=\!0.70\pm0.20\,\mbox{Mpc}$,
and $r_{\mathrm{s},200}^{\mathrm{min}}\!=\!0.85_{-0.10}^{+0.15}\,\mbox{Mpc}$,
relating to WL masses of 
$M_{200}^{\mathrm{CFHT}}\!=\!0.7_{-0.4}^{+0.7}\!\times\!10^{14}\,\mbox{M}_{\sun}$ for
CL\,1701+6414 and 
$M_{200}^{\mathrm{CFHT}}\!=\!0.9\pm0.4\!\times\!10^{14}\,\mbox{M}_{\sun}$
for A~2246, we arrive at lower masses, especially for CL\,1701+6414, but
consistent within the uncertainties of the MMT data.
Using the $S$-peaks as centres yields very similar results. 

Our CFHT data give more plausible best-fit concentration parameters
of $c_{\mathrm{s,NFW}}^{\mathrm{min}}\!=\!7.5_{-3.0}^{+>8}$ for A~2246, and 
$c_{\mathrm{p,NFW}}^{\mathrm{min}}\!=\!4.5_{-4.5}^{+>11}$ for CL\,1701+6414,
although the constraints are poor.
We conclude that a dual-NFW modelling is feasible, but more sensitive to
the choice of cluster centres than a single NFW fit to $r_{200}$ and 
$c_{\mathrm{NFW}}$. 
Adding more cluster components would even increase these interdependencies.
However, the main point here is that the MMT and CFHT analyses agree.

\subsection{CL\,1641+4001} \label{sec:res1641}

The $S$-statistics map of CL\,1641+4001 exhibits several shear peaks which form
a connected structure of $>\!20\arcmin$ extent (Fig.~\ref{fig:cl1641}). 
Located within a plateau of $S>\!3\sigma$ significance, the \textsc{rosat} 
centre of CL\,1641+4001 (big star symbol) is separated by $95\arcsec$ from the 
primary ($S\!=\!4.12$) shear peak and by $125\arcsec$ from
the secondary ($S\!=\!3.95$, orange triangle in Fig.~\ref{fig:cl1641}) shear
peak. The BCG of CL\,1641+4001 can be found between the \textsc{rosat} centre
and primary shear peak.

The $\langle\varepsilon_{\mathrm{t}}(\theta)\rangle$ profile (lower left panel of
Fig.~\ref{fig:cl1641}) centred on the main shear peak profile is flat,
with a positive average in all bins and the most significant positive signal at
$\sim\!\!9\arcmin$ distance from the cluster centre. In the innermost two bins
($\theta\!<\!3\farcm33$), $\langle\varepsilon_{\times}(\theta)\rangle$ is of 
similar amplitude as the tangential component, but consistent with zero at the
$1\sigma$ level. Similar to CL\,1701+6414, the modelling using 
Eq.~(\ref{eq:merit}) finds a very low $c_{\mathrm{NFW}}\!=\!0.1_{-0.1}^{+0.3}$, 
consistent with zero and reflecting the flat shear profile.

The only cluster candidate in the literature besides CL\,1641+4001 is
SDSS-C4-DR3\,3628 at $z\!=\!0.032$, identified in the SDSS Data Release~3,
using the \citet{2005AJ....130..968M} algorithm, but published solely by
\citet{2007MNRAS.379..867V}.
We test a two-cluster model, introducing a second component at the 
redshift of SDSS-C4-DR3\,3628, implying 
$\langle\langle\beta\rangle\rangle\!=\!0.940$. We choose the  
second-highest shear peak as the centre of the 
secondary component. The offset of $\sim\!3\arcmin$ to the coordinates of 
SDSS- C4-DR3\,3628 (small star symbol in 
Fig.~\ref{fig:cl1641}) is justified by the large mask at the latter position.
The two-cluster fit yields a mass of order $10^{14}\,\mbox{M}_{\sun}$ for
both the primary and the secondary component. This estimate is in stark 
disagreement with the absence of a massive, nearby cluster from our MMT
image, which  would have had to be missed by all but one cluster surveys.

At the same coordinates as SDSS-C4-DR3\,3628 and also at $z\!=\!0.032$, 
\texttt{NED} lists CGCG\,224$-$092, a \emph{galaxy pair}, dominated by the
bright elliptical UGC\,10512. These two galaxies are what we see in the
\megacam ~image\footnote{strongly overexposed and therefore masked} 
and also in the SDSS image of the area. 
Inspection of the respective \textsc{Chandra} image shows significant X-ray
emission, whose extent of $\sim\!30\arcsec$ in diameter ($\sim\!20\,\mbox{kpc}$
at $z\!=\!0.032$) is consistent with being caused by a massive elliptical 
galaxy or small galaxy group.
With $\approx\!1.7\times 10^{41}\,\mbox{erg}\,\mbox{s}^{-1}$ in the
$2$--$10$ keV range, its flux is high for a single galaxy, but the low
temperature of $\approx\!0.6\,\mbox{keV}$ (obtained by fitting an absorbed
\texttt{APEC} model) speaks against a galaxy group.
In conclusion, we deem it unlikely that the complex structure in the $S$-map of 
CL\,1641+4001 bears a significant contribution from the $z\!=\!0.032$
structure.

We prefer the hypothesis that the shear is caused by a complex structure at the
redshift of CL\,1641+4001, although its X-ray morphology does not hint at a 
merger \citep{2009ApJ...692.1033V}. Despite its shortcomings, we return to the 
simplest explanation for the time being and model CL\,1641+4001 by a single NFW
component: We obtain a minimum of $\chi^{2}$ for 
$r_{200}^{\mathrm{min}}\!=\!1.06_{-0.26}^{+0.30}\,\mbox{Mpc}$ and
$c_{\mathrm{NFW}}^{\mathrm{min}}\!=\!0.1_{-0.1}^{+0.3}$. These results, entailing a
mass estimate of $2.3_{-1.3}^{+2.6}\times\!10^{14}\,\mbox{M}_{\sun}$ are
illustrated by the filled circle and solid contours in the lower right panel of
Fig.~\ref{fig:cl1641}. Interestingly, choosing the secondary shear peak as a 
centre yields similar cluster parameters to those
found by choosing the primary shear peak. 
This could hint at a major merger of similarly massive substructures,
but more observations are needed to test this hypothesis.

\section{Verification with independent data} \label{sec:cfhtsec}

\subsection{CFHT Observations} \label{sec:cfhtin}

\begin{table}
 \centering
 \caption{Observation dates, final exposure times and seeing values in the
  coadded CFHT/\textsc{MegaCam} data for CL\,1701+6414.}
 \label{tab:cdata}
 \begin{center} 
 \begin{tabular}{cccc} \hline\hline
 Filter & Observation Dates & $T_{\mathrm{exp}}$ & Seeing \\ \hline
 $g'$ & 2006-03-07 & $1601\,\mbox{s}$ & $0\farcs94$ \\
 $r'$ & 2006-05-29 & $7179\,\mbox{s}$ & $0\farcs66$ \\
 $i'$ & 2006-04-26 & $1922\,\mbox{s}$ & $0\farcs84$ \\
 $z'$ & 2006-04-22,  2006-04-26& $1801\,\mbox{s}$ & $0\farcs82$ \\ \hline\hline
 \end{tabular}
 \end{center}
\end{table}
\begin{figure*}
\sidecaption
\includegraphics[width=10cm]{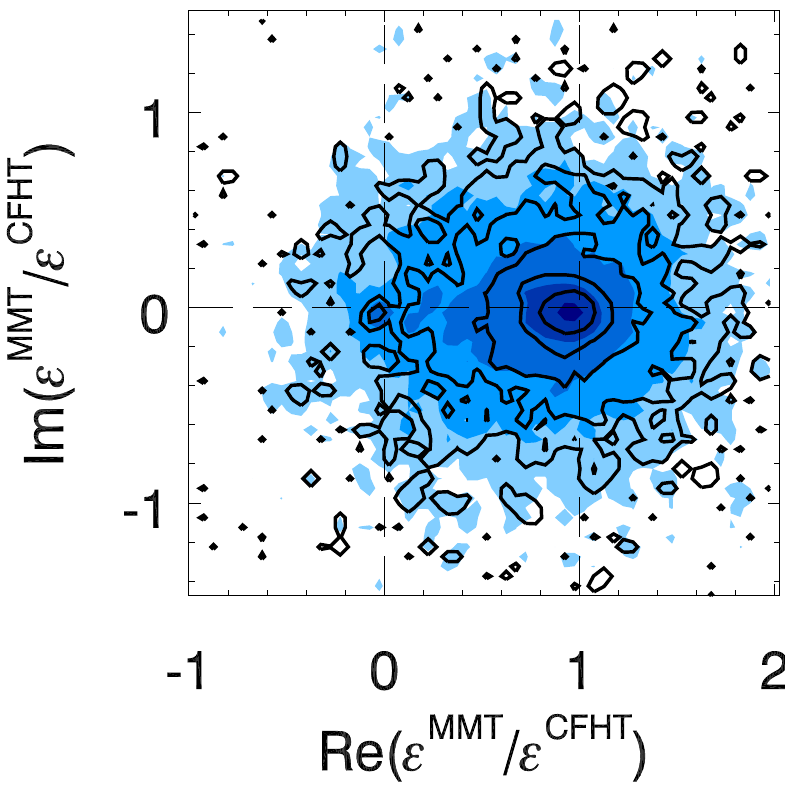}
 \caption{Sample density of the ratio 
 $\varepsilon^{\mathrm{MMT}}/\varepsilon^{\mathrm{CFHT}}$ of the complex ellipticities
 measured for the matched galaxies from the MMT and CFHT $r'$-band catalogues,
 respectively.
 The shaded contours correspond to the logarithmic densities of all galaxies 
 from the MMT lensing catalogue which have a match in the CFHT catalogue. 
 Solid contours give the density of galaxies detected with a signal/noise ratio
 of $\nu>\!15$, the top $32.6$~\%. Note that the normalisation of the 
 $\nu>\!15$ galaxies is scaled up by $1/0.326$ to obtain the same logarithmic 
 contour levels.
 A Gaussian smoothing kernel of full-width half-maximum $0.075$ was applied to
 both contour maps.}
 \label{fig:reimq}
\end{figure*}

CL\,1701+6414 is the only cluster we observed with MMT/\textsc{Megacam} for
which deep, lensing-quality data obtained with another telescope exist. 
It has been observed in the $g'r'i'z'$ filters (P.I.: G.~Soucail, 
Run ID: 2006AF26) using the 
\textsc{MegaPrime/MegaCam} at the Canada-France-Hawaii Telescope 
(CFHT).\footnote{For the sake of clarity, we use ``MMT'' and ``CFHT'' to 
distinguish the data sets.} 
Table~\ref{tab:cdata} lists the specifications of the CFHT data set.
The CFHT data are processed with \texttt{THELI} in the same way as the MMT data,
with a few CFHT-specific modifications to the code 
\citep[cf.][]{2009A&A...493.1197E}, making use of the pre-processing available
for archival CFHT data.
Hence, the results are a suite of coadded and calibrated images in the 
$g'r'i'z'$ passbands, centred on CL\,1701+6414, and with a side length of
$\sim\!1\degr$ each.
From the $g'r'i'$ images, we derived the pseudo-colour images of the centres 
of CL\,1701+6414 and A~2246 in Fig.~\ref{fig:zooms}.

We employ the CFHT data for two kinds of consistency checks with the MMT data:
First, we run the lensing pipeline on the deep CFHT $r'$ band image, applying
the same shape recovery technique to the same objects, but observed with
different instruments. The results of this comparison are detailed in
Sect.~\ref{sec:ellell}.
Second, making use of the CFHT imaging in four bands, we produced a \texttt{BPZ}
\citep{2000ApJ...536..571B} photometric redshift catalogue 
(Sect.~\ref{sec:photozres} and Appendix~\ref{sec:bpz})
with the goal of testing the single-band (magnitude cut) background selection 
in the CL\,1701+6414 MMT lensing catalogue.

\subsection{Comparative shape analysis} \label{sec:ellell}

In this Subsection, we compare shape measurements obtained in the same field
(the one of CL\,1701+6414) using the MMT/\textsc{Megacam} and 
CFHT/\textsc{Megacam} instruments (cf.\ Sect.~\ref{sec:cfhtin}).
Using the same parameter settings for our KSB pipeline, we extracted a KSB
catalogue from the CFHT $r'$-band image.
Subsequently, the CFHT and MMT catalogues were matched, using the 
\texttt{associate} and \texttt{make\_ssc} tools available in \texttt{THELI}.
With the smaller field-of-view of MMT/\textsc{Megacam} defining the location of
possible matches, $68.2$\% of sources in the MMT KSB catalogue are matched to a
CFHT detection. 
Larger masked areas in the CFHT image -- in particular due to reflections
(so-called ghosts) around very bright stars -- are the main cause impeding 
a higher matching fraction. 
Inside the MMT area (measuring at a safe distance from its low-weight edges),
we find $85.5$\% of the CFHT sources to be detected by MMT.

We note that objects in the matched catalogue have comparable 
\texttt{SExtractor} signal-to-noise ratios $\nu$ in both $r'$-band images.
Considering objects with $\nu_{\mathrm{MMT}}\!>\!15$ -- the top quartile of all
objects in the catalogue of matches -- for which selection effects should be 
negligible, we measure 
$\langle\nu_{\mathrm{CFHT}}/\nu_{\mathrm{MMT}}\rangle = 0.832$, with a dispersion of
$0.057$. These values show little dependence on the limiting value of 
$\nu_{\mathrm{MMT}}$, and confirm the visual impression that the $r'$-band 
images are of similar depth.\footnote{While the good $T_{\mathrm{exp}}$ is 
similar for both data sets, the larger mirror area of MMT is probably offset by
the better seeing in the CFHT image.}   

With these preparatory analyses in mind, we investigate the relation between the
ellipticities observed with CFHT and MMT. Figure~\ref{fig:reimq} presents the
ratio $\varepsilon^{\mathrm{MMT}}/\varepsilon^{\mathrm{CFHT}}$ of the complex 
ellipticities measured by KSB on the MMT and CFHT images.\footnote{As the complex
ellipticity is the relevant observable, we prefer considering the components of
the ratio $\varepsilon^{\mathrm{MMT}}/\varepsilon^{\mathrm{CFHT}}$ over the 
ratios for the individual components as measured with the two instruments.}
Shaded contours in Fig.~\ref{fig:reimq} mark lines of equal density of the
distribution of $\varepsilon^{\mathrm{MMT}}/\varepsilon^{\mathrm{CFHT}}$, as measured
from the sources passing the criteria for the MMT galaxy catalogue
(cf.\ Sect.~\ref{sec:bg7}).
Using a grid of mesh size as small as $0.01$ for both the real and imaginary 
axes, we find the density distribution of 
$\varepsilon^{\mathrm{MMT}}/\varepsilon^{\mathrm{CFHT}}$
to scatter around its peak at unity.
Note that the logarithmic scaling in Fig.~\ref{fig:reimq} emphasises the wings
of the distribution.
When repeating the analysis restricted to galaxies detected with 
$\nu_{\mathrm{MMT}}\!>\!15$ -- the top $32.6$~\% of the matched sources contained
in the MMT galaxy catalogue -- the peak at 
$\varepsilon^{\mathrm{MMT}}\!=\!\varepsilon^{\mathrm{CFHT}}$
persists, while the scatter is slightly reduced 
(solid contours in Fig.~\ref{fig:reimq}). 
This can be seen comparing the two outermost solid contours to the shaded
contours, indicating the same levels of number density.

This means, any systematic bias between shear measurements obtained with MMT
and CFHT is smaller than a few percent. We expect a small bias, below the
sensitivity of our measurement, to be present because of the dependence of the
shear calibration factor $f_{0}$ on magnitude and half-light radius $\vartheta$
\citep[cf.\ Appendix~C of][]{2009A&A...504..689H}.
In addition to our results from Paper~I, the consistent galaxy ellipticities
measured with MMT and the well-established CFHT/\textsc{Megacam} mark further
evidence that MMT/\textsc{Megacam} is well-suited for measuring weak
gravitational lensing signals.

\subsection{$S$-statistics from CFHT and MMT} \label{sec:kappakappa}

\begin{figure*}
 \includegraphics[width=16cm]{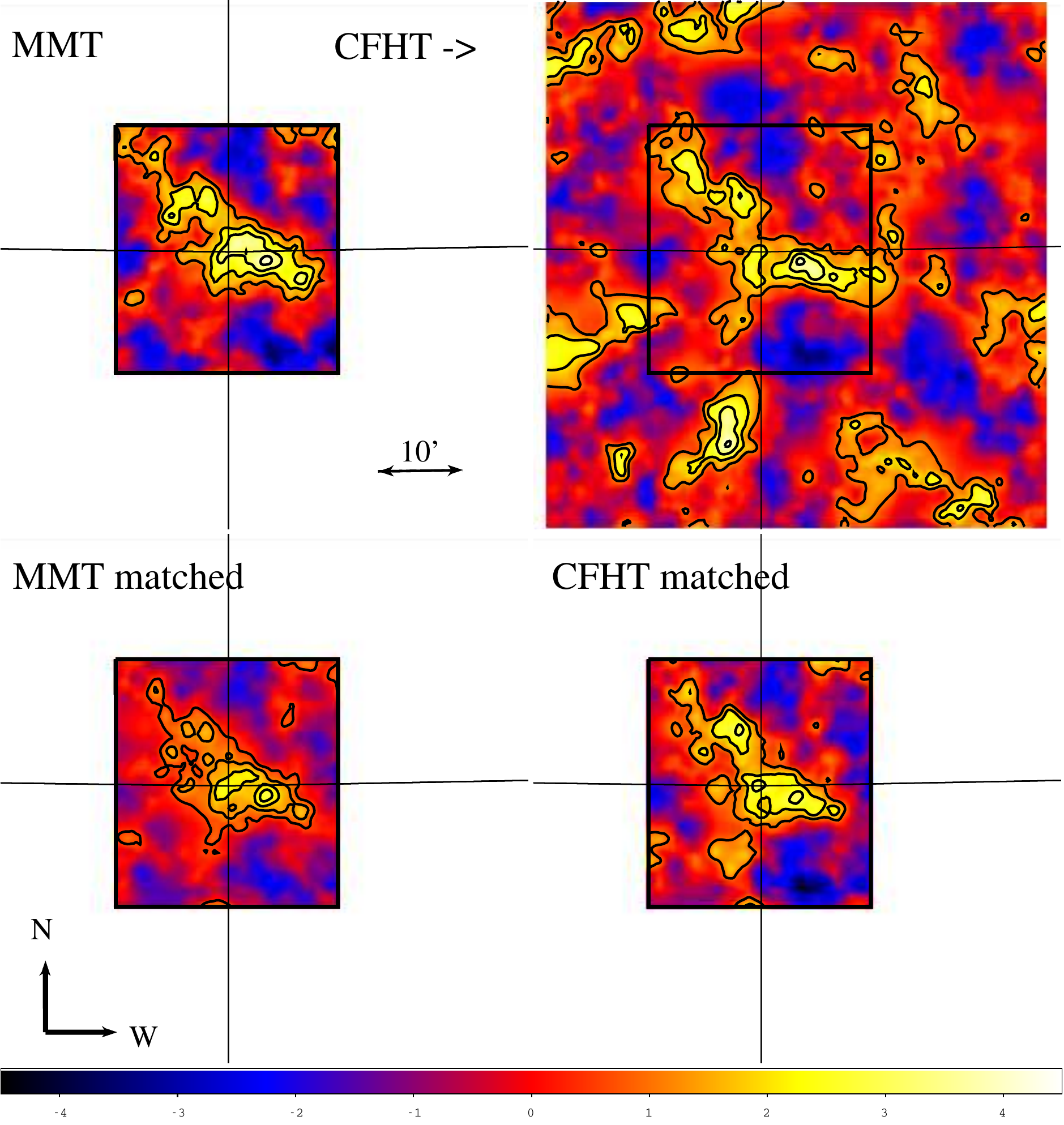}
 \caption{$S$-statistics in the CL\,1701+6414 field drawn from the MMT 
(top left), CFHT (top right), and matched sources catalogues (bottom panels).
The linear colour scale, contours indicating levels of $S\!=\!1$ to $S\!=\!4$,
$\theta_{\mathrm{out}}\!=\!14\farcm5$, and cross-hairs at the position of 
CL\,1701+6414 are the same in all panels.
Thick black squares outline the MMT field-of-view.}
 \label{fig:cfhtmmt}
\end{figure*}

Figure \ref{fig:cfhtmmt} provides a qualitative comparison of the $S$-maps for
CL\,1701+6414 obtained with both CFHT and MMT. Its upper two panels show the
independent shear catalogues drawn from the $r'$ images of both instruments, 
at the respective optimal values $m_{\mathrm{faint}}^{\mathrm{MMT}}\!=\!21.9$ and 
$m_{\mathrm{faint}}^{\mathrm{CFHT}}\!=\!20.2$ for $\theta_{\mathrm{out}}\!=\!14\farcm5$.
The distribution of the $S$-signal in the overlapping region inside the MMT
field-of-view (black square in Fig.~\ref{fig:cfhtmmt}) is astonishingly similar:
Not only do we find the tentative filament from the north-east of VMF~192 to the
south-west of A~2246 (compare Fig.~\ref{fig:cl1701} and the black lines in 
Fig.~\ref{fig:cfhtmmt} indicating the $\alpha_{\mathrm{J2000}}$ and 
$\delta_{\mathrm{J2000}}$ of CL\,1701+6414).
Moreover, also the regions of high $S$ at the eastern and north-western edges of
the MMT field-of-view correspond to peaks in the CFHT $S$-map.
Whereas the detection significance at the peak closest to the position of
CL\,1701+6414 is smaller for CFHT ($S\!=\!2.89$ compared to $S\!=\!3.75$), it is
also more prominent in the sense of a deeper ``valley'' separating it from the
dominant A~2246 peak ($S\!=\!4.30$ in both the MMT and CFHT maps).

The second-most significant ($4.08\sigma$) shear peak in the CFHT
$S$-map is at $\alpha_{\mathrm{J2000}}\!=\!17^{\mathrm{h}}01^{\mathrm{m}}57^{\mathrm{s}}$,
$\delta_{\mathrm{J2000}}\!=\!+63\degr51\arcmin$, outside the southern edge of the
MMT field-of-view, with no known cluster but several brighter ($r'\!<\!20$) 
galaxies in the vicinity.

Can the subtle differences between the MMT and CFHT $S$-maps be attributed to 
shape noise or rather to selection of galaxies at the faint end?
We investigate that by considering the matched-sources catalogue from
Sect.~\ref{sec:ellell} and apply to it the combined selection criteria for the
MMT and CFHT lensing catalogues (e.g.\ both $|\varepsilon^{\mathrm{MMT}}|\!<\!0.8$
and $|\varepsilon^{\mathrm{CFHT}}|\!<\!0.8$).
The resulting $S$-maps derived from the MMT and CFHT ellipticities of the exact
same sources are displayed in the lower left and lower right panels of
Fig.~\ref{fig:cfhtmmt}. 
Naturally, all matched sources are located within the MMT field-of-view.
Qualitatively, the matched $S$-maps again show the same structure, although they
do not appear to be much more similar than the $S$-maps drawn from the 
individual catalogues.
This indicates that galaxy selection plays a relevant role.
We note that based on the CFHT shapes of the matched galaxies, CL\,1701+6414 is
the most significant detection with $S\!=\!3.46$, while the A~2246 peak is
suppressed by $\approx\!1\sigma$ compared to the pure CFHT map.

Quantitatively, the Pearson correlation coefficient of $\varrho\!=\!0.912$ 
between the matched-sources $S$ maps substantiates the visual 
impression of a high correlation. 
When the faintest $\approx\!20$~\% of galaxies are removed from the matched 
catalogue, considering only galaxies brighter than $r'_{\mathrm{cut}}\!=\!24.85$ 
in both the MMT and CFHT images, this value increases to $\varrho\!=\!0.926$.
Removing the faintest $\approx\!40$~\% of galaxies by imposing 
$r'_{\mathrm{cut}}\!=\!24.45$, it further rises to $\varrho\!=\!0.938$. 

The detection of the same shear peaks reassures us that the multi-peaked 
$S$-distribution analysed in detail in Sect.~\ref{sec:cl1701} traces an actual
shear signal and removes any doubts that the $S$-filament across the
MMT field-of-view could be merely an instrument-dependent artefact, 
e.g.\ residuals of improper PSF anisotropy correction.

\begin{figure}
\includegraphics[width=9cm]{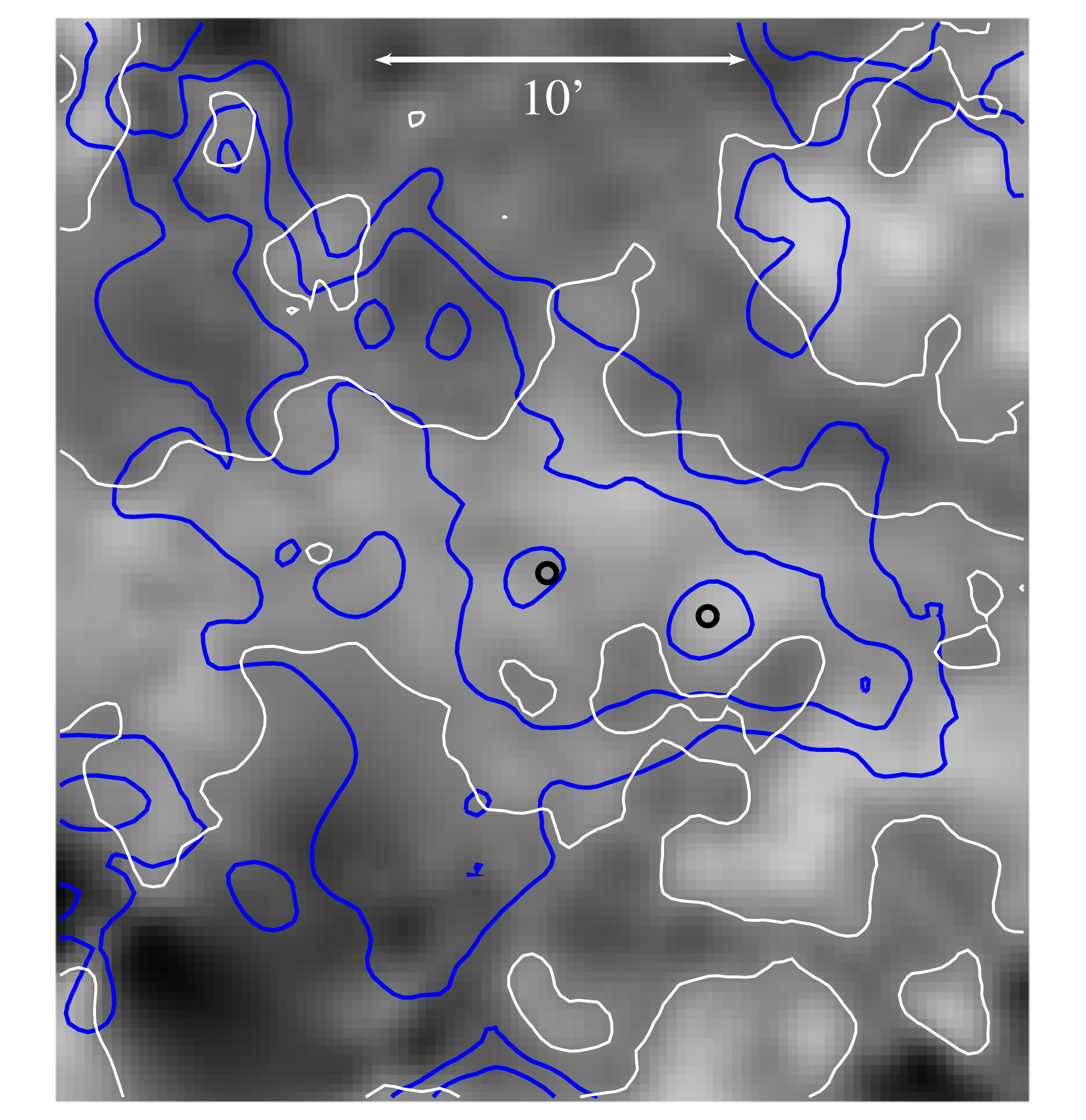}
\caption{Average and difference aperture mass $\mathcal{M}^{\pm}$
(Eq.~\ref{eq:mpm}), measured in the matched MMT-CFHT catalogue. The grey-scale
and white contours give $\mathcal{M}^{-}$, the thicker blue contours show 
$\mathcal{M}^{+}$. The spacing for both contours is in multiples of $0.015$,
starting at $0$. Black circles mark the positions of CL\,1701+6414 and A~2246.}
\label{fig:mpm3}
\end{figure}
As a final test to the hypothesis that we see the same shear signal measured in
the ellipticities from both instruments, we consider the average and difference
aperture mass of the matched sources:
\begin{equation} \label{eq:mpm}
\mathcal{M}^{\pm}\!=\!\pi\theta_{\mathrm{out}}^{2}N(\pmb{\theta}_{\mathrm{c}})^{-1}
\sum_{j}{\varepsilon^{\pm}_{\mathrm{t,}j}
Q_{j}(|\pmb{\theta}_{\!j}\!-\!\pmb{\theta}_{\mathrm{c}}|/\theta_{\mathrm{out}})}
\end{equation}
with $\varepsilon^{\pm}_{\mathrm{t,}j}$ the tangential component of
\begin{equation}
\varepsilon^{\pm}_{j}=
(\varepsilon^{\mathrm{MMT}}_{j}\!\pm\!\varepsilon^{\mathrm{CFHT}}_{j})/2
\end{equation}
and the index $j$ running over all $N(\pmb{\theta}_{\mathrm{c}})$ galaxies within
a distance $\theta_{\mathrm{out}}$ from $\pmb{\theta}_{\mathrm{c}}$.
The outcome of this experiment is shown in Fig.~\ref{fig:mpm3}:
While $\mathcal{M}^{+}$ (blue contours) retrieves the signal of both clusters
(black circles), exhibiting the expected great similarity to the matched-sources
$S$-maps in Fig.~\ref{fig:cfhtmmt}, $\mathcal{M}^{-}$ (grey-scale and white
contours in Fig.~\ref{fig:mpm3}) has a much smaller amplitude. 
Its pattern is not obviously related to the one seen in $\mathcal{M}^{+}$:
Although the main clusters reside in a region of enhanced $\mathcal{M}^{-}$,
they do not correspond to peaks in $\mathcal{M}^{-}$.
The absence of the $\mathcal{M}^{+}$-peaks in $\mathcal{M}^{-}$ is consistent
with the absence of a noticeable shear calibration bias between MMT and CFHT
(cf.\ Sect.~\ref{sec:ellell}).
A possible explanation for the stripe-like pattern in $\mathcal{M}^{-}$ are
differences in the spatially varying anisotropy correction. 
We conclude these effects to be small and no impediment to direct comparisons of
MMT and CFHT WL measurements, which we conclude to be consistent.

\paragraph{The SW peak}

A peculiar feature in the MMT $S$-map of CL\,1701+6414 is the $3.6\sigma$
shear peak at $\alpha_{\mathrm{J2000}}\!=\!17^{\mathrm{h}}00^{\mathrm{m}}05^{\mathrm{s}}$,
$\delta_{\mathrm{J2000}}\!=\!+64\degr11\arcmin00$), south-west of A~2246
(cf.\ Fig.~\ref{fig:cl1701}).
This peak does not correspond to an evident overdensity of galaxies in the
MMT and CFHT $r'$-band images, nor to an extended emission in the 
\textsc{Chandra} X-ray images.
The pure CFHT $S$-map does not show a counterpart to the SW peak detected
with MMT, although the $S$-contours of A~2246 are extended towards its 
direction.
Interestingly, in the matched-sources $S$-maps, we detect a $2.8\sigma$ peak
from the MMT ellipticities and a $2.4\sigma$ peak from the CFHT ellipticities. 
We checked that the SW peak in the MMT $S$-map does not arise from a chance 
alignment of a few galaxies with extraordinary high $\varepsilon_{\mathrm{t}}$ 
resulting from stochastic shape noise.
Considering the observations in the CFHT and matched-catalogue $S$-maps, it
seems likelier that we observe a true ``shear peak'' arising from the superposed
light deflections of line-of-sight structure in the complex and dense environment
of the CL\,1701+6414/A~2246 field.

\subsection{Photometric redshift results} \label{sec:photozres}

Using photo-$z$s based on the available $g'r'i'z'$ observations 
(Table~\ref{tab:cdata}) is challenging because of the small spectral coverage
and shallowness of the data. Nevertheless, a comparison with sources for which
SDSS spectroscopic redshifts are known, revealed a coarse redshift sorting to
be possible, with typical errors of $\sigma(z_{\mathrm{ph}})\!\approx\!0.25$ for
the relevant $z\!\lesssim\!0.5$ redshift range (Appendix~\ref{sec:zszp}).
Matching the photo-$z$ catalogue with the CL\,1701+6414 MMT data set, we can
identify most sources in the galaxy catalogue with a photo-$z$ galaxy, albeit
with low quality for most sources (Appendix~\ref{sec:bpzmmt}).

Drawing the $S$-statistics from this catalogue, whereby galaxies are sorted
based on their CFHT photo-$z$s, we retrieve shear peaks similar to 
Fig.~\ref{fig:cl1701} for the background catalogue, while CL\,1701+6414 does not
show up as a shear peak in the foreground catalogue (Appendix~\ref{sec:rzig}).
From this, we draw two conclusions: First, we likely see an indication of shear
emanating from more than one lens plane, namely CL\,1701+6414 on the one hand
and A~2246 and associated structures on the other hand.
Second, owing to the poor quality of the four-band photo-$z$, a desirable
calibration of single-band shear catalogues lies beyond the grasp of this data
set.

\section{Accuracy of the mass estimates} \label{sec:accuracy}

\begin{table*}
 \caption{Weak lensing masses resulting from our analysis. Given are the WL 
  masses $M_{200,\mathbf{B12}}^{\mathrm{wl}}(r_{200,\mathbf{B12}}^{\mathrm{wl}})$, 
  assuming the B12 mass-concentration relation (cf.\ Table~\ref{tab:massescb}),
  their lower and upper
  statistical ($\sigma^{-}_{\mathrm{stat}}$ and $\sigma^{+}_{\mathrm{stat}}$), 
  systematic ($\sigma^{-}_{\mathrm{sys}}$ and $\sigma^{+}_{\mathrm{sys}}$), and total
  error margins ($\sigma^{-}_{\mathrm{tot}}$ and $\sigma^{+}_{\mathrm{tot}}$). 
  In addition, the corresponding relative errors are presented. 
  All masses are given in units of $10^{14}\,\mbox{M}_{\sun}$.}
 \begin{center}
 \centering
 \begin{tabular}{c|ccccccc|cccccc} \hline\hline
 Cluster & $M_{\mathrm{wl}}(r_{200,\mathrm{wl}})$ & 
 $\sigma^{-}_{\mathrm{stat}}$ & $\sigma^{+}_{\mathrm{stat}}$ & 
 $\sigma^{-}_{\mathrm{sys}}$ & $\sigma^{+}_{\mathrm{sys}}$ & 
 $\sigma^{-}_{\mathrm{tot}}$ & $\sigma^{+}_{\mathrm{tot}}$ &
 $\dfrac{\sigma^{-}_{\mathrm{stat}}}{M^{\mathrm{wl}}}$ & 
 $\dfrac{\sigma^{+}_{\mathrm{stat}}}{M^{\mathrm{wl}}}$ &
 $\dfrac{\sigma^{-}_{\mathrm{sys}}}{M^{\mathrm{wl}}}$ & 
 $\dfrac{\sigma^{+}_{\mathrm{sys}}}{M^{\mathrm{wl}}}$ &
 $\dfrac{\sigma^{-}_{\mathrm{tot}}}{M^{\mathrm{wl}}}$ & 
 $\dfrac{\sigma^{+}_{\mathrm{tot}}}{M^{\mathrm{wl}}}$  \\ \hline
 CL\,0030+2618 & $6.23$ & $1.76$ & $1.87$ & $2.07$ & $1.97$ & $2.72$ & $2.72$ &
                 $28$\% & $30$\% & $33$\% & $32$\% & $44$\% & $44$\%  \\
 CL\,0159+0030 & $4.10$ & $2.07$ & $2.46$ & $1.25$ & $1.31$ & $2.42$ & $2.78$ &
                 $50$\% & $60$\% & $30$\% & $32$\% & $59$\% & $68$\%  \\
 CL\,0230+1836 & $10.23$ & $5.18$ & $6.59$ & $4.68$ & $3.59$ & $6.98$ & $7.51$ &
                 $51$\% & $64$\% & $46$\% & $35$\% & $68$\% & $73$\%  \\
 CL\,0809+2811 & $9.24$ & $3.14$ & $3.46$ & $2.59$ & $2.62$ & $4.07$ & $4.34$ &
                 $34$\% &$37$\% & $28$\% & $28$\% & $44$\% & $47$\%  \\
 CL\,1357+6232 & $2.52$ & $1.34$ & $1.74$ & $0.99$ & $0.89$ & $1.67$ & $1.96$ &
                 $53$\% & $69$\% & $39$\% & $35$\% & $66$\% & $78$\%  \\
 CL\,1416+4446 & $1.71$ & $0.79$ & $0.95$ & $0.60$ & $0.58$ & $1.00$ & $1.12$ &
                 $46$\% & $56$\% & $35$\% & $34$\% & $58$\% & $65$\%  \\
 CL\,1641+4001 & $2.05$ & $1.26$ & $1.51$ & $0.78$ & $0.73$ & $1.48$ & $1.68$ &
                 $61$\% & $74$\% & $38$\% & $36$\% & $72$\% & $82$\%  \\
 CL\,1701+6414 & $2.03$ & $0.98$ & $1.17$ & $0.78$ & $0.70$ & $1.25$ & $1.37$ &
                 $48$\% & $58$\% & $38$\% & $35$\% & $61$\% & $67$\%  
  \\ \hline\hline
 \end{tabular}
  \label{tab:masses}
 \end{center}
\end{table*}
\begin{table*}
 \caption{Components of the statistical error, assuming the B12
   mass-concentration relation. 
   We list all components entering (Eq.~\ref{eq:err}):
   The uncertainties $\sigma^{\pm}_{\mathrm{cali}}$ due to shear calibration, and 
   $\sigma^{\pm}_{\mathrm{geom}}$ from $\langle D_{\mathrm{ds}}/D_{\mathrm{s}}\rangle$, 
   the projectional uncertainty $\sigma^{\pm}_{\mathrm{proj}}$ due to cluster 
   triaxiality, and $\sigma^{\pm}_{\mathrm{LSS}}$ from the projection of
   unrelated LSS. All errors are given in units of $10^{14}\,\mbox{M}_{\sun}$;
   the numbers in parentheses present the relative uncertainties.
   The relative statistical uncertainties are relatively independent 
   of the choice of the concentration parameter.}
 \begin{center}
  \begin{tabular}{c|cc|cc|cc|cc} \hline\hline
   Cluster & $\sigma^{-}_{\mathrm{cali}}$ & $\sigma^{+}_{\mathrm{cali}}$ & 
   $\sigma^{-}_{\mathrm{geom}}$ & $\sigma^{+}_{\mathrm{geom}}$ & 
   $\sigma^{-}_{\mathrm{proj}}$ & $\sigma^{+}_{\mathrm{proj}}$ & 
   $\sigma^{-}_{\mathrm{LSS}}$ & $\sigma^{+}_{\mathrm{LSS}}$ \\\hline
 CL\,0030+2618 & $1.09$ ($18$\%) & $0.26$ ($4$\%) & $0.65$ ($10$\%) & 
                 $0.73$ ($12$\%) & $0.62$ ($10$\%) & $1.00$ ($16$\%) & 
                 $1.52$ ($24$\%) & $1.62$ ($24$\%) \\
 CL\,0159+0030 & $0.38$ ($9$\%) & $0.14$ ($3$\%) & $0.27$ ($7$\%) & 
                 $0.29$ ($7$\%) & $0.41$ ($10$\%) & $0.66$ ($16$\%) & 
                 $1.08$ ($26$\%) & $1.08$ ($26$\%) \\
 CL\,0230+1836 & $3.75$ ($37$\%) & $0.58$ ($6$\%) & $1.56$ ($16$\%) & 
                 $2.36$ ($23$\%) & $1.02$ ($10$\%) & $1.64$ ($16$\%) & 
                 $2.08$ ($20$\%) & $2.08$ ($20$\%) \\
 CL\,0809+2811 & $1.19$ ($13$\%) & $0.39$ ($4$\%) & $0.74$ ($8$\%) & 
                 $0.80$ ($9$\%) & $0.92$ ($10$\%) & $1.48$ ($16$\%) & 
                 $1.97$ ($21$\%) & $1.97$ ($21$\%) \\
 CL\,1357+6232 & $0.58$ ($23$\%) & $0.14$ ($6$\%) & $0.26$ ($10$\%) & 
                 $0.33$ ($13$\%) & $0.25$ ($10$\%) & $0.40$ ($16$\%) & 
                 $0.71$ ($28$\%) & $0.71$ ($28$\%) \\
 CL\,1416+4446 & $0.28$ ($16$\%) & $0.07$ ($4$\%) & $0.11$ ($7$\%) & 
                 $0.12$ ($7$\%) & $0.17$ ($10$\%) & $0.27$ ($16$\%) & 
                 $0.44$ ($29$\%) & $0.44$ ($29$\%)  \\
 CL\,1641+4001 & $0.41$ ($20$\%) & $0.12$ ($6$\%) & $0.24$ ($12$\%) & 
                 $0.27$ ($13$\%) & $0.21$ ($10$\%) & $0.33$ ($16$\%) & 
                 $0.58$ ($29$\%) & $0.58$ ($29$\%) \\
 CL\,1701+6414 & $0.42$ ($21$\%) & $0.12$ ($6$\%) & $0.21$ ($10$\%) & 
                 $0.21$ ($10$\%) & $0.20$ ($10$\%) & $0.32$ ($16$\%) & 
                 $0.58$ ($29$\%) & $0.58$ ($29$\%) \\ \hline \hline
  \end{tabular} 
 \label{tab:erran}
 \end{center}
\end{table*}

\subsection{Error Analysis} \label{sec:err7}

The error analysis of the seven clusters analysed in Sect.~\ref{sec:sec3} 
follows the method described in Paper~I, i.e.\ we apply 
\begin{equation}
\sigma_{\mathrm{tot}}^{2}=\sigma_{\mathrm{stat}}^{2}\!+\!
\sigma_{\mathrm{sys}}^{2}=\sigma_{\mathrm{stat}}^{2}\!+\!
\sigma_{\mathrm{LSS}}^{2}\!+\!\sigma_{\mathrm{proj}}^{2}\!+\!
\sigma_{\mathrm{geom}}^{2}\!+\!\sigma_{\mathrm{cali}}^{2}
\label{eq:err}
\end{equation}
to calculate the total uncertainty in mass for each cluster. 
We will now discuss how we obtain the different terms in Eq.~(\ref{eq:err}). 
The statistical error  $\sigma_{\mathrm{stat}}$ is inferred from the tabulated
$\Delta\chi^{2}$ for the cluster on the grid in $r_{200}$ and $c_{\mathrm{NFW}}$: 
Taking $\Delta\chi^{2}\!=\!1$, we find the upper and lower limits of $r_{200}$ 
and then applying Eq.~(\ref{eq:mdelta}). Table~\ref{tab:masses} 
compares the masses of our eight clusters and their errors.

The components $\sigma_{\mathrm{cali}}$ and $\sigma_{\mathrm{geom}}$, accounting for 
the uncertainties in the shear calibration factor $f_{0}$ and the redshift 
distribution of the source galaxies are likewise determined from the analysis 
of the parameter grid. Assuming the redshift distribution to be well modelled 
by the fits to the CFHTLS \textsl{Deep~1} photo-$z$ catalogue, we vary 
$\langle\langle\beta\rangle\rangle$ by the uncertainties tabulated in 
Table~\ref{tab:defmod}. 
As expected, $\sigma_{\mathrm{geom}}$ increases with redshift because of the higher
relative uncertainty in $\langle\langle\beta\rangle\rangle$.

\subsection{Redshift Distribution}

Comparing the source number counts in the CFHTLS \textsl{Deep~1} field with our
MMT data, we find very good matches to the $r'$-band source counts in the
CL\,0030+2618 and CL\,1641+4001 fields, our observations with the deepest
limiting magnitudes and a high density 
$n_{\mathrm{KSB}}\!\gtrsim\!40\,\text{arcmin}^{-2}$ in the KSB catalogues
(cf.\ Tables~\ref{tab:data} and \ref{tab:shapecats}).
The other cluster catalogues exhibit a completeness limit (peak in the source
count histograms) at slightly brighter $r'$-magnitudes, but follow the  
\textsl{Deep 1} closer than the corresponding, alternative $r^{+}$-band source 
counts from the COSMOS photo-$z$ catalogue \citep{2009ApJ...690.1236I}.
In order to test for a possible bias in $\langle\beta\rangle$ for the shallower 
cluster fields, we repeat the fit to the redshift distributions from
the four \textsl{Deep} fields with the following modification: Introducing a
magnitude cut, we remove all galaxies with $r'\!>r'_{\mathrm{max}}$ from the
CFHTLS catalogues.
Virtually independent of $z_{\mathrm{d}}$, we find the $\langle\beta\rangle$ for 
the cases with and without magnitude cut to agree within mutual error bars for 
$r'_{\mathrm{max}}\!\gtrsim\!25.2$, meaning that 
the variation within the \textsl{Deep} fields has the same amplitude as the 
effect of removing the faintest sources. In our shallowest field, CL\,0230+1836,
we measure a \emph{limiting magnitude} of $r'_{\mathrm{lim}}\!=\!25.1$, with 
$15$\% of galaxies in the galaxy shape catalogue at $r'\!>\!25.2$. 
We thus conclude that no significant bias
in $\langle\beta\rangle$ is introduced by using the \emph{full} 
\citet{2006A&A...457..841I} catalogue as a redshift distribution proxy and the 
dispersion among the four fields as its uncertainty.

\subsection{Dilution by cluster members}

As in the case of CL\,0030+2618, we not only consider the uncertainty of 
$\pm0.05$ we estimate for $f_{0}$, but also take into account the dilution by 
remaining foreground galaxies in the shear calibration error. 
Once again using the CFHTLS \textsl{Deep~1} photo-$z$ 
catalogue as a proxy, we determine the fraction of galaxies at 
$z_{\mathrm{ph}}\!<\!z_{\mathrm{cl}}$ after applying the respective background 
selection. We measure this fraction $\hat{f}_{\mathrm{d}}$ to increase with $z$: 
it varies from $8.7$\% for CL\,0159+0030 to $32.1$\% for CL\,0230+1836 
(Table~\ref{tab:defmod}). 
As can be seen for CL\,0809+2811 and CL\,1416+4446 at the same redshift 
$z\!=\!0.40$, the background selection based on three bands results in a lower
$\hat{f}_{\mathrm{d}}\!=\!10.5$\% that the mere magnitude cut 
($\hat{f}_{\mathrm{d}}\!=\!13.6$\%) for only one band.
Adding the two components of the error in quadrature, the lower limit we 
consider for $f_{0}$ ranges from $0.97$ for CL\,0159+0030 to $0.73$ for 
CL\,0230+1836.

\subsection{Uncorrelated Large Scale Structure}

To calculate the error $\sigma_{\mathrm{LSS}}$ induced by LSS, we need to
extrapolate the findings of \citet{2003MNRAS.339.1155H}, covering only the
cases of $5 h^{-1}$, $10 h^{-1}$, and $20 h^{-1}\times10^{14}\,\mathrm{M}_{\sun}$
to lower masses. (Note that our $M^{\mathrm{wl}}$ estimate for CL\,0030+2618 is 
very close to the first case.) The respective error contributions read from 
Fig.~6 of \citet{2003MNRAS.339.1155H} are $\sim\!1.2 h^{-1}$, $\sim\!1.7 h^{-1}$,
and $\sim\!2.7 h^{-1}\times10^{14}\,\mathrm{M}_{\sun}$.
By assuming that the \emph{relative} LSS error 
$\sigma_{\mathrm{LSS}}/M^{\mathrm{wl}}$ increases linearly towards smaller masses, 
we arrive at the following relation:
\begin{equation} \label{eq:sigmalss}
\sigma_{\mathrm{LSS}}/(10^{14}\,\mathrm{M}_{\sun})\!=\!a M_{14} + b M_{14}^{2}\quad,
\end{equation}
where $a\!=\!0.22 h^{-1}$, $b\!=\!-0.01$, and 
$M_{14}=M^{\mathrm{wl}}/(10^{14}\,\mathrm{M}_{\sun})$. 
We understand Eq.~(\ref{eq:sigmalss}) as an order-of-magnitude estimate for the
LSS error and stress that simulated WL measurements are required to provide a 
better understanding of this important source of uncertainty. In particular,
we expect a larger $\sigma_{\mathrm{LSS}}$ for higher $z_{\mathrm{d}}$ clusters, for
which the existence of intervening massive structure is more likely. 
We notice that the results of \citet{2003MNRAS.339.1155H} are obtained at 
$z\!=\!0.3$, more nearby than our clusters.

In the special case of CL\,1701+6414, the obvious LSS at $z\!\approx\!0.22$ was
taken into account by explicit modelling, in addition to what is described here.
We estimate the error associated with the covariance of the parameters
describing the two clusters in our simultaneous model for CL\,1701+6414 and
A~2264 (cf.\ \ref{sec:cl1701}) by varying the otherwise fixed $r_{200}$ of
A~2264. The additional error from considering 
$r_{\mathrm{s,200}}\!=\!0.8\,\mbox{Mpc}$ and 
$r_{\mathrm{s,200}}\!=\!1.0\,\mbox{Mpc}$, according to the uncertainties from our
four-parameter fit, give a negligible contribution to the statistical error.
Nevertheless, we caution that an additional uncertainty likely arises from our
model choice.

\subsection{Triaxiality Projection Bias}

Applying the \citet{2005ApJ...629..781K} fitting formula for the 
largest-to-smallest axis ratio of a triaxial halo as a function of mass
to all our eight clusters, we arrive at expectation values of 
$0.60\!<\!\eta\!<\!0.64$ for the largest-to-smallest axis ratio. 
Hence, considering the triaxiality biases of \citet{2007MNRAS.380..149C}, we use
$\sigma_{\mathrm{proj}}^{+}\!=\!0.16\,M^{\mathrm{wl}}$ for the error $M^{\mathrm{wl}}$
induced by overestimation and
$\sigma_{\mathrm{proj}}^{-}\!=\!0.10\,M^{\mathrm{wl}}$ for the one induced by
underestimation caused by the projection of triaxial halos.

\section{Discussion} \label{sec:disc}

The statistical, systematic, and total errors for all eight clusters are 
summarised in Table~\ref{tab:masses}, both as absolute masses and as relative 
errors. Table~\ref{tab:erran} provides the details on the composition of the 
systematic error for the eight clusters.
We note that for all our clusters, in particularly the ones with small WL 
masses, the statistical uncertainties are the largest component in the
total error (the second largest usually being the projection of unrelated LSS).
The relative statistical errors range between $\approx\!30$\% and 
$\approx\!60$\%.
The reason for this can be twofold: First, the large statistical uncertainties
\textit{per se} are caused by the small signal-to-noise in the lensing signals 
and thus a consequence of the \emph{low net exposure times in the lensing-band}
images, once we removed frames with high PSF anisotropy (Table~\ref{tab:data}). 
Second, our account of the systematics might underestimate
or neglect contributions to the systematic error. 

For instance, uncertainties in the determination of the centres and the radial 
fitting ranges are not considered in Eq.~(\ref{eq:err}),
which we discuss in Sect.~\ref{sec:centdisc}.
After checking the statistical validity of our cluster detection in
Sect.~\ref{sec:mapdisc}, we evaluate several effects influencing the accuracy
with which we can measure weak lensing masses. 
A reliable quantification of all of these uncertainties is beyond the scope
of this article but very desirable with respect to the constraints on cosmology
at which the next generation of cluster weak lensing projects is aiming.

\subsection{Significance of Cluster Detections} \label{sec:mapdisc}

The $S$-statistics is known to produce spurious shear peaks even at high
significance levels, although as simulations show, false detections above the
$\sim\!4\sigma$ level are rare 
\citep[e.g.,][]{2005A&A...442...43H,2007A&A...470..821D}. 
Still, in principle, there is a nonzero, but small chance for one or the other 
of our detections to be false. Spurious detections are more sensitive against 
changes in the lensing catalogue or $\theta_{\mathrm{out}}$. Our tests with 
different photometric cuts and values for $\theta_{\mathrm{out}}$ found our 
cluster shear peaks to be robust.
Another reaffirmation is the persistence of signals when bootstrapping the 
lensing catalogue, which we performed for CL\,1357+6232, and CL\,1416+4446.

In order to test the interpretation of $S$-values as 
significances, we conducted the following test: 
For each galaxy in the catalogue, we add to the phase $\varphi$ of the
complex ellipticity estimator 
$\varepsilon\!=\!|\varepsilon|\,\exp{(2\mathrm{i}\varphi)}$ an
additional term $\varphi_{\mathrm{rnd}}$ drawn randomly from a uniform 
distribution in the interval $0\!\leq\!\varphi_{\mathrm{rnd}}\!<\!\pi$.
This procedure should completely remove the lensing signal from the
data such that the resulting value of $S$ be normally distributed
around zero, with a standard deviation $\sigma\!=\!1$.

We produced $10^{6}$ realisations of such a randomised catalogue for
each cluster and find the $S$-distributions for all eight cluster
detection to be well represented by a Gaussian distribution.
In all cases, the absolute of the mean value $\mu$ of the fitted Gaussian is 
$|\mu|\!<\!0.002$, and of the same order of magnitude as the
uncertainty in $\mu$ derived from the fit. We do not find a bias to either
positive or negative $S$.
For six of the eight clusters, we find for the standard deviations $\sigma$ of 
the fitted Gaussians values of $|1\!-\!\sigma|\!<\!0.01$, with 
$\sigma\!=\!0.962$ and CL\,1416+4446 and $\sigma\!=\!0.978$ for CL1641+4001,
respectively, the largest measured deviations from the expected $\sigma\!=\!1$.

For only one cluster, we find one $|S|\!>\!5$ event among the $10^{6}$ 
realisations, consistent with the expectation of one such event in 
$1.7\times10^{6}$ realisations of the expected Gaussian distribution $G$. 
Therefore, we conclude that this randomisation test does not find indications
for an overestimation of the significance of our cluster detections, as inferred
from the $S$-statistics. On the contrary, the small standard deviations measured
from the fits to the CL\,1416+4446 and CL1641+4001 correspond to very slightly
\emph{underestimated} significances of these two cluster detections.

\subsection{The Role of Cluster Centres} \label{sec:centdisc}

\begin{figure}
\includegraphics[width=8cm,angle=90]{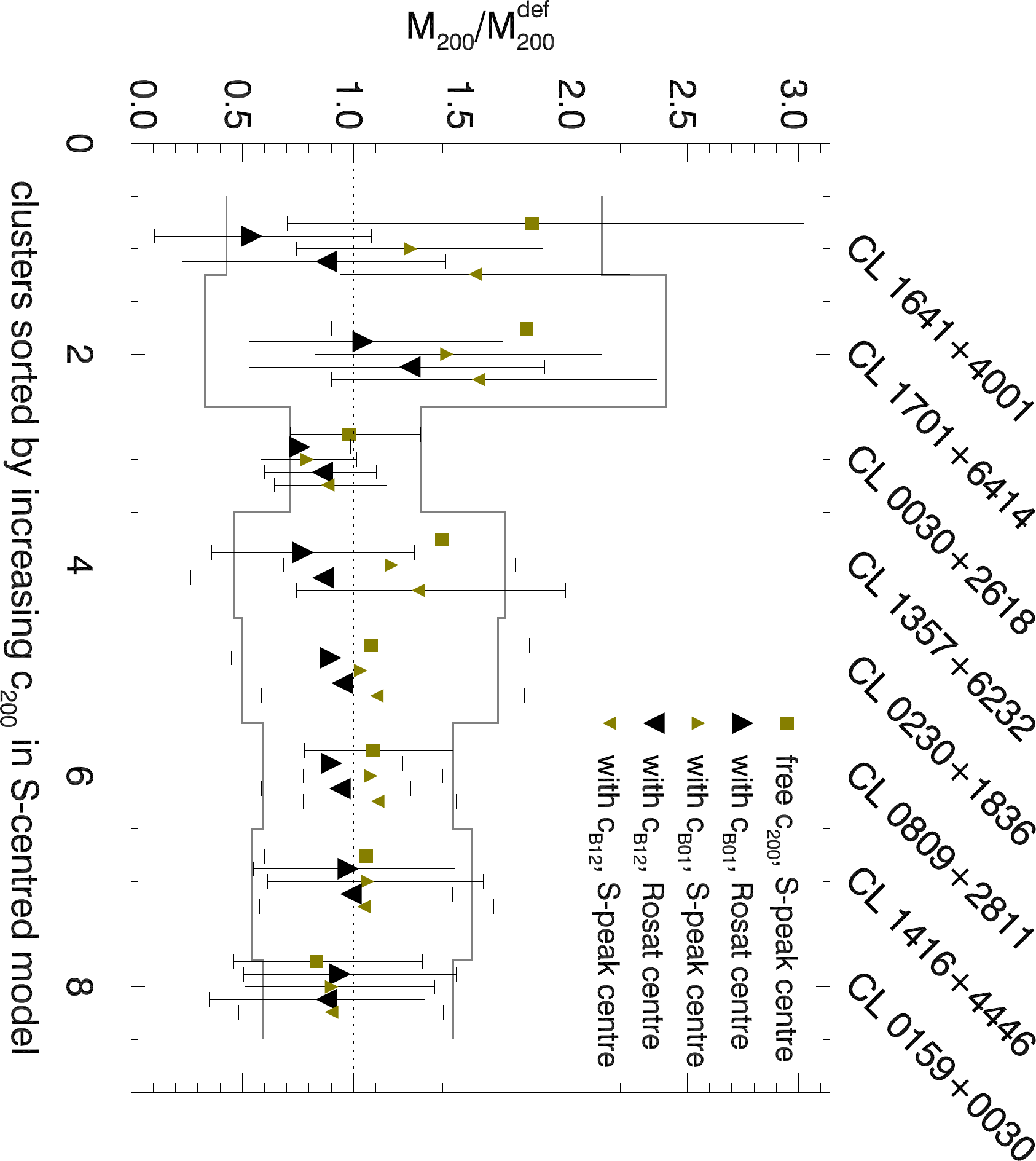}
\caption{Dependence of weak lensing cluster masses on the choice of centre
and mass--concentration--relation. Upward and downward triangles show the
masses using the B01 and B12 mass--concentration--relation relative to the case
of a free $c_{200}$. The square and small triangles present models centred on
the $S$-peaks for the free-$c_{200}$, B01, and B12 cases, respectively.
Clusters are ordered by increasing (free) $c_{200}$ in the $S$--centred model.
Grey lines denote the statistical error bars in the reference model
(\textsc{Rosat}-centred, free $c_{200}$).}
\label{fig:cchoice}
\end{figure}

\citet{2011arXiv1103.4607D} recently demonstrated that the $S$-peak gives a 
robust determination of the cluster centre, showing little susceptibility to
projected large scale structure (LSS). 
However, using $S$-peaks as cluster centres for WL mass estimates is likely to
result in a systematic overestimation of cluster masses, because we pick such
centres that produce the highest masses.
In addition to our default \textsc{Rosat} centres, we partnered all
cluster models with a model choosing the $S$-peak as cluster centre.
While we plan to quantify the centring bias once a more complete sample of our
clusters is available, we can show overall trends based on the eight clusters
discussed here. 

The separations between the shear peaks and \textsc{rosat} centres are 
$<\!3\arcmin$ in all cases and $<\!1\arcmin$ for four of the eight clusters
(Table~\ref{tab:defmod}). This coincidence of X-ray and lensing centres adds 
further significance to the $S$-detections. Two out of the four remaining 
clusters, CL\,0159+0030 and CL\,0809+2811, have their $S$-peaks within larger 
masked areas, reducing the accuracy with which the centres can be
determined. The complicated shear fields in the vicinities of CL\,1701+6414 and
CL\,1641+4001, with separations $>\!1\arcmin$ between lensing and 
\textsc{rosat} centres are discussed in Sects.~\ref{sec:cl1701} and 
\ref{sec:res1641}.
We note that the \textsc{rosat} cluster centres themselves are accurate to 
$\sim\!10\arcsec$.

First, we notice that the $S$-centred masses are biased high for six of
the eight clusters (squares in Fig.~\ref{fig:cchoice}). The median is a $9$~\%
higher mass using shear peak centres ($M_{200}^{\mathrm{S-cen}}$) than with
\textsc{Rosat} centres ($M_{200}^{\mathrm{R-cen}}$). We measure
$\langle M_{200}^{\mathrm{S-cen}}/M_{200}^{\mathrm{R-cen}}\rangle\!=\!1.25$,
with a $0.36$ standard deviation. This indicates the expected presence of a
centring bias, although a small one compared to its scatter.

The most extreme differences are measured for the low-concentration clusters
CL\,1641+4001 and CL\,1701+6414. The sorting in Fig.~\ref{fig:cchoice} by 
increasing $c_{200}$ in the $S$-centred model seems to suggest a trend of a
larger $M_{200}^{\mathrm{S-cen}}/M_{200}^{\mathrm{R-cen}}$ in clusters of low 
concentration. We do not see a similar trend when sorting clusters by
$\Delta\theta$, the separation between the two types of centres 
(cf.\ Table~\ref{tab:defmod}), but caution that we deal with small number 
statistics.
When applying the B01 (B12) mass--concentration relations (big and small
upward and downward triangles in Fig.~\ref{fig:cchoice}), we arrive at similar
results, with higher median biases of $18$\% ($20$\%), but still large scatter.

\subsection{Low concentration clusters}

As Fig.~\ref{fig:cchoice} shows, the relative differences between masses
measured with the six different models for centre and concentration parameter
are most  pronounced for CL\,1641+4001 and CL\,1701+6414, the two clusters
with the lowest intrinsic concentrations $c_{200}^{\mathrm{S-cen}}$.

First we remark, that low values for $c_{200}$ are neither unexpected nor
unheard of in the weak lensing literature 
\citep[cf., e.g.,][]{2010MNRAS.405.2215O,2012arXiv1204.2743P}.
Small concentration parameters as in CL\,0809+2811 or CL\,1357+6232 are not
unique to our analysis method, especially when using a cluster centre 
significantly offset from the shear peak or in fields affected by masks in 
the data. Our analysis highlights the importance of such effects.

 We note that, using an elliptical NFW cluster model,
\citet{2010MNRAS.405.2215O} found four out of the $25$ X-ray selected clusters
they analysed a best-fit $c_{\mathrm{vir}}\!<\!1$.
These four extreme cases are among the clusters \citet{2010MNRAS.405.2215O}
exclude from further analysis for centring problems or obvious misfit of the
assumed model. The authors do not, however, report similar problems for 
three further clusters with $1\!<\!c_{\mathrm{vir}}\!<\!2$.

The more extreme cases of CL\,1641+4001 and CL\,1701+6414 point to the
limitations of describing a cluster shear field using a single, spherically
symmetric function, even if obvious line-of-sight structure like A~2264 is
taken into account.
Nevertheless, tests like the comparison of $\kappa$-maps for CL\,1701+6414
from MMT and CFHT (Sect.~\ref{sec:kappakappa}) demonstrate that the measured
low concentration parameters reflect the reality of the mass distribution and 
are not artifacts of the analysis. Concerning cosmological applications, the 
rejection of such clusters will likely result in a biased mass, which
we prefer to avoid.
We notice that, making different choices in the data analysis, 
\citet{2012arXiv1205.3103H} likewise succeed in measuring masses of 
clusters showing a flat shear profile.

\subsection{Influence of mass--concentration--relations}

Comparing the masses obtained with the three different choices for $c_{200}$
(free parameter, B01 and B12 mass--concentration relations; 
Fig.~\ref{fig:cchoice} and Table~\ref{tab:massescb}), we observe interesting
trends.
Intrinsically, the B12 relation yields lower concentrations for the same
cluster than B01. This relates to the result that 
the mass assuming B01 exceeds the B12 mass, except for CL\,0159+0030.
Despite the overall agreement of the masses within their errors, 
the spread between the two masses seems to decrease with $c_{200}$, 
independent of the choice of centre.

On average, using a mass-concentration relation results in a lower mass 
estimate compared to a free $c_{200}$, with $\sim\,1\sigma$ significance for
the B01 relation 
($\langle M_{200,\mathrm{B01}}/M_{200,\mathrm{def}}\rangle\!=\!0.85\pm0.16$)
and less so for the B12 relation
($\langle M_{200,\mathrm{B12}}/M_{200,\mathrm{def}}\rangle\!=\!0.95\pm0.13$).

For the $S$-peak--centred masses, we obtain a similar picture.
The B01 masses increase relative to the free--$c_{200}$ masses, 
following this trend monotonically despite the 
statistical uncertainties.
In Fig.~\ref{fig:cchoice}, this can be seen looking at the squares and small
upward triangles.
The correlation can be explained by the shape found in the confidence contours
of most of the clusters for which $c_{\mathrm{200}}$ can be constrained 
(e.g.\ CL\,1357+6232; Fig.~\ref{fig:cl1357}): Relatively high values for
$c_{\mathrm{200}}$ only agree with the data for $r_{200}$ smaller than the 
best-fit value.
The underlying reason is the absolute value of the shear signal ruling out
simultaneously higher values for $c_{200}$ and $r_{200}$.

It will be interesting to see if the tentative trends in Fig.~\ref{fig:cchoice}
will persist for a more complete cluster sample.
We note that due to the substructure and non-sphericity of 
$\Lambda$CDM halos, we expect mass--concentration
measured with weak lensing to be biased with respect to the more direct
estimates from the simulations \citep[e.g.,][]{2012MNRAS.421.1073B}.  
At the current level of eight clusters, all mass estimates agree with each 
other within their error margins.
We notice the relative statistical uncertainties to be only weakly dependent
on the choice of $c_{200}$ (free parameter, B01, or B12 mass--$c_{200}$--relations).

\subsection{Masking of bright stars} \label{sec:mmask}

 \begin{table}
  \caption{Three bright stars exacerbating the analysis of the
    CL\,0159+0030, CL\,0230+1836, and CL\,0809+2811 fields, identified by their
    BD and HD designations. We cite
    SIMBAD (\texttt{http://simbad.u-strasbg.fr/simbad/}) for
    stellar positions, $V$ magnitudes and spectral types (Spec.). 
    By $\theta$ we denote the separations between the respective
    \textsc{rosat} cluster centre and star.}
   \begin{center}
   \label{tab:stars}
   \setlength{\tabcolsep}{1.8mm}
    \begin{tabular}{lcccccc}\hline
   BD & HD & $\alpha_{\mathrm{J2000}}$ & $\delta_{\mathrm{J2000}}$ & $\theta$ & $m_{V}$ & Spec.\\\hline\hline
   $-$00 301 & 12134 & 01:59:10.3 & +00:30:24 & $1.94\arcmin$ & $8.28$ & F0 \\
   +18 315 & 15551 & 02:30:30.1 & +18:39:51 & $3.59\arcmin$ & $8.25$ & K0 \\
   +28 1562 & 67543 & 08:09:34.3 & +28:11:46 & $1.51\arcmin$ & $8.60$ & F0 \\\hline
  \end{tabular}
  \end{center}
 \end{table}
In Table~\ref{tab:stars}, we summarise the properties of the magnitude $8$--$9$
stars that impede the analysis in the CL\,0159+0030, CL\,0230+1836, and 
CL\,0809+2811 fields. By coincidence, these three most severe cases among the
sample of $36$ clusters are among our \textsc{Megacam} targets, reminding us
that such fields must not be discarded when analysing a statistically complete
sample.
Using the example of CL\,0809+2811, we study the impact of these
stars and their masking on the $S$-maps and mass estimates.

Removing the masks generated for regions of deviant source density 
(Sect.~\ref{sec:datared}, red squares in Fig.~\ref{fig:cl0809}) 
does not increase the number of usable galaxies
significantly: Where scattered light strongly affects the local background 
estimation, sources are discarded in an early stage of catalogue preparation.
At the position of the \textsc{rosat} centre, there are no detections in the
first place.
Without masking, the $S$-peak of CL\,0809+2811 is shifted by $2\farcm4$ to 
the north-east (closer to the \textsc{rosat} centre) and very slightly lower 
($S_{\mathrm{max}}\!=\!5.27$ instead of $S_{\mathrm{max}}\!=\!5.39$).
Because our default model excises galaxies at $<\!1\farcm5$ separation, 
to avoid the strong lensing regime, the impact on the mass is below $2$\%.

A more important point could be the extra uncertainty in the chosen 
cluster centre,
as we find the largest offsets between WL and X-ray peaks for clusters with
large masks. The $M_{200}^{\mathrm{S-cen}}\!<\!M_{200}^{\mathrm{R-cen}}$
observed for CL\,0159+0030 (Fig.~\ref{fig:cchoice}) is likely due to a 
washed-out lensing peak. Nevertheless, we do not observe similar peculiarities
for CL\,0809+2811.

\subsection{Further sources of uncertainty}

There are several potential sources of uncertainty which
are not considered in Eq.~(\ref{eq:err}), 
for instance, uncertainties in the determination of the centres or
the contamination correction available only for clusters imaged in $g'r'i'$.
A reliable quantification of these errors will require further analysis. 

We also do not consider the uncertainty in the choice of 
$\max{(|\varepsilon|)}$ in the error analysis. 
However, we account for its effect via the shear calibration such that we do 
not expect a significant additional systematic error.
Carefully calibrated simulations of cluster lensing  are necessary to test
our assumptions on the shear calibration factor. 
In Paper~I, we observed in CL\,0030+2618 a counter-intuitive decrease of the 
best-fit value for $r_{200}^{\mathrm{min}}$ with increasing 
$\max{(|\varepsilon|)}$. 
Indeed, only CL\,1357+6232 shows a similar relative decrease in 
$r_{200}^{\mathrm{min}}$. 
Averaging over all eight clusters, these cases
are balanced by CL\,0230+1836 and CL\,1416+4446, for which we
measure $r_{200}^{\mathrm{min}}$ to \emph{increase} with $\max{(|\varepsilon|)}$.
With $\max{(|\varepsilon|)}\!=\!1.0$, we measure for four cases a smaller
$r_{200}^{\mathrm{min}}$ than for $\max{(|\varepsilon|)}\!=\!0.8$, and in four
cases a larger radius. The same holds for $\max{(|\varepsilon|)}\!=\!10^{4}$.
These results suggest that the
uncorrected bias due to $\max{(|\varepsilon|)}$ might be small. 

We notice that the roles of the shear calibration $f_{0}$, considered in 
Eq.~(\ref{eq:err}) as $\sigma_{\mathrm{cali}}$ and the correction $f_{1}(\theta)$
for cluster members cannot be 
completely disentangled. On the one hand, considering the cluster member 
correction separately is justified by the radial dependence of $f_{1}(\theta)$. 
On the other hand, we stress that the uncertainty in $f_{1}(\theta)$
might be large due to the weak detections of the cluster red sequence. 
In addition, the effect of cluster member decontamination on the mass 
estimate ($3$\% to $7$\%; up to $11$\% using shear-peak centres) lies within 
the range of the related systematic error component 
$\sigma_{\mathrm{cali}}^{-}$ which
is significantly smaller than the statistical uncertainty in the mass. 
Hence, a possible plan to consider cluster membership consistently -- also for 
single-band clusters -- would be to include it into the systematic error. 
Again, we suppose performing simulations of cluster WL fields 
to be helpful for the further investigations.

Finally, we note that total $1\sigma$ error intervals consistent with
cluster masses close to zero do not mean these clusters are detected merely
at the $\sim\!1\sigma$ level: Shear calibration, $z_{\mathrm{s}}$-distribution,
and triaxiality errors are multiplicative, such that they do not affect
the detection significance.

\section{Summary and conclusion} \label{sec:summary}

In this study, the second in the series on the \emph{400d} survey WL follow-up,
we reduced and analysed MMT/\textsc{Megacam} observations for seven clusters of
galaxies.
Building on Paper~I, data reduction is performed using \texttt{THELI}, and WL
shear catalogues are extracted using an implementation of the KSB+ algorithm. 

In the three cases, where we have MMT observations in $g'r'i'$, 
we define lensing catalogues based on a refined version of the three-colour
method used in Paper~I. By comparing with the colours observed for a
\citet{2006A&A...457..841I} photo-$z$ field, we exclude sources from regions in 
colour-colour-magnitude space containing a large fraction of foreground galaxies
from the analysis. For clusters with only one MMT band, we apply a cut in
magnitude as background selection.

We detect all of our $0.39\!<\!z\!<\!0.80$ clusters using the aperture mass
method ($S$-statistics) at the $>\!3.5\sigma$ level. Performing a 
\citet{2001A&A...374..740S} mass reconstruction, we find the projected mass to
follow the $S$-statistics closely. The WL masses of our clusters are
determined from NFW modelling of their tangential shear profiles, yielding
masses in the 
$10^{14}\,\mbox{M}_{\sun}\!\leq\!M_{200}^{\mathrm{wl}}\!<\!2\!\times\!\,10^{15}\,\mbox{M}_{\sun}$
interval. 

Two of our clusters are exceptional due to their complicated shear morphology:
For CL\,1701+6414, where several known clusters lie close to one another in
projection, we simultaneously fit the shear of the two strongest $S$-peaks,
identified with A~2246 and our target.
The field of CL\,1641+4001 also exhibits multiple shear peaks, but we find no
evidence for the presence of more than one cluster.

An independent analysis of the CL\,1701+6414 field using archival CFHT data
confirms the superposition of several weak lensing sources.
By matching shear catalogues from MMT and CFHT, we find ellipticities measured
with both instruments to be consistent with the assumption of noisy, but unbiased
measurements of the same quantity. Hence, MMT/\textsc{Megacam} is proven to be 
equally good for WL science as the well-established CFHT/\textsc{Megacam}.
We further produced a photo-$z$ catalogue based on CFHT $g'r'i'z'$ data of the
field. Despite the shallowness of three bands, we are able to devise a coarse
foreground/background selection for CL\,1701+6414. This experiment again
confirms not only shear peaks for several known clusters but also the validity 
of the magnitude-cut selection. Better photo-$z$ data will be needed to
potentially turn this cross-check into a calibration for single-band lensing data.

We find the error budgets for our cluster masses to be dominated by statistical
uncertainties (which can be suppressed by using a large cluster sample), but
with a significant contribution of systematic uncertainties.
Statistical uncertainties are naturally higher for relatively high-$z$ clusters
like ours, but the role of data quality (weather and instrumental conditions
that led to drastic reductions in data depth) can hardly be overestimated.
Our data show weak indications for clusters with low intrinsic
concentration to be more susceptible to model choices, 
such as the concentration parameter or assumed centre.

As the second paper in the \emph{400d} WL series, the main results of this 
study are that:
\begin{itemize}
\item Instrumental effects are well under control. 
\item Reliable masses can be obtained in the presence of bright stars 
close to the cluster centre -- an important finding for a successful follow-up 
of a \emph{complete} sample!
\item Ground-based WL works at least till $z\!\approx\!0.8$.
\item We can correct for extreme cases of massive foreground structures.
\item Clusters of low concentration are a common occurrence in the
high redshift, moderate mass population.
\end{itemize}
Furthermore, we identify areas of possible future improvements of the methods we
applied: As observational constraints will forbid complete homogeneity of the
data analysis to some degree for each survey. Therefore, better methods to
calibrate WL analyses in particular with a different number of available filters
need to be developed. This applies specifically to background selection and
correction for cluster members.

Concerning the modelling of clusters with complex shear morphology, we point
out that low-mass clusters with a strong fraction of disturbed or merging
systems will build the bulk of the population observed by deep and wide
future surveys like \textsc{eRosita} or \textsc{Euclid}.
Hence, the question arises how clusters deviating from a simple NFW mass
distribution can be weighed most accurately.
A possible method is to apply methods that do not assume
radial symmetry, e.g.\ aperture mass techniques.
Nevertheless, profile-fitting methods are well established and increasingly
well understood. Alternatively, by
combining simulational efforts with improved data analysis and
modelling, biases resulting from profile assumptions can be corrected.
Thus, applying profile fits even to low-signal, merging clusters might
prove the best method to measure reliable masses for large cluster samples.

\begin{acknowledgements}
HI likes to thank Frank Bertoldi for support of this work and
Matthias Klein, Reiko Nakajima, Mischa Schirmer, Ismael Tereno, Bharadwaj 
Vijaysarathy, Daniela Wuttke, and Yu-Ying Zhang for helpful discussions.
Partial support for this work has come from the Deutsche Forschungsgemeinschaft
(DFG) through Transregional Collaborative Research Centre TRR 33 as well as 
through the Schwerpunkt Program 1177. THR acknowledges support from the DFG 
through Heisenberg grant RE 1462/5 and grant RE 1462/6.
CLS was supported in part by Chandra grants GO9-0135X, GO9-0148X, and 
GO1-12169X, and by the F. H. Levinson Fund of the Silicon
Valley Community Foundation, which helped support the MMT observations.
We acknowledge the grant of MMT observation time (program 2007B-0046)
through NOAO public access.
\end{acknowledgements}

\bibliographystyle{aa}
\bibliography{hisrael_400dII_proved}

\appendix
\section{Photometric calibration details} \label{sec:augias}

\begin{figure*}
\vspace{-1.5cm}
\includegraphics[width=\textwidth]{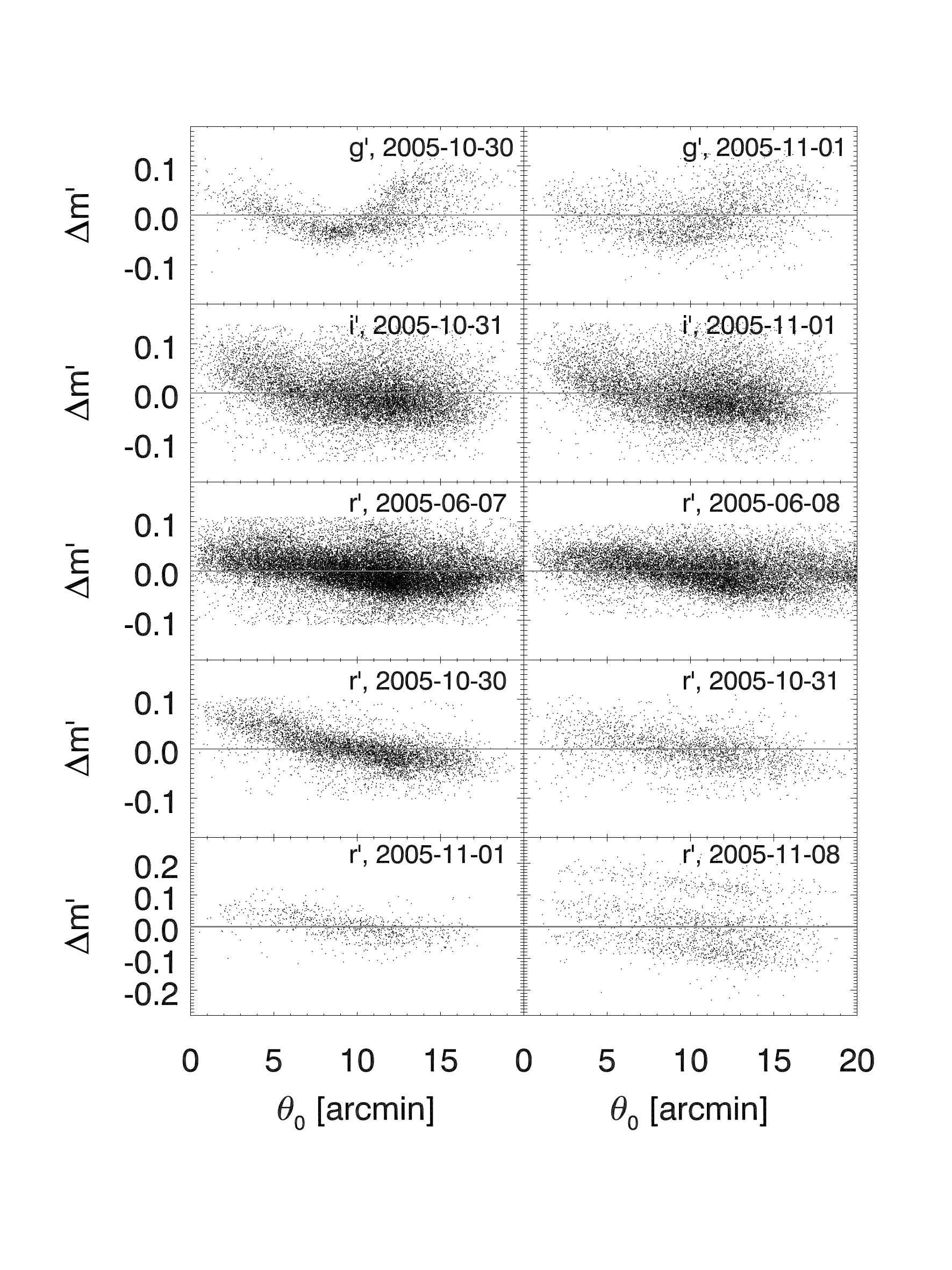}
\caption{Accuracy of the photometric calibration: For the different 
combinations of filters and nights used to calibrate the data sets discussed in
this work and Paper~I, the scatter $\Delta m'$ around the best-fit solution (solid line)
is shown. Each point corresponds to an SDSS standard source for which the 
abscissae give the separation $\theta_{0}$ in arc minutes from the centre of the
pointing. Note that for each panel a maximum $\Delta m'$ has been determined
by iterative $3\sigma$-clipping.}
\label{fig:augias2}
\end{figure*}
\begin{table*}
 \caption{Coefficients of photometric calibration 
  defined by Eq.~(\ref{eq:photofit}) for all \emph{photometric} nights within
  our MMT/\megacam ~\survey ~observations.}
 \begin{center}
 \begin{tabular}{ccccccc}\hline\hline
   Filter & Obs.\ Date & $Z_{\mathrm{f}}^{\dagger}$ & $\beta_{\mathrm{f}}$ & $c_{\mathrm{SDSS}}$ & $\gamma_{\mathrm{f}}$ & $n_{\mathrm{par}}^{\ddagger}$ \\ \hline
   $g'$ & 2005-10-30 & $27.277\pm0.005$ & $0.106\pm0.007$ & $g'\!-\!r'$ & $(-0.15)^{\S}$ & 2 \\
        & 2005-11-01 & $27.286\pm0.005$ & $0.116\pm0.005$ & $g'\!-\!r'$ & $(-0.15)^{\S}$ & 2 \\ \hline
   $i'$ & 2005-10-31 & $26.426\pm0.002$ & $0.124\pm0.002$ & $r'\!-\!i'$ & $(-0.05)^{\S}$ & 2 \\
        & 2005-11-01 & $27.408\pm0.009$ & $0.119\pm0.002$ & $r'\!-\!i'$ & $-0.03\pm0.01$ & 3 \\ \hline
   $r'$ & 2005-06-07 & $26.819\pm0.001$ & $0.040\pm0.001$ & $g'\!-\!i'$ & $(-0.10)^{\S}$ & 2 \\
        & 2005-06-08 & $26.834\pm0.008$ & $0.048\pm0.001$ & $g'\!-\!i'$ & $-0.12\pm0.01$ & 3 \\
        & 2005-10-30 & $26.950\pm0.018$ & $0.046\pm0.002$ & $g'\!-\!i'$ & $-0.10\pm0.02$ & 3 \\
        & 2005-10-31 & $26.959\pm0.004$ & $0.042\pm0.003$ & $g'\!-\!i'$ & $(-0.10)^{\S}$ & 2 \\
        & 2005-11-01 & $26.960\pm0.008$ & $0.048\pm0.004$ & $g'\!-\!i'$ & $(-0.10)^{\S}$ & 2 \\
        & 2005-11-08 & $26.807\pm0.005$ & $0.046\pm0.003$ & $g'\!-\!i'$ & $(-0.10)^{\S}$ & 2 \\ \hline \hline
 \end{tabular}
 \label{tab:photosol}
 \end{center}
 \begin{minipage}{120mm}
  \smallskip
  $^{\dagger}$ Normalised to an exposure time of $1\mbox{s}$ and 
    an airmass $a\!=\!0$.\\
  $^{\ddagger}$ Number of parameters used in the fit.\\
  $^{\S}$ Fixed at the default value.
 \end{minipage}
\end{table*}

Applying the \citet{2006A&A...452.1121H} method, photometric calibration of our
data is established by fitting instrumental ($m_{\mathrm{inst}}$) to reference 
($m_{\mathrm{SDSS}}$) magnitudes for a sample of objects, taking into account 
variable airmass $a$ and a colour term $c_{\mathrm{SDSS}}$ describing the 
transformation between \megacam ~and SDSS filter systems:
\begin{equation} \label{eq:photofit}
m_{\mathrm{inst}}-m_{\mathrm{SDSS}} = \beta_{\mathrm{f}} c_{\mathrm{SDSS,f}}+\gamma_{\mathrm{f}} a+Z_{\mathrm{f}}\quad,
\end{equation}
Depending on the photometric quality of the observations, we fit the zeropoint
$Z_{\mathrm{f}}$ together with the parameters $\beta_{\mathrm{f}}$ or 
$\gamma_{\mathrm{f}}$ in optimal conditions, or keep $\gamma_{\mathrm{f}}$ fixed
at the default value depending on the filter $f$ for poorer conditions. 

The resulting values for the fit parameters, as well as the colour indices
$c_{\mathrm{SDSS,f}}$ for the different filters are presented in 
Table~\ref{tab:photosol}, for all photometric nights of our \megacam 
~runs.\footnote{Note that some values in Table~\ref{tab:photosol} are corrected
w.r.t.\ Table~A.1 in Paper~I. The amount of these corrections is of the order 
of, and in most cases smaller than, the scatter observed in 
Fig.~\ref{fig:augias2}.}
We find the zeropoints $Z_{\mathrm{f}}$ of the photometric nights to agree among 
the $g'r'i'$ filters, with a largest deviation of $\approx\!0.15\,\mbox{mag}$. 
The scatter 
$\Delta m'\!=\!m_{\mathrm{inst}}-m_{\mathrm{SDSS}}+\beta_{\mathrm{f}} c_{\mathrm{SDSS,f}}+\gamma_{\mathrm{f}} a+Z_{\mathrm{f}}$ 
of the individual SDSS standards about the best-fit solution 
(Fig.~\ref{fig:augias2}) has a comparable amplitude. 
The errors of $Z_{\mathrm{f}}$ given in Table~\ref{tab:photosol} are the formal fitting errors. 
Figure~\ref{fig:augias2} presents the data from which the fit parameters have
been determined, applying an iterative $3\sigma$-clipping fit of 
Eq.~(\ref{eq:photofit}).

Comparing the colour terms $\beta_{\mathrm{f}}$ for the different nights, we
find considerable agreement within the values for each of the three bands, 
although the formal errors underestimate the true uncertainties.
We suggest that the large span in values of $\beta_{\mathrm{g}}$
might be caused by the known dependence of the filter throughput on the 
distance to the optical axis.
Plotting the scatter $\Delta m'$ as a function of the separation $\theta_{0}$ of
the source from the optical axis of \megacam ~(Fig.~\ref{fig:augias2}), given by
the pointing position in the \texttt{fits} header, we can confirm trends of 
$\Delta m'(\theta_{0})$ in all filters, most pronounced for the $g'$ band data
taken on 2005 October~30. 
This trend is likely caused by a combination of the sky concentration effect
(position-dependent illumination due to scattering in the telescope optics) 
and the position-dependent transmissivity of the \textsc{Megacam} filters,
which is strongest in the $g'$ band (cf.\ Fig.\ A.3 in Paper~I).
A more conclusive investigation of this issue, requiring full propagation of 
errors on instrumental magnitude, lies beyond the scope of this paper.
Because the radial dependence
observed in Fig.~\ref{fig:augias2} does not exceed the residual scatter for
sources at the same $\theta_{0}$, the global photometric fits 
(Eq.~\ref{eq:photofit}) fulfil the requirements of our analysis.

\section{Details of Background Selection} \label{sec:bgdetails}

\begin{table*}
\caption{Cuts defining the polygons used for background selection for the
$z\!\approx\!0.40$ and $z\!=\!0.80$ clusters, based on the colours of foreground
galaxies (Fig.~\ref{fig:zd040}). We specify the values of $g'\!-\!r$, 
$r'\!-\!i'$, and $s_{\beta}\!=\!\beta (r'\!-\!i')-(g'\!-\!r')$ at the edges of 
the exclusion polygons for $m_{\mathrm{bright}}\!<\!r'\!<\!m_{\mathrm{faint}}$.} 
 \begin{center}
 \begin{tabular}{cccccccc} \hline\hline
 Redshift & $\min{(r'\!-\!i')}$ & $\max{(r'\!-\!i')}$ & $\min{(g'\!-\!r')}$ & 
 $\max{(g'\!-\!r')}$ & $\beta$ & $\min{(s_{\beta})}$ & $\max{(s_{\beta})}$\\\hline
 $z\!=\!0.4$ & $-1.0$ & $0.7$ & $0.1$ & $1.6$ & $2.5$ & $-3.5$ & $0.5$ \\
 $z\!=\!0.8$ & $-1.0$ & $1.2$ & $0.3$ & $1.7$ & $1.5$ & $-3.0$ & $0.7$ \\ \hline
 \end{tabular}  \label{tab:newpoly}
 \end{center}
\end{table*}
\begin{figure*}
\includegraphics[width=16cm,angle=90]{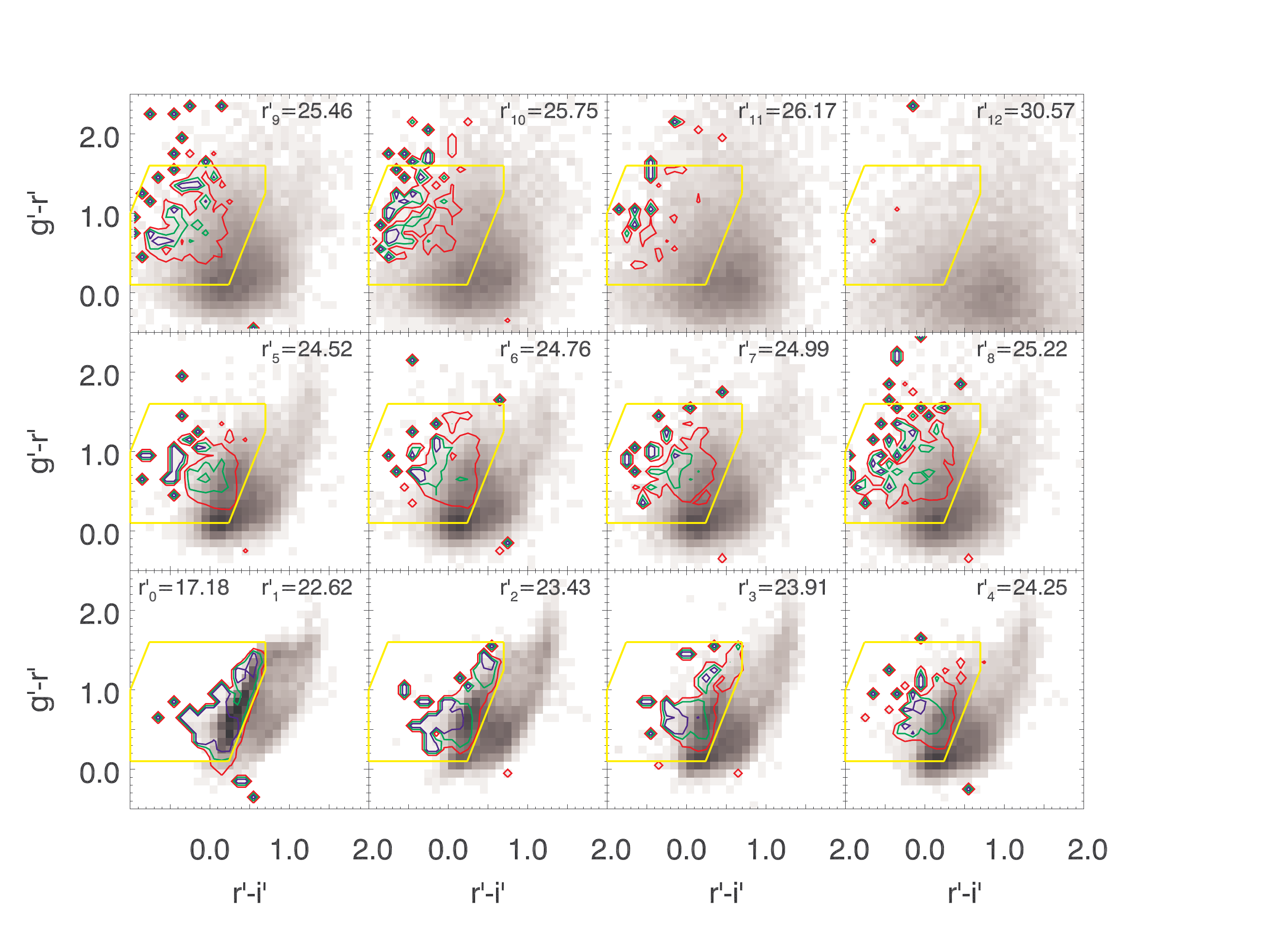}
\caption{
 Fraction of $z_{\mathrm{ph}}\!\leq\!0.40$ galaxies in the \textsl{Deep 1} 
 \citep{2006A&A...457..841I} field as a function of their $g'\!-\!r'$ and 
 $r'\!-\!i'$ colours and $r'$ magnitude. Each panel shows a dodecile of the 
 photo-$z$ catalogue, i.e. one of twelve equally populated magnitude bins, 
 where the $k$-th dodecile includes all galaxies 
 $r'_{k-1}\!\!\leq\!r'\!\!<\!r'_{k}$. In each panel, the number 
 $N_{ij}$ of galaxies within cells of mesh size 
 $\Delta(g'\!-\!r')\!=\!\Delta(r'\!-\!i')\!=\!0.1$ is shown, using the same 
 grey scale $\propto\!\!\sqrt{N_{ij}}$. White grid cells are empty.
 The red, green, and blue contours enclose regions in which $25$\% ($50$\%,
 $75$\%) of galaxies have a $z_{\mathrm{ph}}\!\leq\!0.40$.
 Based on the distribution of $z_{\mathrm{ph}}\!\leq\!0.40$ galaxies in the three
 brightest dodeciles, we define the yellow polygon
 (see Table~\ref{tab:newpoly}), in order to remove 
 foreground galaxies from the CL\,0159+0030 and CL\,0809+2811 fields.}
\label{fig:zd040}
\end{figure*}
In the intermediate magnitude range 
$m_{\mathrm{bright}}\!\leq\!r'\!\leq\!m_{\mathrm{faint}}$, our lensing catalogues
for three-band clusters
include galaxies selected from $g'\!-\!r'$ versus $r'\!-\!i'$ colour-colour-
diagrams (Sect.~\ref{sec:bg7}). We find our method justified by considering the
\citet{2006A&A...457..841I} photo-$z$ catalogue:

Figure~\ref{fig:zd040} presents the galaxy numbers and the fraction of
$z_{\mathrm{ph}}\!\leq\!0.50$ sources in the \textsl{Deep~1} photo-$z$ catalogue 
as a function of the $r'$ magnitude and $g'\!-\!r'$ and $r'\!-\!i'$ colours. 
First, the catalogue is divided into its dodeciles in $r'$, i.e.\ twelve
magnitude bins of equal population are defined where the $k$-th bin consists of
the galaxies $r'_{k-1}\!\!\leq\!r'\!\!<\!r'_{k}$. By $r'_{k}$, we denote the 
magnitude of a source such there is a fraction of $k/12$ of brighter galaxies in
the catalogue. 
Second, for each dodecile, we show the number $N_{ij}$ of galaxies falling into
grid cells of mesh size  $\Delta(g'\!-\!r')\!=\!\Delta(r'\!-\!i')\!=\!0.1$, 
using a grey scale.
Figure~\ref{fig:zd040} highlights that at bright $r'$, only a narrow strip in
the colour--colour space spanned by $g'\!-\!r'$ and $r'\!-\!i'$ is populated,
while the locus of galaxies becomes much more diffuse towards fainter $r'$.
Third, for each grid cell, we determine the fraction of galaxies we define as
foreground sources, i.e. the sources with a redshift estimate 
$z_{\mathrm{ph}}\!\leq\!0.40$. The red, green, and blue contours in 
Fig.~\ref{fig:zd040} mark regions of the colour--colour space populated by
$25$\%, $50$\%, and $75$\% of foreground galaxies compared to the 
$z_{\mathrm{d}}\!=\!0.40$ clusters, CL\,0159+0030 and CL\,0809+2811.
The contours are defined such that $f_{\mathrm{fg}}$ exceeds the respective
threshold in all grid cells enclosed by the contour.

As expected, $f_{\mathrm{fg}}$ generally decreases towards fainter magnitudes,
with only a few $z_{\mathrm{ph}}\!\leq\!0.40$ sources at $r'\!>\!26.0$. For all
magnitudes, foreground sources with $r'\!-\!i'\!>\!0.5$ are rare. In the
brightest three dodeciles, a well-defined region with a distinctive edge towards
redder $r'\!-\!i'$ colours exists\footnote{Towards very blue $r'\!-\!i'$ 
colours, few galaxies are found in the CFHTLS~D1 catalogue, basically all of 
them at low $z\!<\!0.4$ redshift. This can be seen from the contours in 
Fig.~\ref{fig:zd040} which follow the irregular shape of the point cloud. 
We choose a conservative $\min{(r'\!-\!i')}\!=\!-1.0$ limit for the selection 
polygons.}
Although the preferred locus of 
$z_{\mathrm{ph}}\!\leq\!0.40$ galaxies depends little on the $r'$ magnitude, the
zone populated by low-$z$ objects becomes more diffuse for fainter sources.
The insignificant role of foreground galaxies $r'\!>\!25.0$ justifies that our
background selection includes all galaxies fainter than $m_{\mathrm{faint}}$.
Furthermore, the secondary role of $m_{\mathrm{bright}}$ compared to
$m_{\mathrm{faint}}$ becomes clear from Fig.~\ref{fig:zd040}, noticing the small
number of $r'\!<\!20$ galaxies.

Calculating $f_{\mathrm{fg}}$ for a cluster redshift of $z_{\mathrm{d}}\!=\!0.80$,
the regions in colour--colour space where a given value of
$f_{\mathrm{fg}}$ is exceeded extend as well towards fainter $r'$ as towards
redder $g'\!-\!r'$ and $r'\!-\!i'$ colours. For $z_{\mathrm{d}}\!=\!0.80$, only
a small number of \emph{background} sources remain in the first dodecile, while
there are significant foreground objects even in the 
$r'_{10}\!<\!r'\!<\!r'_{11}$ bin.

We adjust our background selection polygon to $z_{\mathrm{d}}\!=\!0.40$ and 
$z_{\mathrm{d}}\!=\!0.80$ by defining criteria based on $f_{\mathrm{fg}}\!>\!0.25$ 
contours in the three brightest dodeciles ($r'\!<\!23.91$) of the 
\textsl{Deep~1} photo-$z$ catalogue (Table~\ref{tab:newpoly} and yellow polygon
in Fig.~\ref{fig:zd040}). We exclude galaxies at 
$m_{\mathrm{bright}}\!\leq\!r'\!\leq\!m_{\mathrm{faint}}$ and matching these criteria
from the lensing catalogues.
Performing a cross-check for $z_{\mathrm{d}}\!=\!0.50$, we confirm the background
selection in Paper~I to be sensible, although not optimal. In fact, more
``self-calibrations'' can be achieved by combining three-colour photometry with
photo-$z$ catalogues (Klein et al.\ in prep.).

\section{Details of photo-$z$ analysis} \label{sec:bpz}

\begin{figure*}
\sidecaption
\includegraphics[width=10cm,angle=90]{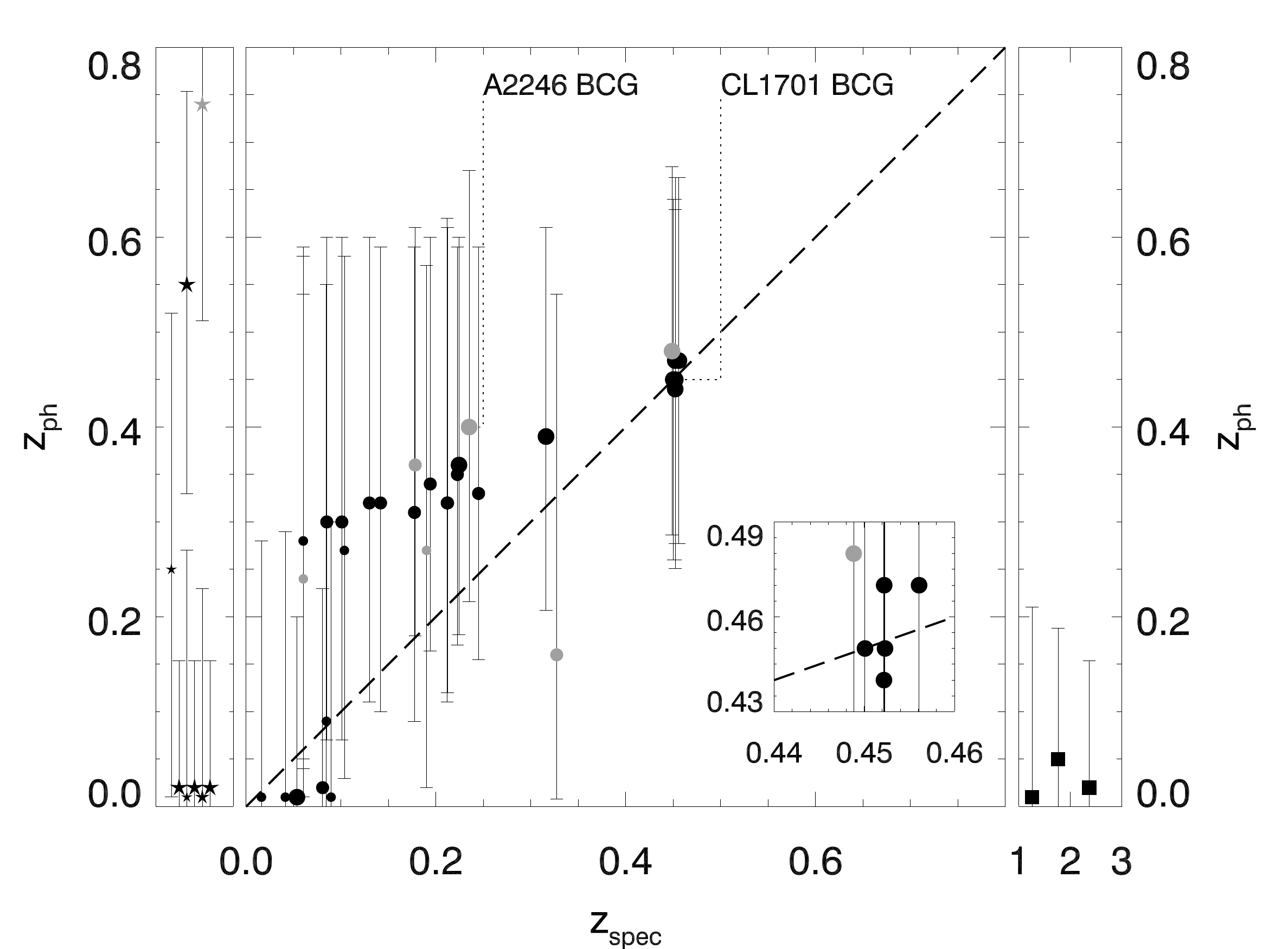}
\caption{CFHT photometric redshifts plotted against spectroscopic redshifts
from SDSS. The left panel (star symbols) shows the $z_{\mathrm{ph}}$ for objects
identified as stars in SDSS, the middle panel (filled circles) for normal 
galaxies, and the right panel (filled squares) for QSOs. A small inlay gives 
a zoomed version for $z_{\mathrm{spec}}\!\approx\!0.45$ galaxies. 
The size of the symbols marking the photo-$z$ estimate correspond
to the quality parameter $o$ (\texttt{ODDS}): big symbols for 
$o\!>\!0.9$, medium-sized symbols for $0.8\!<\!o\!<\!0.9$, and small symbols for
$0.7\!<\!o\!<\!0.8$.
Error bars for the $z_{\mathrm{ph}}$ give the uncertainty interval defined by the
\texttt{BPZ} \texttt{Z\_B\_MIN} and \texttt{Z\_B\_MAX} parameters. 
Objects inside a flagged region of one of the CFHT images 
(\texttt{CANDMASK}$=\!1$) are shown in grey.}
\label{fig:zszp}
\end{figure*}

\subsection{Spectroscopic calibration} \label{sec:zszp}

Because our photo-$z$ catalogue for the CFHT CL\,1701+6414 field was distilled
from only four bands, three of which have rather shallow exposure time
(Table~\ref{tab:cdata}), we tested its quality by comparison with publicly
available SDSS spectroscopy redshifts of the same field.
SDSS spectra are only available for a selection of the brightest 
($r'\!\lesssim\!19$) sources, with a total of $88$ matches for the 
$\approx\!270000$ object photo-$z$ catalogue.
Out of the $58$ sources flagged as good by \texttt{BPZ} in all four filters,
$45$ are identified as normal galaxies by SDSS, nine are identified as stars, 
and four as QSOs.
Figure~\ref{fig:zszp} displays $z_{\mathrm{ph}}$ as a function of $z_{\mathrm{spec}}$
of the matched sources for which the \texttt{BPZ} quality parameter 
\texttt{ODDS} is $o\!>\!0.7$. The size of the symbols in Fig.~\ref{fig:zszp}
(star symbols for stars, filled circles for normal galaxies, and filled squares
for QSOs) corresponds to the value of $o$. 

Generally, the photo-$z$ uncertainties (given by the \texttt{BPZ} 
\texttt{Z\_B\_MIN} and \texttt{Z\_B\_MAX} parameters) are large, although the
sources in Fig.~\ref{fig:zszp} rank among the brightest in the catalogue.
Nevertheless, the $z_{\mathrm{ph}}$ estimates for normal galaxies seem to follow a
remarkably narrow and monotonic function of $z_{\mathrm{spec}}$, in the range
$0\!<\!z_{\mathrm{spec}}\!\lesssim\!0.45$ probed by the SDSS spectral targets in
Fig.~\ref{fig:zszp}. The step at $z_{\mathrm{spec}}\!\approx\!0.10$, below which
galaxies get assigned $z_{\mathrm{ph}}\!\approx\!0.0$ and above which they are
overestimated to be at $z_{\mathrm{ph}}\!\approx\!0.3$, can be explained by the
lack of a $u'$ filter crucial for detecting the $400\,\mbox{nm}$ break at these
redshifts. In particular, this applies to galaxies in the $z\!\approx\!0.22$
structures in the foreground to CL\,1701+6414, as exemplified by the
$z_{\mathrm{ph}}\!\approx\!0.40$ for one of the A~2246 BCG candidates at
$z_{\mathrm{spec}}\!\approx\!0.235$.

Photo-$z$ estimates for the highest redshift ($z\!\approx\!0.45$) galaxies with
SDSS spectra are stunningly accurate, despite the large uncertainties.
The overall trend seen in Fig.~\ref{fig:zszp} is consistent with the results
of the \textsl{CFHTLS-Archive-Research Survey} \citep{2009A&A...493.1197E}.
In their analoguous comparison of CFHT \texttt{BPZ} photo-$z$s to SDSS spectra,
they find a turnover to $z_{\mathrm{ph}}\!<\!z_{\mathrm{spec}}$ for 
$z_{\mathrm{spec}}\!\gtrsim\!0.45$ (for a small absolute number of such galaxies).

It is a lucky coincidence that the $z_{\mathrm{ph}}$--$z_{\mathrm{spec}}$--relation
intersects the dotted equality line precisely at the redshift of our cluster of
interest.
There are six galaxies $0.448\!<\!z_{\mathrm{spec}}\!<\!0.457$ among the SDSS
spectral targets (inlay in Fig.~\ref{fig:zszp}), while there are none in the 
$0.36\!<\!z_{\mathrm{spec}}\!<\!0.44$ range.
These six include the BCG of CL\,1701+6414 at 
$z_{\mathrm{spec}}\!=\!0.4523\pm0.0001$, for which \texttt{BPZ} returns
$z_{\mathrm{ph}}\!=\!0.45\pm0.19$. 
However, even the closest of the other five is separated by $12\farcm0$ or 
$4.0\,\mbox{Mpc}$ in projection and thus not part of or closely interacting 
with CL\,1701+6414.
Still, its $z_{\mathrm{spec}}\!=\!0.4522\pm0.0001$ indicates they might belong to
the same large-scale structure.

The photo-$z$ estimates for QSOs (right panel of Fig.~\ref{fig:zszp}) are
clearly off, which doesn't come as a surprise as their is no QSO spectrum
among the templates employed by \texttt{BPZ}. 
Although five out of eight stars in Fig.~\ref{fig:zszp} (left panel) get 
assigned $z_{\mathrm{ph}}\!\leq\!0.02$, there are also three cases in which our
photo-$z$ catalogue suggests higher $z_{\mathrm{ph}}$ for
objects that by visual inspection and SDSS classification clearly stellar:
The available CFHT photometry \emph{alone} does not allow for an accurate 
star/galaxy classification.

\begin{figure}
\includegraphics[width=7.5cm,angle=90]{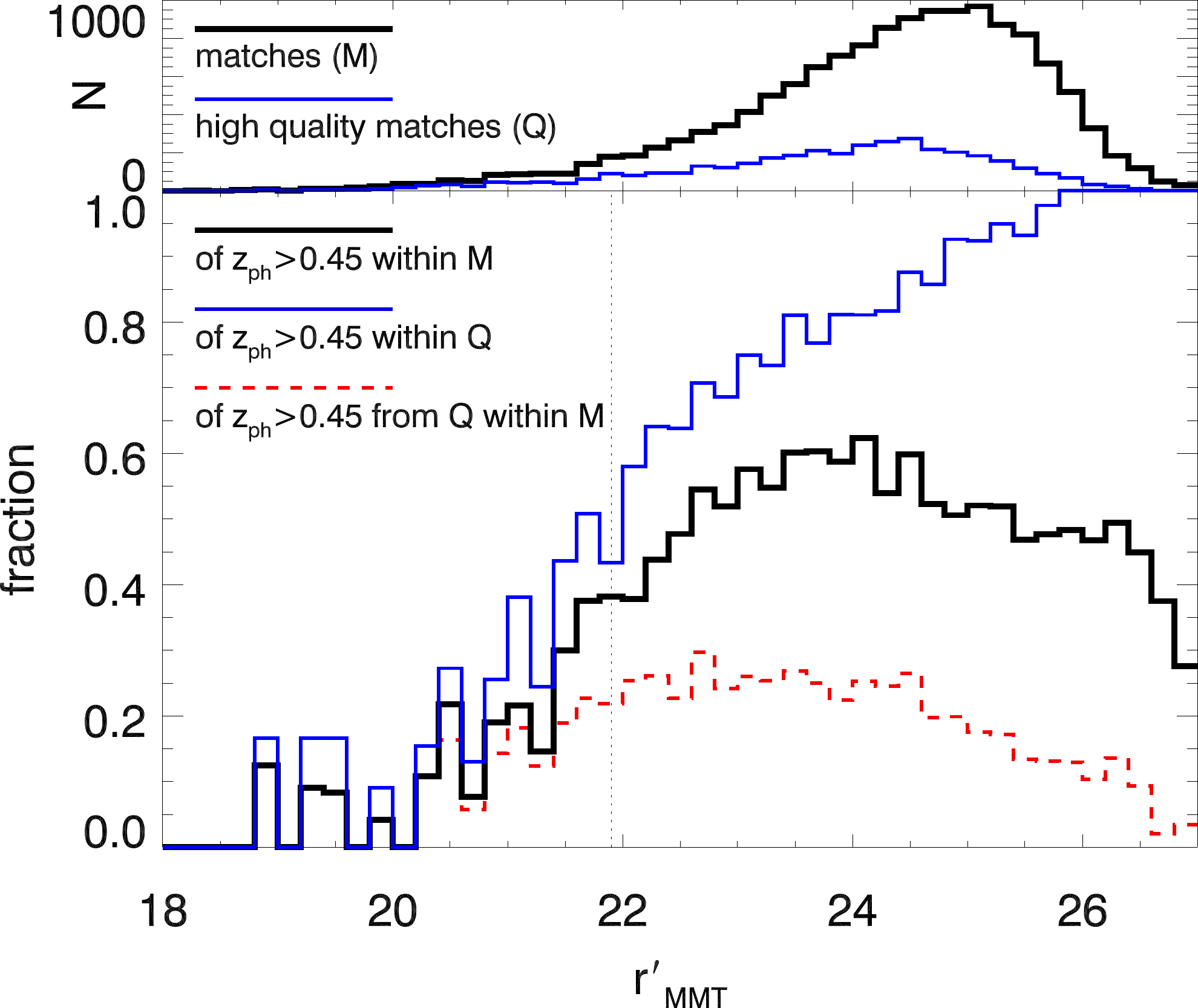}
\caption{\textit{Upper panel:} Histograms of the MMT-photo-$z$ matched 
catalogue as a function of MMT $r'$ magnitude: Plotted are all matches 
(thick black line) and the subset of matches with photo-$z$s passing all quality
criteria (``high quality'', blue line).
\textit{Lower panel:} Fractions of galaxies $z_{\mathrm{ph}}\!>\!0.45$ within: 
all MMT-photo-$z$ matches (thick black line), the ``high quality''matches 
(blue line), and of high-quality matches $z_{\mathrm{ph}}\!>\!0.45$ within all
(dashed red line).
A thin dotted line gives the magnitude cut at $m_{\mathrm{faint}}\!=\!21.9$.}
\label{fig:zbgfrac}
\end{figure}

\subsection{Matching with MMT} \label{sec:bpzmmt}

To investigate what benefit the four-band photo-$z$s yield, once the usual 
selection of galaxies by magnitude and half-light radius (Sect.~\ref{sec:bg7})
is applied, we now match the photo-$z$ catalogue with the MMT KSB catalogue.
We first notice that although $95.6$\% of galaxies in the MMT KSB
catalogue are matched bijectively to a CFHT photo-$z$ source, only 
$20.4$\% of the MMT lensing sources satisfy the quality criteria of four 
usable CFHT bands (\texttt{NBPZ\_GOODFILT}$\!=\!4$), $o\!>\!0.8$, and no masking 
(\texttt{CANDMASK}$\!=\!0$). 
We call these estimates \textit{high-quality} photo-$z$s.

Counter-intuitive at the first glance, $76$\% of the high-quality matches have
$z_{\mathrm{ph}}\!>\!0.5$, i.e.\ are likely background galaxies to CL\,1701+6414.
However, this effect can be traced back to the near-absence of high-quality
estimates of $z_{\mathrm{ph}}\!<\!0.3$.
This is consistent with our expectation from the comparison to the spectroscopic
redshifts (Fig.~\ref{fig:zszp}), where the lack of a $u'$ filter systematically
offsets $z_{\mathrm{ph}}$ for $z_{\mathrm{spec}}\!<\!0.45$ galaxies to higher values.
The $z_{\mathrm{ph}}$ distribution derived from the four-band CFHT data deviates 
far from the redshift distributions known from well-studied photo-$z$ fields
\citep[also see Fig.~B.6 in Paper~I]{2006A&A...457..841I,2009ApJ...690.1236I}.
This holds in particular for the high-quality sub-catalogue which, containing
brighter galaxies on average, traces a different population than our MMT
lensing catalogue.

The upper panel of Fig.~\ref{fig:zbgfrac}, showing the magnitude distributions
of the high-quality photo-$z$ catalogue (solid blue line) compared to all 
matches (thick black line) demonstrates that high-quality photo-$z$s tend to
belong to brighter galaxies.
This can be seen from the modes of the histograms and is not surprising given
the necessary detection in the shallow $g'i'z'$ images.
While $98.5$\% of the sources in the MMT galaxy catalogue get matched to a
photo-$z$ galaxy, only for $26.7$\% the photo-$z$ passes all quality cuts.

In particular, the decline with magnitude of the fraction of high-quality
matches affects the fraction of background galaxies with respect to 
CL\,1701+6414 at $z\!=\!0.45$ using our photo-$z$s: 
Considering high-quality matches only (solid blue curve in the lower panel of
Fig.~\ref{fig:zbgfrac}), the fraction of ``photometric background''
($z_{\mathrm{ph}}\!>\!0.45$) increases strongly with MMT $r'$ magnitude.
In fact, \emph{all} of the few $r'_{\mathrm{MMT}}\!>\!25.8$ high-quality matches
show $z_{\mathrm{ph}}\!>\!0.45$.
With respect to the complete catalogue (thick black line), however, the fraction
of photometric background galaxies peaks at $r'_{\mathrm{MMT}}\!\approx\!24$ and 
$\sim\!0.6$ and decreases towards fainter $r'_{\mathrm{MMT}}$.
The fraction of high-quality $z_{\mathrm{ph}}\!>\!0.45$ galaxies compared to all
matches (red dashed line in Fig.~\ref{fig:zbgfrac}) runs rather flat with
$r'_{\mathrm{MMT}}$, never exceeding $0.3$ and subsuming only $15.9$\% of all
matches.

We conclude that the quality of the CFHT data entering the photo-$z$ estimation
makes possible a rough estimation of a normal galaxy's redshift, i.e.\ to
decide if it is more likely to be in the foreground or in the background, but
not a precise redshift distribiution from which $\langle\beta\rangle$ could be
inferred more precisely than using a proxy photo-$z$ catalogue of high quality
(Sect.~\ref{sec:spm}).

\subsection{A photo-$z$ shear catalogue} \label{sec:rzig}

\begin{figure*}
 \sidecaption
 \includegraphics[width=12.5cm,angle=90]{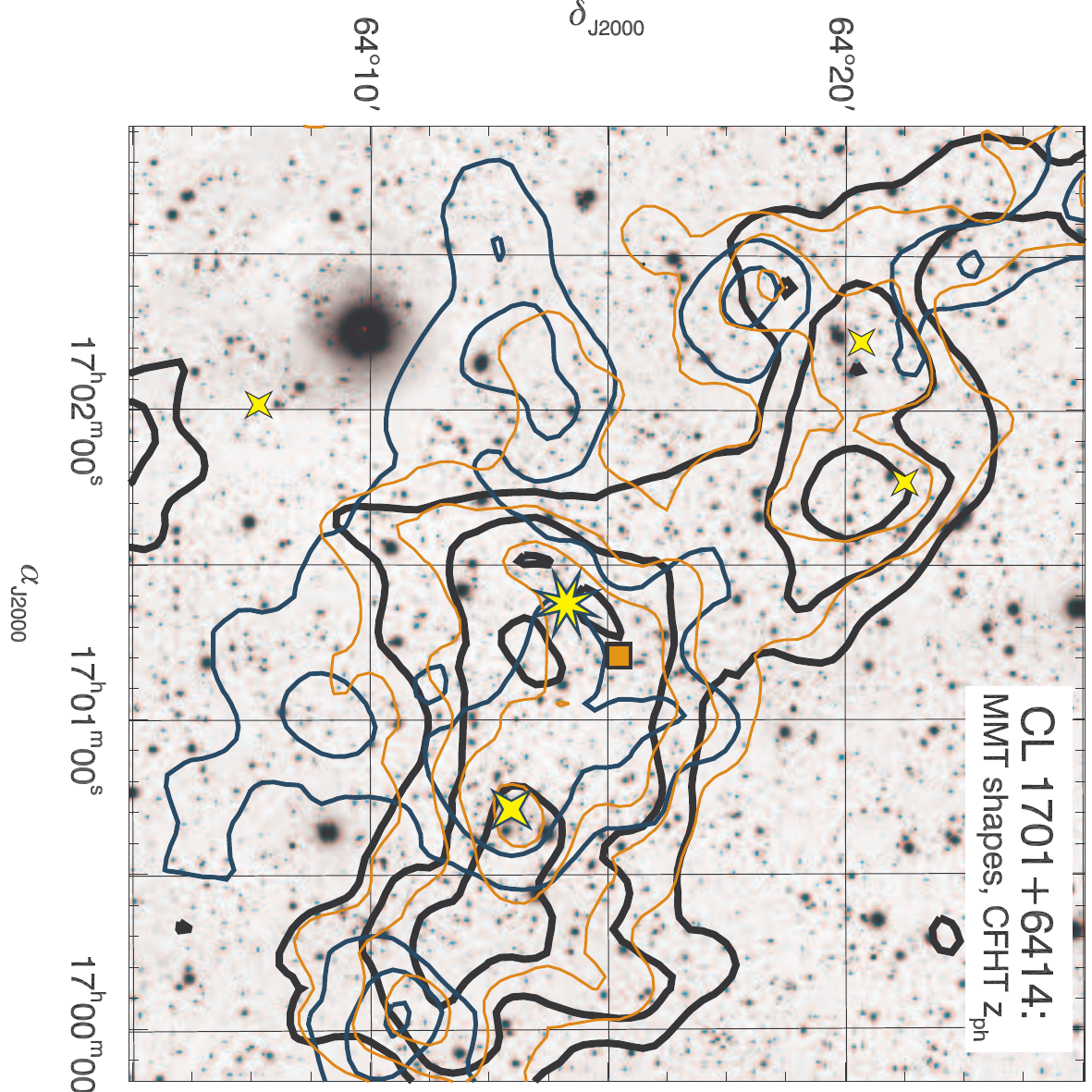}
 \caption{$S$-statisics in the CL\,1701+6414 field using the CFHT photo-$z$ 
catalogue. Thick black contours mark the shear signal from the 
$z_{\mathrm{ph}}\!<\!0.45$ catalogue, medium-thick blue contours are derived from
the complementary $z_{\mathrm{ph}}\!\geq\!0.45$ sources. Thin orange contours show
the signal from the complete MMT galaxy catalogue. All contours start at 
$S\!=\!1$ and are spaced by $\Delta S\!=\!1$. The underlying image and other 
contours are the same as in Fig.~\ref{fig:cl1701}.}
 \label{fig:rzig}
\end{figure*}

The photo-$z$s drawn from the CFHT $g'r'i'z'$
bands provide us with a rough redshift estimate. 
In order to test whether this information can be used to disentangle the shear
signals of CL\,1701+6414 and the foreground structure, in particular A~2246,
we divide the MMT galaxy catalogue: Galaxies with $z_{\mathrm{ph}}\!<\!0.45$ 
are sorted into the ``photo-$z$ foreground'' catalogue, galaxies with
$z_{\mathrm{ph}}\!\geq\!0.45$ are sorted into the ``photo-$z$ background''
catalogue. Because of the poor quality of most photo-$z$ estimates 
we expect only a crude selection.

Figure~\ref{fig:rzig} shows the $S$-maps resulting from these two catalogues,
overlaid on the MMT $r'$-image in the same fashion as for Fig.~\ref{fig:cl1701}.
Thick black contours denote iso-$S$-contours from the photo-$z$ background
catalogue, including $49.9$\% of the lensing catalogue 
($10.3$ galaxies/arcmin$^{-2}$). Solid blue contours in Fig.~\ref{fig:rzig} are
drawn from the complementary photometric foreground catalogue; the signal from
the complete galaxy catalogue is shown as thin orange contours.

The morphology of the $S$-peaks in the photo-$z$ background map follows in its
main features the complete catalogue, as we expect from a sample of true
$z\!>\!0.45$ galaxies. With $S_{\mathrm{max}}\!=\!3.14$, the peak to be associated
with CL\,1701+6414 is nearly as strong as for the full catalogue, and closer to
the cluster's \textsc{rosat} position. The A~2246 peak shows a similar high
fraction of the complete catalogue signal, but the two clusters appear to be
better separated. 
The  photo-$z$-foreground $S$-morphology bears little resemblance to 
Fig.~\ref{fig:cl1701}: Although we still measure $S\!\approx\!2$ close to the
position of CL\,1701+6414, it can not be seen as a distinct peak.
A~2246 is detected just below $3\sigma$, with $S\!>\!3$ only measured for the
``SW peak''. This is consistent with our expectations: As A~2246 is at lower
redshift, some signal should persist in a true foreground catalogue.

Re-defining the photo-$z$ catalogue such that it only contains high-quality
photo-$z$s $>\!0.45$ (cf.\ Sect.~\ref{sec:bpzmmt}) results in a good resolution 
between the $S$-signals of the two main clusters but such catalogue suffers 
from the sparsity of sources ($4.8$ galaxies/arcmin$^{-2}$).

Despite the outcome of this experiment matching our expectations, we keep in
mind the typical uncertainty of $\sigma(z_{\mathrm{ph}})\!\approx\!0.25$ even for
the high-quality photo-$z$s (Fig.~\ref{fig:zszp}), similar to the redshift
separation of CL\,1701+6414 and A~2246. Hence, the photo-$z$ selection using the
available data is not inherently better than the Sect.~\ref{sec:cl1701} 
magnitude cut. Nonetheless, the CFHT photo-$z$s and lensing measurements confirm
the detection of CL\,1701+6414 as a shear source distinct from A~2246 and
give credibility to its mass estimate, the aim of our investigations.

\section{Notes on individual clusters} \label{sec:indivcl}

\begin{figure*}[b]
 \includegraphics[width=17cm]{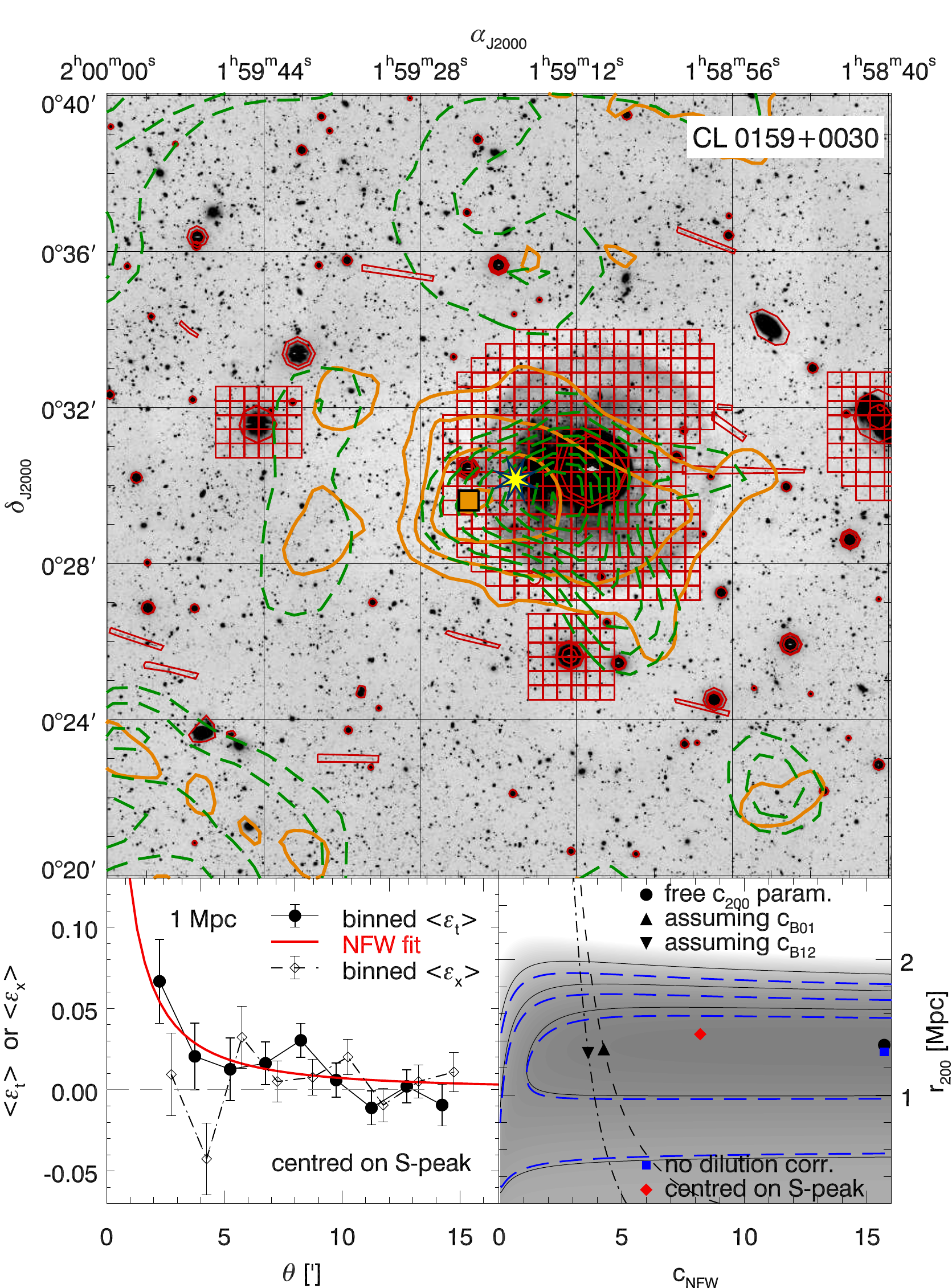}
 \caption{Like Fig.~\ref{fig:cl1357}, but for CL\,0159+0030. In the plot of
  $r_{200}$ against $c_{\mathrm{NFW}}$, a square and dashed contours 
  denote the model minimising
  Eq.~\ref{eq:merit} if no dilution correction is assumed.}
 \label{fig:cl0159}
\end{figure*}

\begin{figure*}
 \includegraphics[width=17cm]{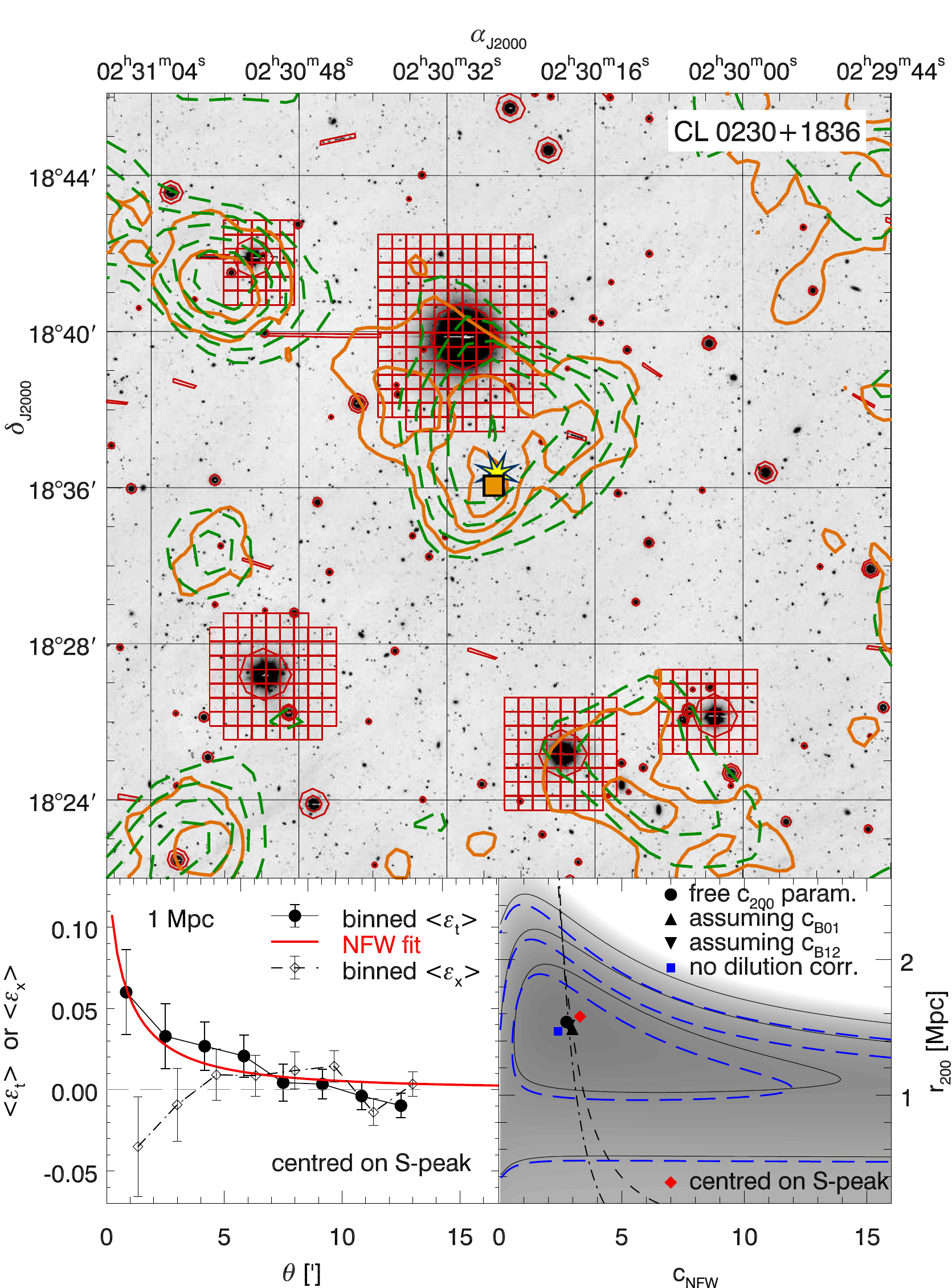}
 \caption{Like Fig.~\ref{fig:cl0159}, but for CL\,0230+0030. Note that,
  in the lower right panel, the filled circle and downward triangle, 
  denoting the best parameters for the free-$c_{\mathrm{NFW}}$ abd B12-models 
  are almost coincident.}
 \label{fig:cl0230}
\end{figure*}

\begin{figure*}
 \includegraphics[width=17cm]{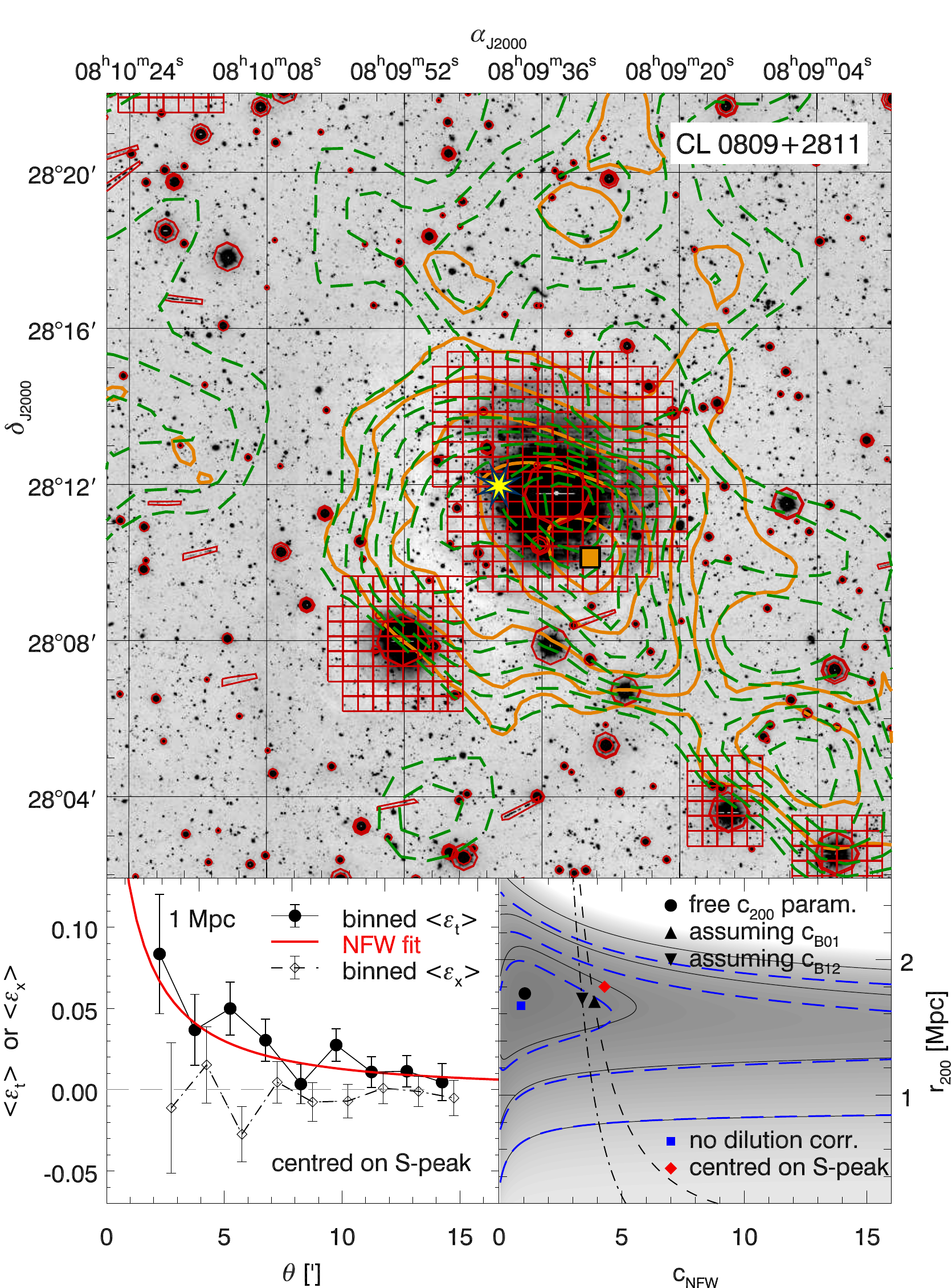}
 \caption{Like Fig.~\ref{fig:cl0159}, but for CL\,0809+2811.}
 \label{fig:cl0809}
\end{figure*}

\begin{figure*}
 \includegraphics[width=17cm]{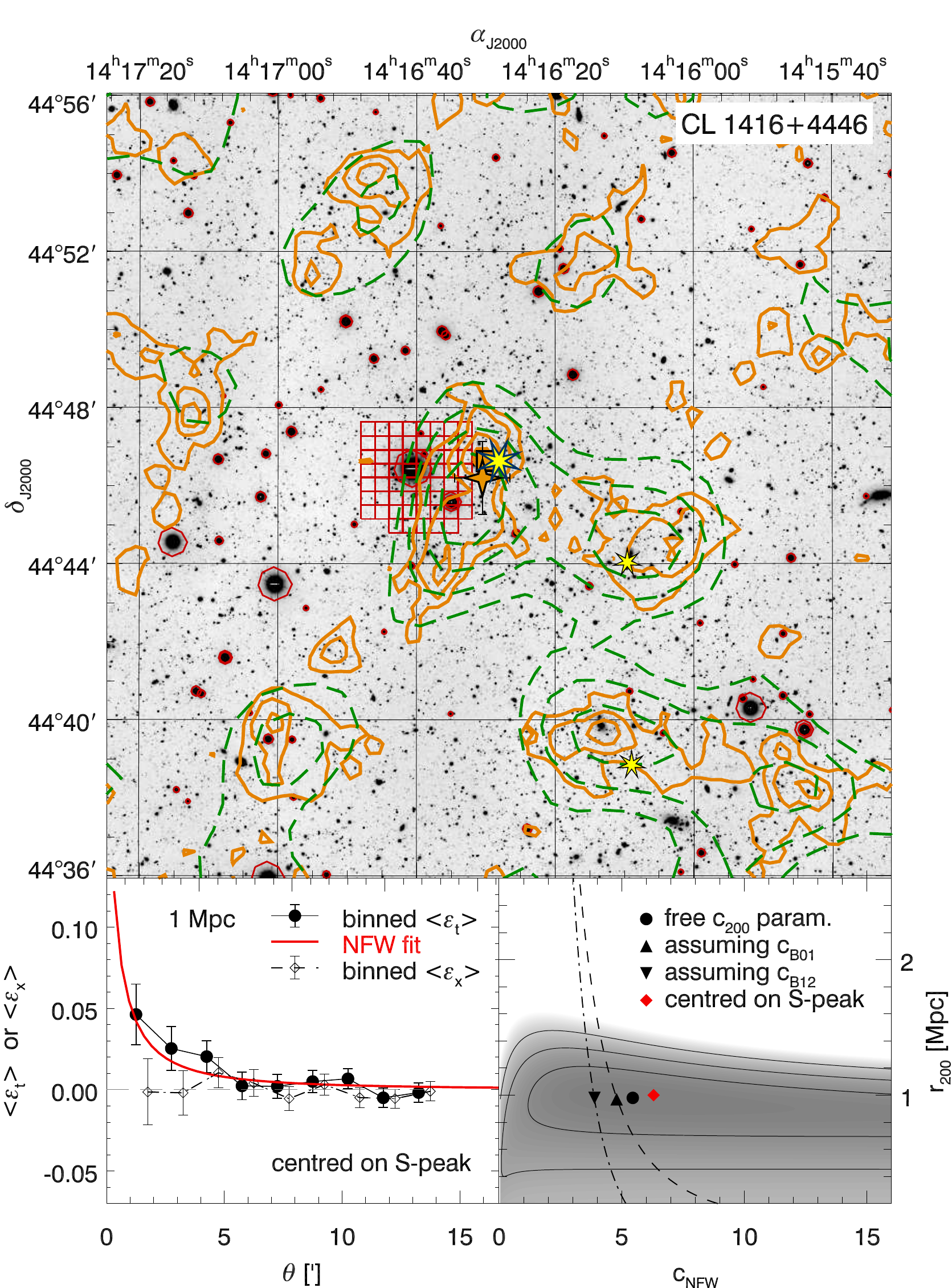}
 \caption{Like Fig.~\ref{fig:cl1357}, but for CL\,1416+4446. Small star symbols
  indicate the positions of further clusters in the field, which might be in
  physical connection to CL\,1416+4446  as parts of a super-cluster.}
  \label{fig:cl1416}
\end{figure*}

\subsection{CL\,0159+0030}

Being located in the SDSS equatorial strip, CL\,0159+0030 has been detected by
\citet{2002AJ....123.1807G} in the SDSS commissioning data, using their 
photometric ``cut and enhance'' cluster finder. 
\citet{2005ApJ...622L..17P} followed up \citet{2002AJ....123.1807G} cluster 
candidates using archival \textsc{XMM-Newton} observations. 
From the $3800\,\mbox{s}$ PN observation \citet{2005ApJ...622L..17P} 
analysed, only a $3\sigma$ upper flux limit of 
$2.1\times10^{-14}\,\mbox{erg}\,\mbox{cm}^{-2}\,\mbox{s}^{-1}$ in the 
$0.5$--$2.0\,\mbox{keV}$ energy range could be inferred. This non-detection 
disagrees both with the flux of 
$3.3\pm0.4\times10^{13}\,\mbox{erg}\,\mbox{cm}^{-2}\,\mbox{s}^{-1}$ 
\citet{2009ApJ...692.1033V} measure for CL\,0159+0030 with \textsc{rosat}
and with their \textsc{chandra} flux of 
$3.6\times10^{13}\,\mbox{erg}\,\mbox{cm}^{-2}\,\mbox{s}^{-1}$ in the same band.

\subsection{CL\,0230+1836}

For CL\,0230+1836, there are neither detections of the cluster itself,
independent from the \emph{400d} survey, nor other galaxy clusters
within a $20\arcmin$ radius listed in NED. To our knowledge, we are the first 
to study this high-$z$ cluster with deep optical observations.

\subsection{CL\,0809+2811}

We hypothesise that CL\,0809+2811 is identical to ZwCl\,0806.5+2822 at 
$\alpha_{\mathrm{J2000}}\!=\!08^{\mathrm{h}}09^{\mathrm{m}}34^{\mathrm{s}}$, 
$\delta_{\mathrm{J2000}}\!=\!+28\degr13\farcm1$, a position $1\farcm9$ off the 
CL\,0809+2811 \textsc{rosat} centre and at similar distance to the bright star 
in the field, where we do not see a concentration of galaxies. Neither do we 
observe an overdensity of galaxies at the position of a secondary shear peak 
with $S\!=\!2.9$ (Fig.~\ref{fig:cl0809}). It is located at
$\alpha_{\mathrm{J2000}}\!=\!08^{\mathrm{h}}09^{\mathrm{m}}08^{\mathrm{s}}$,
$\delta_{\mathrm{J2000}}\!=\!+28\degr05\arcmin22\arcsec$. No cluster within 
$3\arcmin$ of this position is known to NED.

\subsection{CL\,1357+6232}

\citet{2004AJ....128.1017L} conducted a cluster survey on digitised Second 
Palomar Observatory Sky Survey plates, using a Voronoi tesselation technique. 
In their catalogue, they quote a cluster of galaxies at
$\alpha_{\mathrm{J2000}}\!=\!13^{\mathrm{h}}57^{\mathrm{m}}22^{\mathrm{s}}$, 
$\delta_{\mathrm{J2000}}\!=\!+62\degr33\arcmin11\arcsec$, where there is no source 
in the \megacam ~image. Using the relation found between $r'$ magnitude, 
$g'\!-\!r'$ colour and $z_{\mathrm{spec}}$ for a subsample of clusters with 
spectroscopic redshifts, \citet{2004AJ....128.1017L} assign $z\!=\!0.19$ to 
their detection. (NSCS J135722+623311, their \#7243).
Noting that the position of NSCS J135722+623311 is only $16\arcsec$ from the 
\textsc{rosat} centre of CL\,1357+6232, we speculate that it might be the 
result of a confusion of CL\,1357+6232 with two bright galaxies to its east, one
of which (SDSS\,J135723.83+623246.1) has a measured redshift of $z\!=\!0.078$.

\subsection{CL\,1416+4446}

In addition to CL\,1416+4446, we detect two other shear peaks at $>\!3\sigma$
significance to the west and south-west of CL\,1416+4446.
\citet{2004AJ....128.1017L} list a cluster NSCS\,J141623+444558 in their 
catalogue which, by NED, is identified with CL\,1416+4446.
Furthermore, \citet{2004AJ....128.1017L} detected a cluster of galaxies at
$\alpha_{\mathrm{J2000}}\!=\!14^{\mathrm{h}}16^{\mathrm{m}}09^{\mathrm{s}}$ and  
$\delta_{\mathrm{J2000}}\!=\!+44\degr38\arcmin51\arcsec$, with a redshift of 
$z\!=\!0.39$.
Less than $2\arcmin$ north-east of these coordinates we find the south-western 
shear peak which coincides with the $g'\!=\!20.1$ galaxy 
SDSS\,J141613.33+443951.3. For this source, SDSS \citep{2008ApJS..175..297A} 
quotes a spectroscopic redshift of $z\!=\!0.397$.
Note that the brighter galaxy SDSS\,J141603.01+443725.1, located $2\arcmin$ 
further to the south-west from the \citet{2004AJ....128.1017L} cluster position
has an SDSS $z_{\mathrm{spec}}\!=\!0.310$ and does not correspond to an $S$-peak.

\citet{2006ApJ...645..955B} detected a galaxy cluster at 
$\alpha_{\mathrm{J2000}}\!=\!14^{\mathrm{h}}16^{\mathrm{m}}09\fs6$,  
$\delta_{\mathrm{J2000}}\!=\!+44\degr44\arcmin02\farcs4$, coincident with the
western shear peak, comparing archival \textsc{Chandra} data to optical $g'r'i'$
observations in the \textsc{Chandra} Multiwavelength Project.
They assign a redshift $z\!=\!0.427$ to the cluster, designated BGV~50.
In the same \textsc{Chandra} observation, \citet{2006ApJ...645..955B} 
identified another cluster, BGV~53 at
$\alpha_{\mathrm{J2000}}\!=\!14^{\mathrm{h}}16^{\mathrm{m}}27\fs6$ and  
$\delta_{\mathrm{J2000}}\!=\!+44\degr52\arcmin44\farcs4$ and a redshift of
$z\!=\!0.452$, which does not correspond to a bright galaxy in the \megacam 
~image or a peak in the $S$-statistics.

Out of the three confirmed clusters in the field, CL\,1416+4446 not only is the
only \emph{400d} X-ray cluster and the strongest lensing detection, but also 
appears to be the optically richest system in the \megacam ~$r'$-band image. 
Therefore, CL\,1416+4446 possibly presents the most massive system in a 
physically interacting super-structure, indicated by the $z\!\approx\!0.40$ 
redshifts of all mentioned clusters.
Judging by the X-ray morphology, \citet{2009ApJ...692.1033V} classify 
CL\,1416+4446 as a relaxed system, though. We conclude that the CL\,1416+4446 
field qualifies as an interesting candidate for further investigation 
concerning the existence of a super-cluster.

\subsection{Cross-identifications of CL\,1701+6414} \label{sec:detcl1701}

\citet{1998ApJ...502..558V}, on whose \emph{160d} \textsc{rosat} catalogue the 
\emph{400d} sample builds \citep[cf.][]{2007ApJS..172..561B} detect four
clusters in the field: VMF~189 (A~2246), VMF~190 (CL\,1701+6414), VMF~191, and 
VMF~192.
In an independent \textsc{rosat} analysis, \citet{2002ApJ...569..689D} detect
these same four clusters plus RX\,J1702+6407, which we do not detect in WL.
The redshift of $z\!=\!0.7$ found for the  
\citet{2002ApJ...569..689D} optical counterpart of CL\,1701+6414 deviates from 
the redshift of $z\!=\!0.45$ measured by \citet{2007ApJS..172..561B} and all 
other references. CL\,1701+6414 is further listed as RX\,J1701.3+6414 in the 
\textsl{Bright Serendipitous High-Redshift Archival \textsc{rosat} Cluster} 
sample \citep[Bright SHARC,][]{2000ApJS..126..209R}.

\end{document}